\documentclass[]{rsos}

\def\aj{Astron. J}%
\def\apj{Astrophys J}%
\def\apjl{Astrophys. J}%
\def\apjs{Astrophys. J. Suppl. Ser}%
\def\apss{Astrophys. Space Sci}%
\def\aap{Astron. Astrophys}%
\def\aaps{A\&AS}%
\def\mnras{Mon. Not. R. Astron. Soc}%
\def\pasa{PASA}%
\def\pre{Phys.~Rev.~E}%
\def\pasp{Publ. Astron. Soc. Pacific}%
\def\pasj{PASJ}%
\def\ssr{Space~Sci.~Rev}%
\def\zap{Zeitschrift f\"ur Astrophysik}%
\def\nat{Nature}%
\begin{document}

\title{Chemical element transport in stellar evolution models}

\author{
Maurizio Salaris$^{1}$ and Santi Cassisi$^{2,3}$}

\address{$^{1}$Astrophysics Research Institute, Liverpool John Moores University, IC2 Liverpool 
Science Park, 146 Brownlow Hill, L3 5RF, Liverpool, UK\\
$^{2}$INAF -- Osservatorio Astronomico Collurania, Via Mentore Maggini, 64100 Teramo, Italy\\
$^{3}$Instituto de Astrof{\'i}sica de Canarias, Via L{\'a}ctea s/n. E38205 - La Laguna, Tenerife, Spain}

\subject{Astrophysics/Stars}


\corres{Maurizio Salaris\\
\email{M.Salaris@ljmu.ac.uk}}

\begin{abstract}
Stellar evolution computations provide the foundation of several methods applied to study the evolutionary 
properties of stars and stellar populations, both Galactic and extragalactic. 
The accuracy of the results obtained with these techniques 
is linked to the accuracy of the stellar models, and in this context the correct  
treatment of the transport of chemical elements is crucial. Unfortunately, in many respects calculations of 
the evolution of the chemical abundance profiles in stars 
are still affected by sometime sizable uncertainties. 
Here, we review the various mechanisms of element transport included in 
the current generation of stellar evolution calculations, how they are implemented, the free parameters and uncertainties involved, 
the impact on the models, and the observational constraints. 
\end{abstract}


\begin{fmtext}
\section{Introduction}
Almost a century ago Eddington wrote \cite{eddington}:

\lq{It is reasonable to hope that in a not too distant future we shall be competent to understand so simple a thing as a star}\rq.

During this time the theory of stellar evolution has been developed and established, and its main predictions confirmed by a large 
number of empirical tests that have involved photometric, spectroscopic and asteroseismic observations 
(and solar neutrino flux measurements, after the discovery of neutrino oscillations). Results obtained from stellar model computations 
are nowadays widely used to develop a vast array of techniques to estimate distances, ages, star formation histories and chemical evolution of 
star clusters and galaxies, both resolved and unresolved \cite{sc:05}. 

\end{fmtext} 
\maketitle

Obtaining this kind of information from techniques rooted in stellar evolution calculations
is a crucial step to address problems like understanding the mechanisms 
that drive the formation and evolution of galaxies.
The accuracy of results gathered from stellar population analyses is tied to the accuracy of the current generation 
of stellar models; in this respect one particularly thorny issue is how to treat the transport of chemical elements 
in stellar evolution calculations. The problem is that mixing and element transport do not arise 
from the solution of the equations of stellar structure and evolution, instead they have to be \lq{added}\rq \ following 
recipes that often --as we will see-- involve a number of free parameters and/or are subject to sizable uncertainties.

On the other hand, the temporal evolution of the chemical abundance profiles within stellar models 
is a main evolutionary driver, and can be in principle tested through spectroscopic observations 
of photospheric abundances of key elements, asteroseismic observations (study of non-radial pulsations of stars, 
that can test thermodynamical properties of stellar interiors that are also affected by the chemical composition), 
and more indirectly through the effect on star counts 
(sensitive to evolutionary timescales) and evolutionary paths in
colour-magnitude-diagrams (CMDs) or 
Hertzsprung-Russell diagrams (HRDs), that are all affected by the internal chemical profiles.

The aim of this review is to discuss the various mechanisms of chemical element transport included in 
the current generation of stellar models, 
their effect of the evolutionary properties of the models, and the various prescriptions found in the literature, that 
often produce very different results. 
This will allow the reader to appreciate the main uncertainties involved, and what properties of stellar models 
are affected the most.

We start in Sect.~\ref{equations} with a brief summary of the equations of stellar structure and evolution, 
and an overview of the main evolutionary properties of models of different initial masses, 
to set the stage for the discussion that follows. 
The next sections discuss the implementation in stellar modeling, the associated uncertanties and 
the observational constraints of processes like 
convection (Sect.~\ref{convection}), semiconvection  (Sect.~\ref{semiconvection}),  
thermohaline mixing (Sect.~\ref{thermohaline}), atomic diffusion (Sect.~\ref{diffusion}), phase separation 
upon crystallization (Sect.~\ref{wds}), rotationally induced element transport 
(Sect.~\ref{rotation}). 
Section~\ref{synergy} discusses briefly an example of how a combination of several of the 
mechanisms discussed in the previous sections can explain the puzzling trend of 
photospheric Li abundances with effective temperature in open clusters, and conclusions follow in Sect.~\ref{conclusions}. 

\section{Stellar model computation}
\label{equations}

With {\sl stellar modelling} we mean here the calculation of the run of physical (i.e. luminosity, 
temperature, density, specific heats) and chemical quantities, from 
the centre to the photosphere of a star of a given initial mass and chemical composition, and their evolution with time.
Despite tremendous advances in computing power and computational techniques during the last decades, a full detailed 
modelling of a star by solving the equations of radiation hydrodynamics in 3D 
is still unfeasible, and this will be the case for the foreseeable future. 
This inability to model stars with multidimensional radiation-hydrodynamics is a consequence of the extreme range of spatial and 
temporal scales\footnote{For example main sequence core convection of massive short-lived models involves 
$\approx 10^9$ turnover times to complete the exhaustion of fuel. The Reynolds number in stars is $Re > 10^9- 10^{10}$, 
much larger than the effective Reynolds number $\sim 10^4$ in the current best-resolved simulations.} 
that need to be resolved when calculating full evolutionary models covering all stages from the pre-main sequence 
(pre-MS) to the last white dwarf (WD) or pre-supernova phases. Current hydrodynamics computations are however starting  
to be able to provide some guidelines about mixing processes \cite{am:16}, as highlighted in the following sections.

For these reasons complete stellar evolution computations still have to rely basically on the 
following \lq{classical}\rq\ set of 1D equations for spherical, non 
rotating and non-magnetic stars 
(equation of continuity of mass, hydrostatic equilibrium\footnote{In case of pre-supernova fast evolutionary phases the equation of 
hydrostatic equilibrium usually contains an extra term $-(1/4 \pi r^2) (\partial^2 r/\partial t^2)$ called the 
{\sl acceleration term}.}, energy generation and energy transport, respectively) 
and Raphson-Newton solution methods \cite{henyey}:

\begin{equation}
\frac{\partial r}{\partial m}= \frac{1}{4 \pi r^{2} \rho} \label{eqss1}
\end{equation}

\begin{equation}
\frac{\partial P}{\partial m}= -\frac{G m}{4 \pi r^4} \label{eqss2}
\end{equation}

\begin{equation}
\frac{\partial L}{\partial m}=\epsilon_n - \epsilon_{\nu} + \epsilon_g
\label{eqss3}
\end{equation}

\begin{equation}
\frac{\partial T}{\partial m}=-\frac{G m T}{4 \pi r^4 P} \nabla 
\label{eqss4}
\end{equation}

where the independent variable $m$ is the mass enclosed 
within radius $r$, and $T$, $L$, $P$ and $\rho$ are temperature, luminosity, pressure 
and density at the layer specified by the value of $m$. 
The coefficient $\epsilon_{\nu}$ denotes the energy per unit time and unit mass carried away by neutrinos 
(that do not interact with the stellar gas),  
$\epsilon_n$ the energy per unit time and unit mass produced by nuclear reactions. For a generic 
nuclear reaction $n_A A + n_b b \rightarrow $products involving elements $A$ and $b$ 
with mass fractions $X_A$ and $X_b$ and atomic weights $A_A$ and $A_b$      

$$\epsilon_n=R_{Ab} Q_{Ab}$$

where  $Q_{Ab}$ is the amount of energy released by a single reaction, and $R_{Ab}$ is the number of reactions per unit mass and 
unit time, given by 

\begin{equation}
R_{Ab} = \rho^{n_A + n_b
-1} \frac{X_A^{n_A} X_b^{n_b}} {A_A^{n_A} A_b^{n_b}}
\frac{<\sigma v>_{Ab}}{m_H^{n_A+n_b} n_A ! n_b !}
\end{equation}

where $<\sigma v>_{Ab}$ is the reaction cross section\footnote{In case $A$ and $b$ are identical particles, the factor  
$n_A ! n_b !$ has to be replaced with $(n_A + n_b) !$
}.

The coefficient $\epsilon_g$ represents the 
so-called {\sl gravitational energy} produced per unit time and unit mass, and is given by

\begin{equation}
-\epsilon_g=\left(\frac{\partial U}{\partial v}\right)_{T, \mu} \frac{\partial v}{\partial t}
+ \left(\frac{\partial U}{\partial T}\right)_{v, \mu} \frac{\partial T}{\partial t} +
\left(\frac{\partial U}{\partial \mu}\right)_{T, v} \frac{\partial \mu}{\partial t}+P
\frac{\partial v}{\partial t}
\label{epsg}
\end{equation}

where $U$ is the internal energy per unit mass, $v={\rm 1/\rho}$ the specific volume, and $\mu$ 
the mean molecular weight of the stellar matter. 
The term $(\partial U/\partial \mu)_{T, v} (\partial \mu/\partial t)$ arises from the variation
of $U$ at constant temperature and volume due to the change of chemical abundances. Its contribution to the energy 
budget is negligible when nuclear reactions are efficient,
but is very important in case of WDs, where nuclear burnings are inefficient.
When integrated over the whole stellar structure, 
$\epsilon_g$ is equal to the time variation of the internal energy plus the gravitational potential energy of 
the star \cite{sc:05, kww}.

For the case of radiative plus electron conduction energy transport, 
the gradient $\nabla \equiv (d \ ${\rm ln}$(T)/d \ ${\rm ln}$(P))$ is set to $\nabla_{rad}$ 
\begin{equation}
\nabla_{rad}=\frac{3}{16\pi acG} \frac{\kappa L P}{m T^4} \label{eq6d}
\end{equation}

where $a$ is the radiation density constant, $c$ the speed of light, $G$ the gravitational constant, 
$\kappa$ the Rosseland opacity, including also the contribution of electron conduction when appropriate.
In case of convective energy 
transport a theory of convection is needed to calculate the appropriate temperature 
gradient $\nabla_{\rm conv}$ (see Sect.~\ref{convection}).

These equations are complemented by a set of $I$ equations
($s=1, .., I$) for the change of the mass fraction of the $I$ 
chemical elements considered at the layer specified by $m$.
Consider first the changes due just to nuclear reactions, 
an element $s$ is produced by $w$ reactions of the type 
$$n_h h + n_k k \rightarrow  n_p \ s$$ 
and destroyed by $l$ reactions 
$$n_d \ s + n_j j \rightarrow n_z z$$ 
This provides the following equation for the variation of the abundance of $s$\footnote{A term $-X_s/\tau_d$ has to be added 
to the right-hand side of Eq.~\ref {eqss5} in case element $s$ decays with decaying constant $\tau_d$.}

\begin{eqnarray}
\frac{\partial X_{s}}{\partial t} = \sum_w \rho^{n_h + n_k
-1} n_p \frac{X_h^{n_h} X_k^{n_k} A_{s}} {A_h^{n_h} A_k^{n_k}}
\frac{<\sigma v>_{hk}}{m_H^{n_h+n_k-1} n_h ! n_k !} - \nonumber \\
\sum_l \rho^{n_d + n_j -1} n_d \frac{X_{s}^{n_d} X_j^{n_j} A_{s}}
{A_{s}^{n_d} A_j^{n_j}} \frac{<\sigma v>_{sj}}{m_H^{n_d+n_j-1}
n_d ! n_j !} \label{eqss5}
\end{eqnarray}

\begin{figure}[!h]
\centering\includegraphics[width=5.0in]{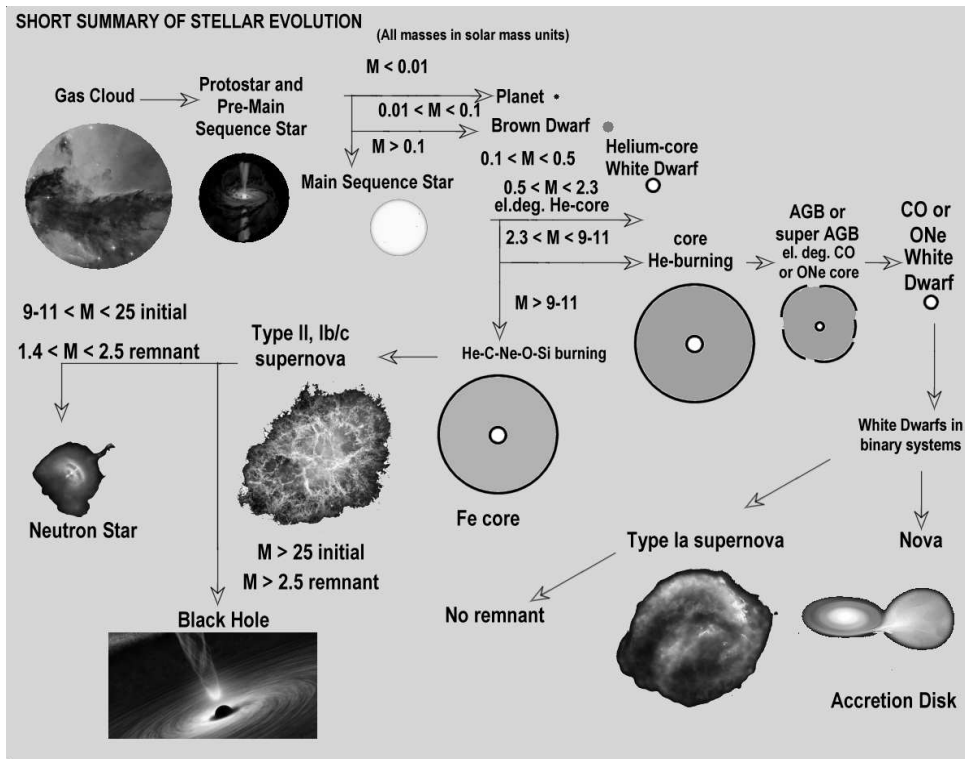}
\caption{General evolutionary paths of single stars with different initial masses. The exact values of the mass ranges depend on the 
initial chemical composition and the details of the adopted element transport mechanisms. We show also some final product of 
the evolution of interacting binaries (Novae, Type Ia supernovae). He-core WDs from single stars can form only on timescales much longer 
than a Hubble time, but are produced nowadays by interacting binary systems. There are observational indications 
that massive stars with initial mass above $\sim$20$M_{\odot}$ might not explode as supernovae, rather collapse directly to black hole 
\cite{smartt}.}
\label{stellarev}
\end{figure}

The complete system of equations (Eqs.~\ref{eqss1}, \ref{eqss2}, \ref{eqss3}, \ref{eqss4} and $I$-times Eq.~\ref{eqss5}) 
is solved at a given time $t$ 
considering the \lq{Lagrangian}\rq\ independent variable $m$, with $r, L, P, T$ as unknowns, once the stellar mass and initial chemical 
composition are specified, 
and prescriptions for the equation of state of the stellar gas, Rosseland mean  
opacities, nuclear reaction cross section and energy generation rates, and neutrino production rates are provided.
The chemical composition is usually denoted 
by $X$, $Y$, $Z$, that correspond to the mass fractions of H, He and all other elements collectively called 
\lq{metals \rq}, respectively. A relative distribution of the metal abundances needs also to be specified.

Notice that the chemical abundance profile enters explicitely Eqs.~\ref{eqss5} and $\epsilon_n$, 
and affects the coefficients $\epsilon_{\nu}$, $\epsilon_g$, 
the opacities $\kappa$, and the equation of state, that all depend on the chemical composition of the stellar matter.  

Figure~\ref{stellarev} shows an overview of the evolutionary paths of single stars with different initial masses (plus the main 
byproducts of interacting binary evolution), as derived from complete stellar evolution models. This general evolutionary framework is 
solid and does not depend on the details of element transport modelling although the precise values 
of the various mass ranges do (and they depend also on the initial chemical composition). It constitutes a reference guideline for 
the discussions presented in the following sections.

It is clear that chemical element transport does not arise naturally from the equations of stellar structure and evolution. 
This makes it necessary to first identify all possible mechanisms --in addition to the nuclear reactions-- 
able to change the chemical abundances at a given mass layer, and then to develop 
formalisms for their implementation in the 1D equations (a major difficulty).
The element transport mechanisms described below will add extra terms to the right-hand side of 
Eqs.~\ref{eqss5}, in addition to the terms describing the effect of nuclear reactions.

\section{Convection}
\label{convection}

Besides radiation and electron conduction, convection is the third fundamental mechanism for energy transport in stars.  
It involves organized large scale motions of matter, that in addition to carrying energy 
are also a very efficient source of mixing. It can be envisaged as a flux of matter from deeper --hence hotter-- stellar layers 
moving vertically outward into cooler layers, and material from
cooler outer layers flowing down to hotter inner layers. 

The implementation of this element (and energy) transport in stellar models requires first a criterion for the onset of the convective instability, 
then a mathematical treatment to predict the main physical properties of convective regions.

\subsection{Instabilities in non-rotating stars} 

Matter inside the stars is never at rest, but usually the gas is locally subject to small random perturbations around equilibrium
positions. Under certain conditions, these small random perturbations can trigger large scale motions that involve sizable fractions of the total 
stellar mass. These large scale motions are
called convection, and are the equivalent of the motion of water elements in a kettle heated from below. 

The treatment of convection in stellar interiors is extremely complicated and requires the 
introduction of various approximations.
This stems from the fact that the flow of gas in a stellar convective region is highly turbulent, forcing us 
to adopt simplified models that provide only mean 
approximate values for the properties of the flow of gas. The so-called {\sl mixing length theory} (MLT) is the 
local, time-independent convection model almost universally used in stellar evolution calculations, that we will discuss below. 
Firstly we show how a simplified linear analysis is sufficient to determine the main criteria for the onset of mixing (not only convective) 
in stellar interiors \cite{kato:66}.

\begin{figure}[!h]
\centering\includegraphics[width=5.0in]{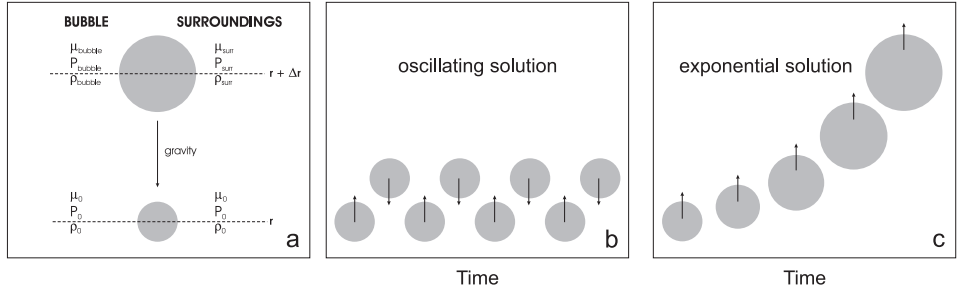}
\caption{{\sl Panel a}: Set-up of the simple analysis for the onset of instabilities (see text for details). 
{\sl Panel b}: Stable situation. A blob of gas displaced vertically by a small amount oscillates around its equilibrium position. 
{\sl Panel c}: Unstable situation. After the initial displacement the blob of gas continues to rise as time goes on. }
\label{conv1}
\end{figure}

We consider a gas element at rest at a distance $r$ from the star centre. This gas bubble will have a
pressure $P_0$, temperature $T_0$, density $\rho_0$ and mean molecular weight $\mu_0$ 
(the molecular weight is the mean mass of the gas particles in atomic mass units) equal to those of the environment, supposed to be in 
radiative equilibrium (in this
section \lq{radiative}\rq\ actually means \lq{radiative plus conductive}\rq) as shown in Fig.~\ref{conv1}. If random motions displace the bubble by a small 
amount $\Delta
r$ away from the equilibrium position, the equation of motion for an element of unit volume can be written as (assuming the viscosity is negligible)
$$\rho \frac{d^2 \Delta r}{dt^2}=-g \Delta \rho$$ where ${\rm \Delta \rho}$ is the difference ${\rm (\rho_{bubble}-\rho_{surr})}$ 
between the bubble 
(supposed to have
constant density) and the surroundings, and $g$ is the local acceleration of gravity. One reasonable assumption is that the motion 
of the bubble is fast enough that
all time derivatives of the mean stellar properties are equal to zero.

As a consequence of the displacement from its equilibrium position, two distinct physical situations can occur, as shown in Fig.\ref{conv1}. 
If the region is convectively stable, the displaced gas parcel experiences a restoring force that moves it back towards the original position, 
as illustrated in the intermediate panel of Fig.\ref{conv1}. Indeed, the blob is subject to a stable oscillation around an equilibrium position  
with a frequency named the Brunt-V{\"a}is{\"a}l{\"a} frequency (see below).
If the bubble of material at position $r+\Delta r$ is less dense than the surrounding material, it will continue to be pushed upwards by 
buoyancy forces and the region is then said to be {\sl convectively unstable}, as shown in panel c of Fig.\ref{conv1}. Similarly, if 
the bubble displaced at a radial distance $r-\Delta r$ is denser than the surrounding material, its motion will continue because 
it is heavier than the local environment.

To determine which solution applies to a given layer within a star we 
assume that along the displacement $\Delta r$ the bubble is always in pressure equilibrium with the surroundings, that is, 
$\Delta P=(P_{bubble}-P_{surr})$=0, and that the molecular weight of the bubble 
$\mu_{bubble}$ is always equal to its initial value $\mu_0$ (there is no matter exchange with the surroundings, hence the gas parcel retains its identity).
The assumption of pressure equilibrium means that the motion of the bubble has to happen with a speed lower than the local sound speed. 
When tere is a molecular weight gradient $d \mu/d
\rho$ throughout the region, the difference $\Delta \mu=(\mu_{bubble}-\mu_{surr})$ will be equal to 
$\Delta \mu=\mu_0-[\mu_0+(d \mu/dr)\Delta r]$, hence ${\rm \Delta \mu=-\frac{d \mu}{dr} \Delta r}$.

Using the relationships $d ${\rm ln}$(P)=(1/P)dP$ and $d ${\rm ln}$ (\mu)=(1/\mu)d\mu$ one gets 

$$\Delta \mu=-\mu (d {\rm ln}(\mu)/d
{\rm ln}(P)) (d {\rm ln}(P)/dr) \Delta r$$
 
Differentiating with respect to time one obtains

$$\frac{d \Delta \mu}{dt}=-\mu \frac{d {\rm ln}(\mu)}{d {\rm ln}(P)} \frac{d
{\rm ln}(P)}{dr}\frac{d \Delta r}{dt}$$

The temperature difference $\Delta T=(T_{bubble}-T_{surr})$ depends on the difference between the temperature gradients in the bubble and 
in the surroundings 
and the rate of temperature change due to energy losses from the bubble (due for example to thermal diffusion)  
whose efficiency will depend on a parameter we denote as $\zeta$, 
hence 

$$\Delta T=\left[ \left(\frac{dT}{dr}\right)_{ad}
-\left(\frac{dT}{dr}\right)_{rad}\right] \Delta r-\zeta \Delta T
dt$$

By introducing the notation ${\rm \nabla \equiv (d {\rm ln}(T)/d{\rm ln} (P))}$ and differentiating with respect 
to time, one obtains

$$\frac{d \Delta T}{dt}= T \frac{d{\rm ln}(P)}{dr}(\nabla_{\rm ad}- \nabla_{\rm rad})\frac{d \Delta r}{dt}-\zeta \Delta T$$


If ${\rm \Delta P}$=0, and in the assumption that the differences ${\rm \Delta T}$, ${\rm \Delta \rho}$ and ${\rm \Delta \mu}$ are
small, we obtain from the equation of state

$$\chi_{\rho}\frac{\Delta \rho}{\rho}+\chi_T \frac{\Delta T}{T}
+\chi_{\mu} \frac{\Delta \mu}{\mu}=0$$

where

$$\chi_{\rho}=(d {\rm ln}(P)/d {\rm ln} (\rho))_{T, \mu}, 
\chi_{T}=(d {\rm ln}(P)/d {\rm ln} (T))_{\rho, \mu}, \chi_{\mu}=(d
{\rm ln}(P)/d {\rm ln} (\mu))_{\rho, T}$$


We have derived in this way the following set of 4 homogeneous
equations for the four unknowns $\Delta T$, $\Delta
\rho$, $\Delta \mu$ and $\Delta r$:

\begin{equation}
\rho \frac{d^2 \Delta r}{dt^2}+g \Delta \rho=0 \label{eq1}
\end{equation}
\begin{equation}
\frac{d \Delta \mu}{dt}+\mu \frac{d {\rm ln}(\mu)}{d {\rm ln}(P)} \frac{d
{\rm ln}(P)}{dr}\frac{d \Delta r}{dt}=0 \label{eq2}
\end{equation}
\begin{equation}
\frac{d \Delta T}{dt} +T \frac{d{\rm ln}(P)}{dr}(\nabla_{\rm rad}-
\nabla_{\rm ad})\frac{d \Delta r}{dt}+\zeta \Delta T=0 \label{eq3}
\end{equation}
\begin{equation}
\chi_{\rho}\frac{\Delta \rho}{\rho}+\chi_T \frac{\Delta T}{T}
+\chi_{\mu} \frac{\Delta \mu}{\mu}=0 \label{eq4}
\end{equation}

One can search for solutions of the form $\Delta x=A
e^{nt}$. By inserting into the respective equations this
functional dependence for $\Delta T$, $\Delta \rho$,
$\Delta \mu$ and $\Delta r$, a non trivial
solution is found when the determinant derived from the coefficients of
$A_T$, $A_{\rho}$, $A_{\mu}$ and $A_{r}$
is equal to zero, giving

\begin{equation}
n^3+\zeta n^2 + \left[g \frac{\chi_{T}}{\chi_{\rho}}
\frac{d{\rm ln}(P)}{dr}
\left(\nabla_{\rm rad}-\nabla_{\rm ad}+\frac{\chi_{\mu}}{\chi_{T}}
\nabla_{\mu}\right)\right] n +\left(\zeta g
\frac{\chi_{\mu}}{\chi_{\rho}}
\frac{d{\rm ln}(P)}{dr}\nabla_{\mu}\right)=0 \label{eq5}
\end{equation}

where we have defined $\nabla_\mu \equiv d{\rm ln}(\mu)/d{\rm ln}(P)$

The condition that at least one of the $n$ is real and positive or imaginary with a positive real part (unstable solutions) 
is given by the Hurwitz criterion, resulting in at least one of the following conditions to be satisfied 
(we recall that the pressure $P$ always increases towards the star centre, $\chi_{\mu}$ is negative, ${\rm \chi_T}$ 
and $\chi_\rho$ positive)

\begin{equation}
\Delta_\mu < 0 \label{eq7}
\end{equation}
 \begin{equation}
\nabla_{\rm rad} > \nabla_{\rm ad}-\frac{\chi_{\mu}}{\chi_{T}}\nabla_{\mu} \equiv \nabla_{\rm L} \label{eq8}
\end{equation}
\begin{equation}
{\rm\nabla_{\rm rad} > \nabla_{\rm ad}} \label{eq9}
\end{equation}

If $\Delta_\mu < 0$ ($\mu$ increasing towards the surface) the medium is always unstable. 
When the gas parcel is displaced upwards (downwards) by a small distance $\Delta r$, its density will be lower (higher) 
than the environment, and will continue to be pushed upwards (downwards) by buoyancy. The temperature difference between the displaced mass element 
and its surroundings suppresses or favours this displacement, depending upon the difference between $\nabla_{\rm rad}$ and $\nabla_{\rm ad}$ (more below).
For the more common case 
of $\Delta_\mu \geq 0$ (heavier elements are usually synthesized by nuclear reactions in the central regions) we should consider 
Eq.~\ref{eq8} --the so-called 
\lq{Ledoux criterion \rq}-- and Eq.~\ref{eq9} --the so-called 
\lq{Schwarzschild criterion \rq}. The second term on the right-hand-side of Eq.~\ref{eq8} is positive for a positive $\nabla_{\mu}$ 
(the composition gradient has a stabilizing effect) 
so that if the gradients satisfy the Ledoux criterion, the Schwarzschild criterion is automatically satisfied, hence 
the Schwarzschild criterion determines the presence of an unstable medium. 

These instability criteria are \lq{local \rq}, in the sense that they can be applied layer-by-layer without accounting for 
non-local effects that can be however relevant when dealing with convective mixing. 

\begin{figure}[!h]
\centering\includegraphics[width=3.0in]{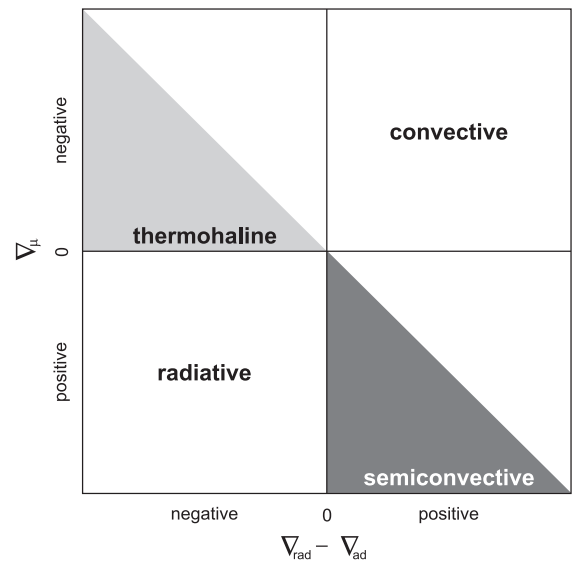}
\caption{Sketch of the $\nabla_\mu - (\nabla_{\rm rad} - \nabla_{\rm ad})$ stability plane with different regimes labelled (see text for details). The diagonal line 
dividing into half the top left and bottom right diagrams denotes $\nabla_{\rm L}$}.
\label{nablas}
\end{figure}

Figure~\ref{nablas} displays a qualitative sketch in the $\nabla_{\mu}$-$(\nabla_{\rm rad}-\nabla_{\rm ad})$ diagram of the region where 
instabilities occur, divided into four quadrants. The only stable region in this diagram is the bottom left quadrant, where 
$\nabla_\mu \geq 0$ and $\nabla_{\rm rad} < \nabla_{\rm ad}$. All other quadrants are unstable regions, although the type of mixing 
involved depends on the exact values of the gradients. If $\nabla_\mu \geq 0$ and $\nabla_{\rm ad} < \nabla_{\rm rad} < \nabla_L$ 
(lower right quadrant) the instability is called \lq{semiconvection \rq}. 

The linear analysis in the 
semiconvective regime shows that if the gas bubble is displaced upwards and $\zeta$=0 
its internal temperature will be slightly larger compared to the surroundings, hence its lower density will favour a continuation of the 
displacement. On the other hand $\mu$ of the raising bubble will be larger than the environment and in the semiconvective regime of the gradients 
this effect will prevail, causing the return of the mass 
element to its starting position. When $\zeta>0$ the temperature of the element returning from above is smaller than its initial value at $r$, therefore 
the restoring force is larger upon returning than when the bubble was leaving its starting position. As a consequence the bubble returns back to 
a radial location $r$ with a larger velocity than when it left for the upwards motion. On the downward excursion the same effect (with an opposite sign) 
will operate and the amplitude of these oscillation will progressively increase due to the increasing radiative losses during the oscillatory motion.

The growth of these oscillations (overstability) will lead to chemical mixing and thus decrease or destroy the stabilizing gradient $\nabla_{\mu}$. 
Results from numerical simulations and laboratory experiments show a more complex picture \cite {merryfield:95, w:13}, 
with in some cases the formation of well-mixed fully convective layers, separated by stratified interfaces, 
but the linear analysis suffices to highlight the peculiarity of the mixing associated with semiconvection. 

The efficiency (timescale) of semiconvective mixing is difficult to estimate and depends on the efficiency of the bubble energy losses;  
an analysis of the growth rate in the framework of the linear analysis shows that as a result of semiconvective mixing 
$\nabla_{\rm rad} \sim \nabla_{\rm ad} $ \cite{langer:83} and this is another approach to treat semiconvective mixing 
in stellar evolution calculations, i.e. to impose a mixing efficiency such that $\nabla_{\rm rad} \sim \nabla_{\rm ad}$ (actually 
this was the original way in which semiconvection was implemented \cite{sh:58}) 
in a semiconvective region (see Sect.~\ref{semiconvection}).

If $\nabla_{\mu} < 0$ and $\nabla_{\rm rad} < \nabla_{\rm L}$ (top left quadrant) the instability is 
usually called a \lq{thermohaline \rq} instability. If $\zeta$=0 a gas bubble displaced downwards will have a larger 
$\mu$ than the environment, but also a larger temperature than the radiative environment. The combined effect is that the bubble density is  
lower than that of the environment, and buoyancy will push back the displaced gas element. However, if $\zeta >0$, the energy loss 
will eventually decrease the bubble temperature enough to induce a further displacement downwards due to the effect of the larger $\mu$.

This instability is controlled by the heat leakage of the displaced element, and it is observed for example when 
a layer of hot salt water lies over a layer of cold fresh water. 
Upon displacing downwards a blob of hot salt water, due to the fact that heat transfer by molecular collisions is 
typically faster than the motion of chlorine and sodium ions that cause the composition to equilibrate, the sinking blob 
will be able to come into thermal equilibrium with the surrounding medium faster than achieving composition equilibrium. 
Given that salt water is 
heavier than fresh water at the same temperature, the blob will continue to sink in the surrounding fresh water. As this motion continues, the 
medium develops \lq{fingers}\rq\ of salt water reaching down into the fresh water. 
Inside a star, the role of salt can be played by a heavier element like helium, in a hydrogen-rich medium.
This is an example of so-called \lq{doubly diffusive instabilities}\rq, because it involves the diffusion of two different components (particles and heat).  

In general, a convectively unstable region mixes matter on very short timescales compared with evolutionary timescales 
(either nuclear or Kelvin-Helmholtz timescales), and 
the chemical profile in convective layers can always be assumed uniform to a very good approximation (instantaneous mixing approximation), 
with abundances of individual elements equal to values averaged over the whole convective region. 
If the convective region
extends from mass layer $m_1$ (inner boundary) to mass layer
$m_2$ (outer boundary) within the star, inside this region
the abundance $X_s$ of a generic element $s$ is 
constant. At the boundaries
(one or both of them) one may have a discontinuity between the
homogeneous convective chemical profile and the profile in the 
radiative regions, for example due to nuclear
reactions (previous and/or present). Due to these effects and just to give an example, the
time evolution of $X_s$  (in the approximation of instantaneous mixing) within an expanding convective 
shell is to a first order given by

\begin{equation}
\frac{dX_s}{dt}=\frac{1}{\Delta
m}\left[\int_{m_1}^{m_2}\frac{dX_s}{dt}dm +
\frac{dm_2}{dt}(X_{s2}-X_s)
-\frac{dm_1}{dt}(X_{s1}-X_s)\right] \label{eqchemev3}
\end{equation}

where $\Delta m=m_2-m_1$, $X_{s1}, X_{s2}$ are the
abundances on the radiative side of the discontinuities at,
respectively, the inner and outer boundary of the convective
region. The first term in the integral describes the
variation due to the nuclear burnings (if efficient), whereas the other two terms 
describe the change in composition when the boundaries
of the convective zone move into surrounding regions of -- in
principle -- inhomogeneous composition.

Exceptions to the validity of the instantaneous mixing in convective regions 
are the advanced evolutionary phases of massive stars about to explode as Type II supernovae 
(see Fig.~\ref{stellarev}),
the production of Li due to the Cameron-Fowler mechanism in the envelopes of red giant branch stars \cite{cw:71},
proton ingestion episodes into the intershell convection zone of low metallicity asymptotic giant branch (AGB) stars \cite{cct96, sdl11}.
In this case 
the nuclear burning timescales in the convective regions are comparable to convective mixing timescales. This case is 
usually treated with a time-dependent convective mixing 
discussed in the next subsection.  

Before closing this section we just introduce a quantity that will appear often in the rest of the paper. 
In case of $\zeta$=0, gas elements in a stable region will oscillate around their equilibrium position with 
a frequency called Brunt-V{\"a}is{\"a}l{\"a} frequency, usually denoted with $N$. 
This can be derived easily from Eq.~\ref{eq5}, and is equal to

\begin{equation}
N^2 = -g \frac{d {\rm ln}P}{dr} \frac{\chi_T}{\chi_{\rho}} \left( \nabla_{\rm rad} - \nabla_{\rm ad}+\frac{\chi_{\mu}}{\chi_{T}}\nabla_{\mu} \right)
\end{equation}

Another expression for this frequency (that we will use in the rest of the paper) involves the derivatives  
$\delta = -(\partial \ln \rho / \partial \ln T)_{P,\mu}$ and $\phi = (\partial \ln \rho / \partial \ln \mu)_{P,T}$, 
and the pressure scale height $H_P \equiv -dr/d{\rm ln}(P)=P/(\rho g)$ (where $g$ is the gravitational acceleration), and is  

\begin{equation}
N^2 = N^2_T + N^2_\mu = \left( \frac{g \delta }{H_P}(\nabla_{\rm ad} -\nabla) +  \frac{g \phi}{H_P} \nabla_\mu \right)
\end{equation}

Using $\delta$ and $\phi$, the gradient $\nabla_L$ (Ledoux gradient) can be rewritten as

\begin{equation}
\nabla_{\rm L} \equiv (\nabla_{\rm ad} + \frac{\phi}{\delta} \nabla_\mu)
\end{equation}

\subsection{The mixing length theory of convection}

From the point of view of the evolution of chemical abundances, the instantaneous mixing approximation does not 
require a model for stellar convection. However, this is necessary for calculating 
convective velocities and fluxes, and when 
a time-dependent description of convective mixing is required.
The formalism almost universally used in stellar evolution calculations is the MLT \cite{prandtl}, 
a simple, local, time independent model,
firstly applied to stellar modelling by \cite{bie:32}. The formulation by \cite{bv:58} is 
usually employed in modern stellar evolution calculations. 

\begin{figure}[!h]
\centering\includegraphics[width=2.0in]{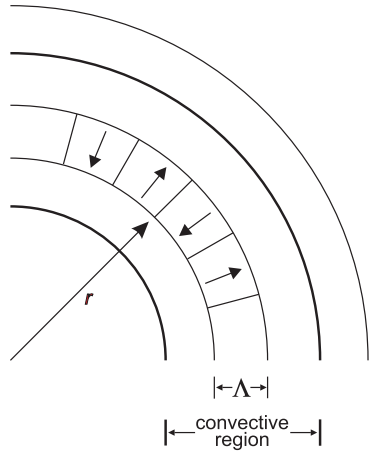}
\caption{Schematic illustration of the MLT approximation to convective motion. The mixing length $\Lambda$ corresponds to the characteristic radial distance scale over 
which rising and falling convective elements move before merging with the surrounding medium.}
\label{mlt}
\end{figure}

The basic idea of the MLT is to assume that the stellar fluid is composed of identifiable convective elements 
that move vertically in the gravitational field between regions of 
higher and lower temperature. Indeed, there is no net mass flow, but the effect is an outward transport of energy. The MLT assumes a characteristic distance 
over which bubbles rise before dissipating, the so-called mixing length $\Lambda$ (see Fig.~\ref{mlt}), and describes the motion of these bubbles 
over the characteristic scale $\Lambda$ under some general assumptions:

\begin{itemize}
\item{All bubbles have the same characteristic size that is of the same order as $\Lambda$;}
\item{$\Lambda$ is much smaller than any other length scale of physical significance in the star;}
\item{the physical properties, i.e. temperature, density, pressure and chemical composition, of the bubbles differ only slightly from the surrounding medium;}
\item{Pressure equilibrium with the environment is maintained. This means that the velocities of the convective elements 
are small compared with the local sound speed in the local environment.}
\end{itemize}

$\Lambda$ is assumed to be equal to a multiple of the local pressure scale height $H_p$, i.e. $\Lambda=\alpha_{MLT} H_P$, with $\alpha_{MLT}$ 
--the {\sl mixing length parameter}-- being a constant to be empirically
calibrated (usually by reproducing the solar effective temperature at the solar age with a solar model). 

The need for a mean free path in the simple framework of the MLT can be easily explained as follows. Let's consider the
cross sections of the rising and falling gas columns; if originally in a given convective layer the cross sections were the same, the rising gas 
(always in pressure
equilibrium with the surroundings) will expand by a factor $e$ after a distance equal to $H_P$. 
This means that at this point within the star there is much less space available for the
falling gas. On the other hand, the amount of falling material must be the same as the rising one, otherwise the star would either dissolve or concentrate 
all the mass in the interior, thus
violating the hydrostatic equilibrium condition. The only solution is that after a distance $\Lambda$ of the order of $H_P$ part of the material 
stops and inverts its motion.

By relying on the previous assumptions, the MLT provides the following equations 
for the velocity of the convective elements $v_{\rm c}$, the convective flux $F_{\rm c}$ and the convective efficiency $\Gamma$ (defined as 
the ratio between the excess heat content of a raising convective bubble just before its 
dissolution, and the energy radiated during its lifetime) \cite{tas:90}

\begin{equation}
v_c^2=\frac{a \Lambda^2 g \delta (\nabla_{\rm c} - \nabla^{'})}{H_p}
\label{eqml21}
\end{equation}
\begin{equation}
F_c=\frac{b \rho {\rm v}_c c_p T \Lambda (\nabla_{\rm c} - \nabla^{'})}{H_p}
\label{eqml22}
\end{equation}
\begin{equation}
\Gamma\equiv \frac{\nabla_{\rm c}-\nabla^{'}}{\nabla^{'}-\nabla_{ad}}=\frac{4 c_p \rho^2 \Lambda v_{\rm c} \kappa}{c^2 a T^3}
\label{eqml23}
\end{equation}

where $\nabla^{'}$ is the temperature gradient of a rising (or falling) element of matter within the convective 
region, and $\nabla_{\rm c}$ is the average temperature gradient of all the matter at a given level 
within the convective zone (the quantity needed to solve the stellar structure equations). 
There are three additional free parameters besides $\alpha_{MLT}$, i.e.  $a$, $b$ and $c$ that are usually fixed a priori and 
define the MLT \lq{flavour}\rq\ \cite{sc:08} \footnote{The values assumed in the classical and widely employed MLT flavour 
\cite{bv:58} are $a$=1/8, $b$=1/2, and $c$=24. Solar models calculated with this choice of the MLT flavour usually require 
$\alpha_{MLT}$ of the order of 1.6-2.0, the exact value depending on the specific physics inputs and treatment of the outer boundary conditions 
in the calculations. Remarkably enough, different choices of these three parameters found in the literature provide the same results 
when $\alpha_{MLT}$ is recalibrated on the Sun.}
There are 4 unknowns, namely $v_{\rm c}$, $F_{\rm c}$, $\nabla_{\rm c}$, $\nabla^{'}$ and the three previous equations plus an additional equation 
arising from the fact that the total flux to be transported by radiation plus convection is known from the solution of the 
stellar evolution equations

\begin{equation}
\frac{L}{4 \pi r^2}=\frac{4acgT^4 m}{3\kappa P r^2} \nabla + F_{\rm c}
\label{eq11}
\end{equation}

If convection is efficient in the deep stellar interiors, the MLT 
provides $\nabla \rightarrow \nabla_{ad}$ and velocities of
the order of 1 -- 100 ${\rm m\,s^{-1}}$, many orders of magnitude
smaller than the local sound speed. On the contrary, in convective
layers close to the surface the gradient is strongly
superadiabatic and velocities are much larger, of the order of 1
-- 10 ${\rm km\,s^{-1}}$, close to the local sound speed.

External or inner regions of a stellar model are convective in the following cases:

\begin{itemize}

\item{Large values of the opacity $\kappa$. Given that $\nabla_{\rm rad} \propto \kappa F$, where 
$F$ is the energy flux, $\nabla_{\rm rad}$ tends to increase above $\nabla_{ad}$.   
The radiative opacity generally increases with decreasing temperature --for a given chemical composition-- hence this 
situation occurs most commonly in the cooler outer layers of stars;}

\item{If the energy generation rate in the star is very sensitive to the temperature, then the energy flux $F$ rises rapidly 
as $r$ approaches zero in the stellar centre. This large heat flow can eventually cause 
$\nabla_{\rm rad}$ to increase above $\nabla_{ad}$. 
This situation occurs only in stellar cores, when the nuclear energy generation rate is very 
sensitive to temperature, as is the case of the H-burning CNO-cycle, He-burning and more advanced nuclear burning stages;}

\item{In ionisation zones the adiabatic gradient 
can decrease below the typical value $\nabla_{\rm ad}\sim 0.4$. 
Also, in these regions the radiative opacity usually increases. Due to these effects, one can expect the ionisation zones 
located in the outer layers to be convective.}

\end{itemize}

When the convective mixing timescales are comparable to nuclear burning timescales, a diffusive approach is usually employed to 
follow the chemical evolution \cite{ce:76}. This means that an extra-term is added to the right-hand side of Eq.~\ref{eqss5} 
for a given element $s$, i.e. the right-hand side of the following diffusion equation 

\begin{equation}
\frac{\partial X_s}{\partial t}\bigg|_{M_r} = \frac{1}{\rho r^2}\frac{\partial}{\partial r}\Big(
D_{\rm c}\rho r^2\frac{\partial X_s}{\partial r}\Big),
\label{overdiff}
\end{equation}

where the diffusion coefficient associated to the convective transport is taken to be $D_{c}=\frac{1}{3} \alpha_{MLT} \ v_{\rm c} \ H_P$, 
using the value of $v_{\rm c}$ derived from the MLT (see \cite{weaver:78, langer:85, vc:05} for examples). 

The MLT is very appealing for its mathematical simplicity (despite the free parameters involved) and the fact that it relies just on 
local quantities, hence it is easy to include in stellar evolution codes and does not affect the stability of the 
numerical solution of the stellar evolution equations.   
Once $\alpha_{MLT}$ is calibrated on the Sun, the resulting HRDs (and CMDs) 
of the models generally reproduce the observations satisfactorily within the current errors\footnote{But see also 
\cite{bonaca} who claim that $\alpha_{MLT}$ should vary with stellar properties, using data from the ${\sl Kepler}$ satellite.}. 
Comparisons of the effective temperature evolution of models calculated with a solar $\alpha_{MLT}$ and 
with a calibration of $\alpha_{MLT}$ obtained from 3D radiation hydrodynamics simulations of stellar envelopes \cite{tr:14}  
display an agreement within 30-50~K \cite{sc:15}\footnote{Another calibration of $\alpha_{MLT}$ from 
an independent suite of 3D radiation hydrodynamics simulations of stellar envelopes \cite{mwa:15} has not been implemented yet in stellar evolution calculations.}.
On the other hand, helioseismic data (the study of non-radial oscillations of the Sun  
\cite{cd:02}) clearly point to shortcomings of the MLT 
description of the physical structure of convective regions \cite{pate:93}. 

An alternative local description of convection included in stellar evolution calculations with the ATON code (\cite{vdm:08}) 
is presented in \cite{cm:91}. The main difference from the MLT is that the convective elements have a spectrum of sizes,  
and the scale length of the convective motions is set to be equal to the distance to the closest convective boundary. 
Stellar evolution tracks calculated with this convection model show a different evolution of the effective temperature $T_{eff}$ 
compared to calculations with the MLT, especially along the red giant branch phase. Differences increase with increasing initial metallicity.
This convection model also suffers from shortcomings when compared with helioseismic results \cite{pate:93}.

There are very sophisticated non-local Reynolds stress models that describe not only convection, but also (see below) 
semiconvection and overshooting \cite{canuto:92, canuto:93, xiong:97, ly:07, yl:07, canuto:11}. They 
introduce a large number of equations to be coupled to the stellar structure equations and several free parameters to be calibrated 
observationally, and are generally not included in current stellar evolution modelling.

\begin{figure}[!h]
\centering\includegraphics[width=3.0in]{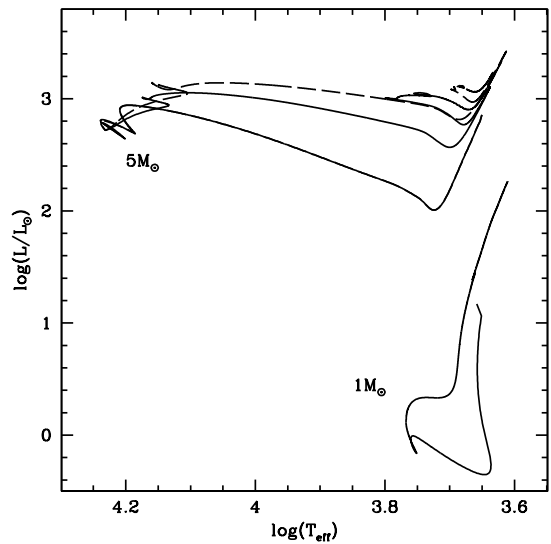}
\caption{HRD of $1M_{\odot}$ and $5M_{\odot}$ stellar evolution models with initial solar chemical composition. 
The $5M_{\odot}$ model is computed until the end of He-core burning. The $1M_{\odot}$ model stops during the RGB evolution. 
The dashed line displays the HRD of the same  $5M_{\odot}$ model calculated with
MS core overshooting (0.2$H_p$ -- see text for details).}
\label{tracksconv}
\end{figure}

\begin{figure}
  \resizebox{0.5\linewidth}{!}{\includegraphics{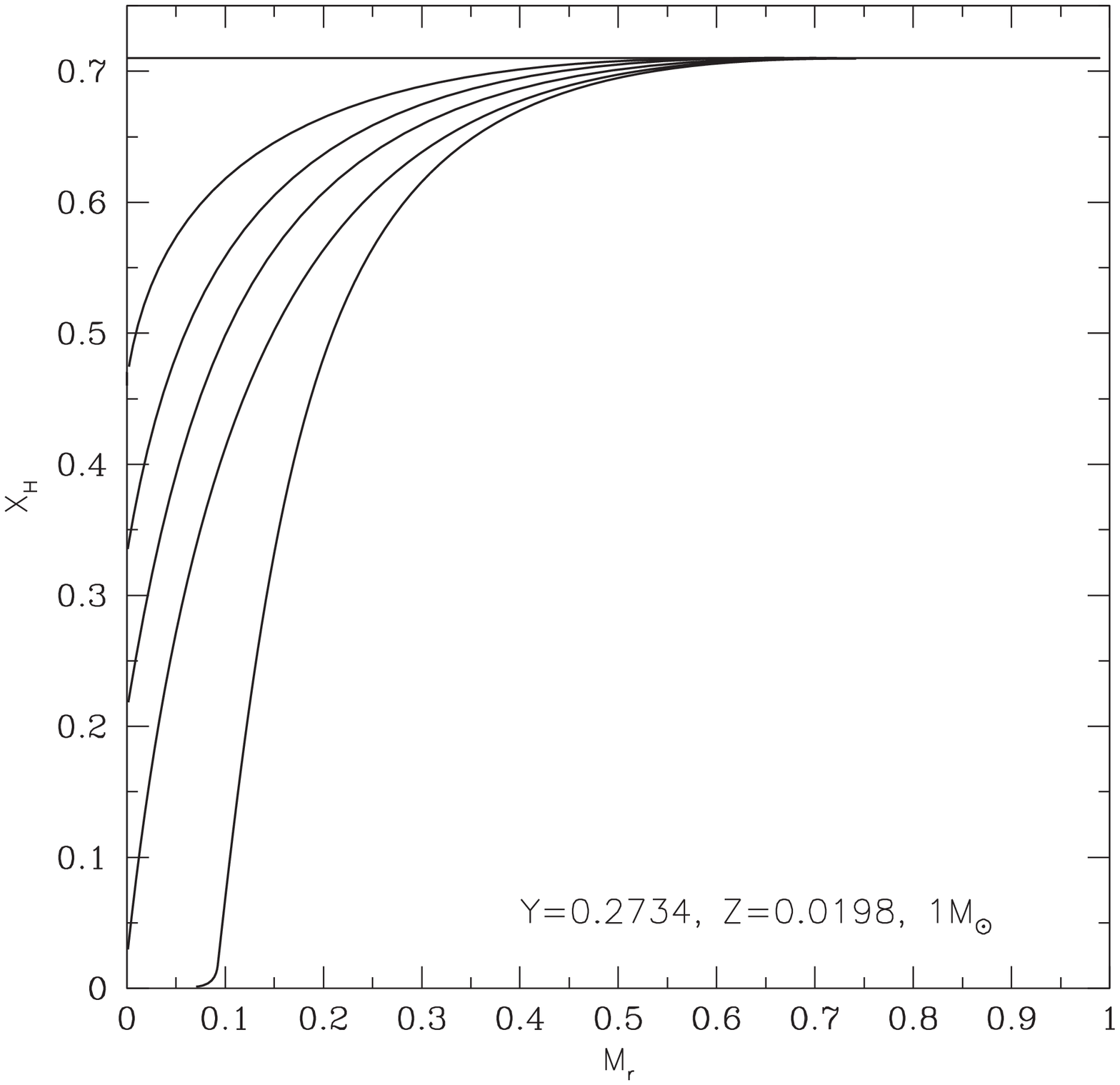}}
  \resizebox{0.5\linewidth}{!}{\includegraphics{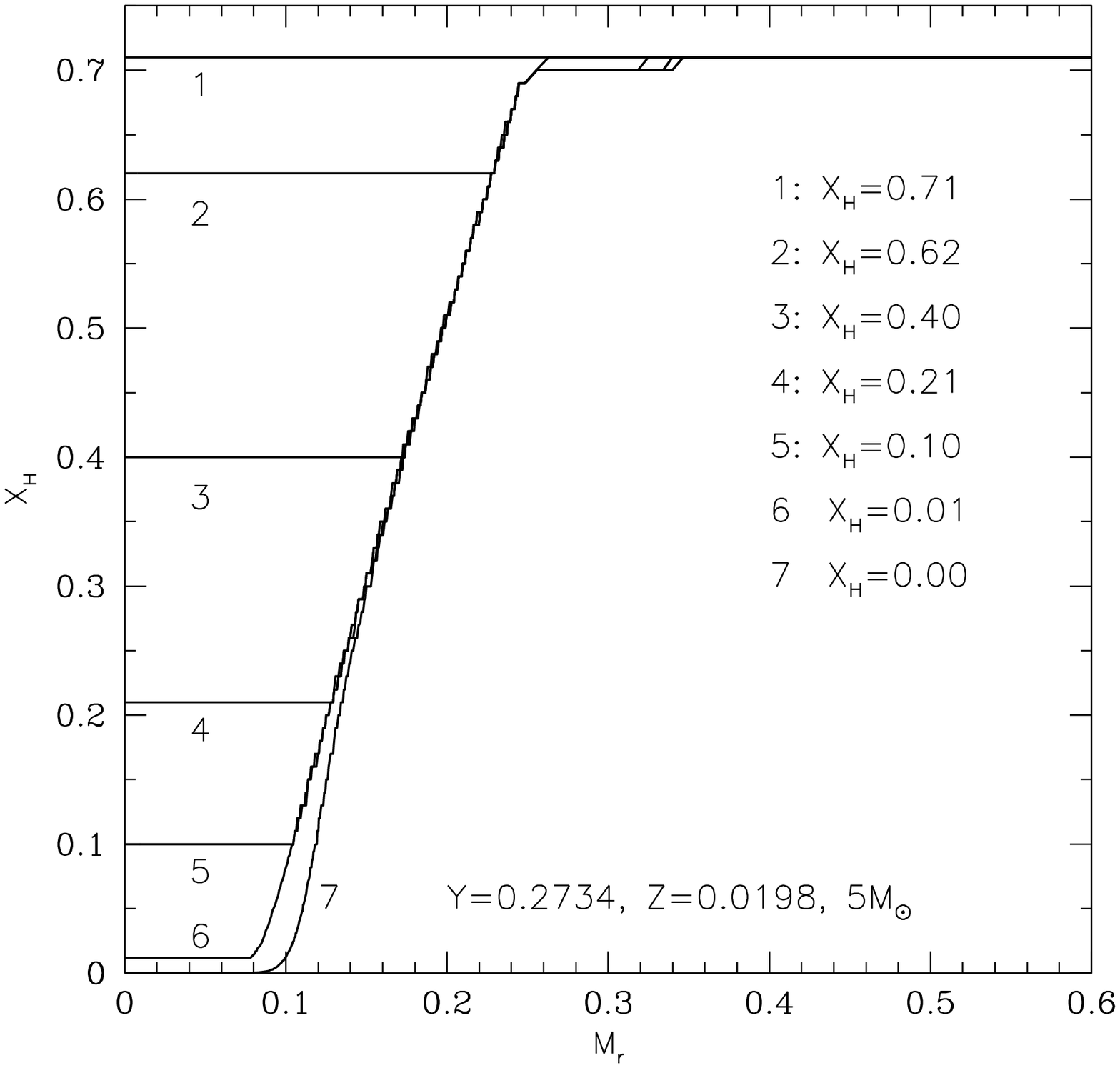}}
\caption{$Left:$ The evolution with time of the H-abundance profile (H-abundance as a function 
of the local fractional mass) within the $1M_{\odot}$ model of Fig.~\ref{tracksconv}, during the MS phase.
$Right:$ The same, but for the $5M_{\odot}$ model without core overshooting. 
The different numbers label a temporal sequence with decreasing values of the H-abundance in the mixed core.}
\label{Hprofiles}
\end{figure}

Convection plays a major role in determining the evolutionary properties of stellar models. Figure~\ref{tracksconv} shows the HRD 
of models with (the $5M_{\odot}$ model) and without (the $1M_{\odot}$ model) convective cores during the MS. The shape of this evolutionary 
phase in the HRD is completely different between the two tracks, because of the different time-evolution 
of the H-profile in the burning region, as shown by Fig.~\ref{Hprofiles}. In the inner radiative layers of the $1M_{\odot}$ model the H-abundance profile changes smoothly 
during the MS evolution, and at any given time the H-abundance increases gradually from the core  
towards the more external layers where the burning becomes progressively less efficient. 
In the convective inner layers of the $5M_{\odot}$ model --where the burning takes place-- the H-abundance profile is uniform, and the 
progressive retreat with time of the outer border of the convective core, produces at a given time $t$ a H-abundance profile flat 
in the innermost regions, then increasing outwards.

Surface convection (present everywhere along the HRD evolution, apart from the 
MS phase of the  $5M_{\odot}$ model) is very important when comparing the surface abundances measured from spectroscopy with the models, 
because of the dredge-up phenomenon, whereby the fully mixed convective envelopes reach layers processed by nuclear burnings, 
hence altering their chemical composition during the red giant branch (RGB) and AGB phases \cite{sc:05}.

\subsection{Convective overshooting}

As already mentioned, the criteria for the onset of convection are local. The boundaries of the convective regions are fixed 
at the layer where the random motions of the gas do not get amplified.
At this convective border the acceleration of the gas elements is zero, but 
not the velocity. One expects therefore the chemically mixed region to extend into the formally stable layers in case of 
both core and envelope convection. 
In real stars mixing beyond the formal boundary most likely results from the interplay of several physical processes 
\cite{as:13, as:15, viallet:15}, 
grouped in stellar evolution modelling under the \lq{umbrella}\rq\ term {\sl overshooting} (or sometimes {\sl convective boundary mixing}). 
These overshooting processes 
are modelled very crudely by introducing free parameters to be calibrated on observations of eclipsing binaries \cite{stancliffe:15, valle16} 
(typically 
comparisons with masses and radii of both components under the assumptions that they are coeval) and open clusters \cite{dsg94}(the shape of the 
MS-turn off region).

The standard approach is to assume that overshooting into the stable layers does not affect their thermal structure, hence 
the temperature gradient stays radiative. The composition is then mixed instantaneously between the formal convective border 
and layers at a distance $\lambda H_P$ from this border, where $\lambda$ ($\lambda<1$) is a free parameter to be observationally 
calibrated, and $H_P$ is the 
pressure scale height at the convective border. This is the approach taken for example in the BaSTI \cite{basti}, DSEP \cite{dsep}, Yale-Yonsei 
\cite{yy} codes. 

A similar approach of instantaneous mixing is taken in \cite{vandenberg:06}. 
The following integral criterion is employed to fix the extent of the convective core overshooting, inspired 
by a constraint on the maximum possible extension of the overshooting region developed by \cite{roxb:89}

\begin{equation}
\int_0^{r_{\rm cc}}F_{\rm over}(L_{\rm rad}-L_{\rm N})\frac{1}{
 T^2} \frac{d\,T}{d\,r} d\,r + \int_{r_{\rm cc}}^{r_{\rm ov}}(2-F_{\rm
 over})(L_{\rm rad}-L_{\rm N})\frac{1}{ T^2} \frac{d\,T}{d\,r} d\,r = 0~
\label{eq:rox1}
\end{equation}

where $F_{\rm over}$ is a free parameter (between 0 and 1) to be calibrated against 
observations. The two luminosities 
$L_{\rm rad}=-((16 \pi a c T^3 r^2)/(3 \kappa \rho)) dT/dr$ and $L_{\rm N}$ represent, respectively, the local 
radiative luminosity and the total luminosity produced by nuclear reactions within radius $r$, whilst 
$r_{cc}$ is the radius of the convective core boundary, 
and $r_{\rm ov}$ the radius of the outer boundary of the overshooting region, to be determined through this equation.

The various releases of Padua stellar evolution models \cite{padua} and the PARSEC 
code \cite{parsec} consider instead an adiabatic gradient for the overshooting region (this is the case of 
{\sl penetrative convection} according to \cite{zahn:91}, because the overshooting material is assumed to be able 
to change the entropy stratification), and 
the spatial size of the overshooting region is determined as follows \cite{bressan:81}.
Starting from each radial distance $r_i$ from the centre inside the formally convective region, the following equation (based on the MLT) 
is integrated outwards, 
up to $r=r_i+l$, where $l= \lambda H_p$, $H_p$ being the pressure scale height at the convective boundary, and $\lambda$ ($\lambda<1$) 
a free parameter 

\begin{equation}
\frac{1}{3}\frac{\partial v_r^3}{\partial r}=\frac{g}{\kappa T} \frac{\chi_T}{\chi_{\rho}} \frac{F_{\rm c}}{c_P \rho}-\frac{g}{\mu} \frac{\chi_{\mu}}{\chi_{\rho}} 
\Delta \mu  \ v_r
\end{equation}

with $\Delta \mu=\mu(r)-\mu(r_i)$. 
The convective flux $F_{\rm c}$ is determined as $F_{\rm c}=F-F_{\rm rad}$ where $F=L/(4 \pi r^2)$ is the total energy flux and 
$F_{\rm rad}=-((4 a c T^3)/(3 \kappa \rho)) dT/dr$. In the overshooting region, where the actual gradient is set to adiabatic, $F_{\rm rad}>F$ 
hence $F_{\rm c}<0$, mimicking the fact that the convective elements penetrating into the stable layers are formally cooler 
than the surrounding medium.

At any given $r$ the convective elements originated in the range between $r-l$ and $r$ will display a range of velocities, and the maximum 
value $v_{m}$ attained at each $r$ is taken to derive the run of $v_m(r)$ as a function of $r$. The border of the instantaneously mixed region is 
then taken at the radial distance for which $v_m$=0.

In all these approaches the free parameters that determine the spatial extent of the convective core overshooting region needs to 
decrease in the regime of small convective cores. The reason is that ${\rm H_P \rightarrow\infty}$ as $r \rightarrow 0$, hence 
the smaller the size of the formally convective core, the larger the size of the actual mixed core (including overshooting region), 
with no smooth transition 
of the shape of MS evolutionary tracks from the convective core to the radiative core case. Observations of the MS turn off (TO) 
of open clusters of various 
ages confirms the need to decrease to zero the extent of the overshooting regions when the stellar mass decreases, in 
the regime of small convective cores (masses below $\sim1.5 M_{\odot}$ --see, e.g., \cite{vandenberg:06}). This means 
that the assumed trend of $\lambda$  (or $F_{\rm over}$)  with mass, 
for masses with small convective cores, introduces an additional degree of freedom, albeit affecting only a 
restricted range of stellar masses.
In this context, following an integral constraint on the maximum possible extent of overshooting regions \cite{roxb:89, roxb:92}, 
it has been proposed to limit the extent of the overshooting region to 15\% of the radius of the formally convective core 
to generate a smooth transition from models with convective cores to models with radiative cores on the MS \cite{wd:01}. 

Additionally, one can find in the literature two very different approaches to the mixing beyond the formal convective boundary.

\begin{enumerate}

\item{A diffusive approach  
\cite{herwig:00} used for example in the STARS \cite{schroder:97}, MESA \cite{mesa11, mesa13} and GARSTEC \cite{garstec} codes.
The element transport beyond the formal convective boundary is described as a diffusive process (avoiding the instantaneous mixing approximation)  
based on results of 2D radiation-hydrodynamics simulations of shallow stellar surface convection zones \cite{freytag:96}

\begin{equation}
\frac{\partial X_i}{\partial t}\bigg|_{M_r} = \frac{1}{\rho r^2}\frac{\partial}{\partial r}\Big(
D_{\mathrm ov}\rho r^2\frac{\partial X_i}{\partial r}\Big)
\label{overdiff2}
\end{equation}

The diffusion coefficient $D_\mathrm{ov}$ is given by

\begin{equation}
	D_\mathrm{ov} = D_{\rm c} \exp\left(- {2z\over f \mathrm{H_p}}\right)
\label{overcoeff}
\end{equation}

where $D_{\rm c}$ is the diffusion coefficient inside the convective region 
($D_{\rm c}=(1/3) \alpha_{MLT} v_{\rm c} H_P$), $z$ 
is the distance from the convective boundary, $H_P$ is the pressure scale height at the convective boundary and $f$ is a 
dimensionless free parameter.
Typical calibrated values of $f$ are $\sim0.01-0.02$.}

\item{Modelling of the mixing as \lq{ turbulent entrainment \rq}, following the simulations by 
\cite{meakin:07}, as employed in the stellar evolution calculations by \cite{staritsin:13}. 
The motion of matter in the zone of convective instability is turbulent, and the rising 
turbulent flow spreads horizontally near the boundary of the turbulent region. 
The interface between the convective region and the stable layers moves through the 
stable region due to the continuous involvement of
new layers in the turbulent motion. The velocity $V_e$ of the penetration of the convective turbulent 
boundary into the stable layers is determined by the \lq{turbulent-entrainment \rq} law written as

\begin{equation}
\frac{V_e}{V_t}= A Ri_B^{-n}
\end{equation}

where $V_t$ is a typical turbulent velocity at the boundary (that can be taken for example from the MLT), 
$A$ and $n$ are parameters that characterize the entrainment 
($A\sim 0.027$ and $n\sim 1.05$ according to the simulations by \cite{meakin:07}), $Ri_B$ the so-called 
bulk Richardson number defined as

\begin{equation}
Ri_B=\frac{l \Delta b}{V_t^2}
\end{equation}

with $l$ denoting the typical size of the eddies doing the entrainment (some fraction of the pressure scale height $H_p$ at the boundary 
of the mixed region) and

\begin{equation}
\Delta b = \int_{\Delta h} N^2 dr
\end{equation}

This integral is performed across a region of thickness $\Delta h$ that contains the convective boundary, $N^2$ is the 
the Brunt-V\"ais\"al\"a frequency.
When $V_e$ is determined, the distance $d$ over which this boundary shifts during an
evolutionary time step $\Delta t$ is given by
$d = V_e  \Delta t$.}
\end{enumerate}

Despite the uncertainties involved in its parametrization, the treatment of overshooting is very important because it affects 
the evolutionary properties of the models. For example core overshooting during the MS produces brighter models, an increased MS lifetime (because more 
fuel is available), larger He-core masses that induce a brighter and shorter lived He-burning phase, and also less extended loops in the HRD 
(see Fig.\ref{tracksconv}). Overshooting below convective envelopes can alter the surface abundances after the dredge-up episodes \cite{spp:15}, 
affect the luminosity of the RGB bump (see Sect.~\ref{thermohaline}), and also the 
extension of the loops in the HRD (in case of the 5$M_{\odot}$ model in Fig.\ref{tracksconv}, the loop during the core He-burning phase 
would become more extended with overshooting also from the convective envelope).
Age estimates of young-intermediate age clusters are obviously affected by the amount of overshooting included in the models, i.e. the larger the 
overshooting region the older the age estimate of a given cluster.
Overshooting also plays an important role during C-burning in super AGB stars, affecting the propagation of the carbon burning flame \cite{doherty}. 
Too large overshooting at the base of the convective C-burning region 
can prevent the flame from reaching the centre, thus producing an hybrid CONe core, and eventually a CONe WD.

It has also been shown that the diffusive approach to mixing in the overshooting region provides different results in terms of evolutionary times 
compared to instantaneous mixing, because of a slower addition of extra fuel from the overshooting region when this scheme is implemented  
\cite{vc:05}.

\section{Semiconvection}
\label{semiconvection}

After convection, semiconvection (called \lq{ double-diffusive convection}\rq\ in oceanography) is 
probably the most significant element transport mechanism in non-rotating stellar evolution models. Below are the two major cases 
where semiconvective transport is efficient, and how this is implemented in stellar models\footnote{Semiconvection can also be important during 
Si-burning \cite{whw:02} in massive stars.}.

\subsection{H-burning phase with convective cores}

\begin{figure}[!h]
\centering\includegraphics[width=3.0in]{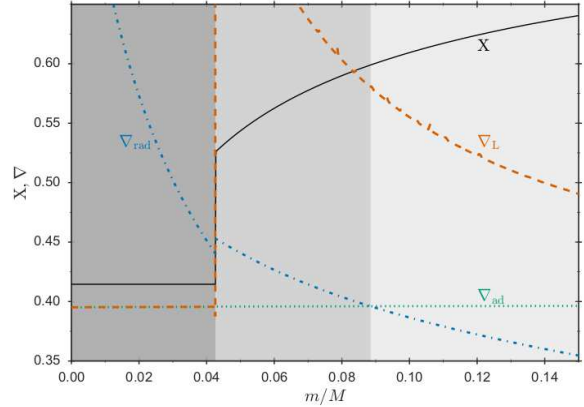}
\caption{Profiles of the hydrogen mass fraction $X$ and the radiative, adiabatic and Ledoux gradient 
as a function of the fractional mass coordinate, within a 1.5$M_{\odot}$ model during the MS phase. 
From left to right, different tones of grey mark convective, semiconvective, 
and radiative layers, respectively (courtesy of V. Silva-Aguirre).}
\label{trackssemiconv}
\end{figure}

There is a large body of literature that addresses the issue of semiconvection in massive stars with convective cores during the MS 
\cite{sh:58, chiosi:70, stothers:70, simpson:71, langer:83}. In a nutshell, layers left behind by shrinking convective cores during the 
MS will have a hydrogen abundance that increases with increasing radius (the chemical composition of each layer is determined by 
the composition of the convective core at the moment the
layer has detached from the retreating mixed region). They are characterized by $\nabla_{\rm ad} < \nabla_{\rm rad} < \nabla_{L}$, 
and a treatment of semiconvective mixing is needed (see Fig.~\ref{trackssemiconv}). 
There is also a narrow range of masses around $1.5 M_{\odot}$ (the values  
depending on the initial chemical composition) with increasing mass of the convective core during part of the MS, where a narrow 
semiconvective region forms right above the fully mixed core \cite{sag:11}.

The efficiency of element transport in this semiconvective region has important consequences for the morphology of the TO of the tracks --very  
efficient semiconvection mimics the case of overshooting beyond the formal convective boundary and no semiconvection\cite{sag:11}--  
and the post-MS evolution of massive stars \cite{sc:75, sc:76} because it changes the H-profile above the H-exhausted region.
To provide a guideline, for massive stars inefficient mixing favours core He-ignition on the cool (red) side of the HRD, whilst efficient mixing 
favours the ignition on the hot (blue) side of the HRD (and increases MS evolutionary timescales). 
One could think that comparisons with the observed ratio of blue to red supergiants (B/R ratio) could provide strong constraints on the efficiency of 
semiconvective mixing in massive stars. However, other factors like the extent of the overshooting region (that acts in the direction 
to reduce the size of the semiconvective layers), and rotation will affect 
the HRD location (and eventually loops) during the He-burning phase \cite{langer:95}. At the moment no set of theoretical models seem to be able to match 
observational constraints on the B/R ratio, that is also strongly affected by mass transfer in binaries, which make up most
of the massive star population.

Semiconvective mixing is included in these evolutionary calculations following various recipes. A traditional approach was to 
iterate the composition in the individual semiconvective layers until $\nabla_{\rm rad}=\nabla_{\rm ad}$ locally \cite{sh:58,sc:75,sc:76}.
More modern calculations adopt again a diffusive approach, including a chosen semiconvective diffusion coefficient $D_{\rm SC}$. 
The coefficient from \cite{langer:83} is implemented in the codes STERN \cite{yl05}, MESA, GARSTEC. It is derived from a 
linear local stability analysis \cite{kato:66}, assuming the MLT to determine the velocity of the gas elements 
(using $\alpha_{MLT}$=1.5), and is given by 

\begin{equation}
	D_\mathrm{SC} = \frac{\alpha_{SC} K}{6 c_p \rho} \frac{\nabla-\nabla_{\rm ad}}{\nabla_{\rm L}-\nabla}
\label{sclanger83}
\end{equation}

where $K=(4acT^3)/(3 \kappa \rho)$ is the thermal conductivity, $\nabla$ is the actual temperature gradient, 
and the free parameter $\alpha_{SC}$ determines the mixing timescale (larger $\alpha_{SC}$ correspond to shorter mixing timescales). 
Values of $\alpha_{SC}$ currently used are of the order of $\sim 0.04-0.1$, calibrated on empirical constraints \cite{langer:85, langer:95}.
The value of $\nabla$ in the semiconvective region is determined from

$$ L=L_{\rm rad} (1 + (L_{SC}/L_{\rm rad})) $$

\begin{equation}
\frac{L_{SC}}{L_{\rm rad}} = \alpha_{SC} \frac{\nabla-\nabla_{\rm ad}}{2 \nabla (\nabla_L-\nabla)} \left[ (\nabla-\nabla_{\rm ad}) 
- \frac{\beta (8-3\beta)}
{32-24\beta-\beta^2} \nabla_{\mu} \right]
\end{equation}

where $\beta$ is the ratio of the gas pressure to the total pressure (gas plus radiation), 
$L$ is the total luminosity, $L_{\rm rad}$ the radiative luminosity that can be written as

$$ L_{\rm rad}= \frac{16}{3} \frac{\pi a c G m T^4}{\kappa P}  \nabla $$

Larger values of $\alpha_{SC}$ produce semiconvective temperature gradients  $\nabla$ increasingly close to $\nabla_{\rm ad}$. 

Another expression for $D_\mathrm{SC}$ has been derived in the assumption of layering of the semiconvective region, 
with nearly uniform composition in each layer, separated by thin boundary layers within which the chemical elements are transported 
by molecular diffusion alone \cite{spruit:92}:

\begin{equation}
	D_\mathrm{SC} = \sqrt{D_s K_T}(\frac{4}{\beta}-3) \frac{\nabla_{\rm rad}-\nabla_{\rm ad}}{\nabla_\mu}
\label{spr}
\end{equation}

where $K_T=K/(\rho c_P)$ is the thermal diffusivity, 
$\beta$ is the ratio of the gas pressure to the total pressure (gas plus radiation), and $D_s$ the diffusion coefficient 
of He due just to the He abundance gradient (see Sect.~\ref{diffusion}). For a chemical composition of essentially two elements 
(in our case H and He) with atomic numbers $Z_1$ and $Z_2$, masses $m_1$ and $m_2$ and number densities $n_1$ and $n_2$ 

$$D_s = \frac{3}{16n} \left(\frac{2 K_B T}{\pi m} \right)^{1/2} \left( \frac{2 K_B T}{Z_1 Z_2 e^2} \right)^2  \frac{1}{{\rm ln}(\Lambda)}$$

where $m=m_1 m_2/(m_1+m_2)$, $n=n_1+n_2$, $K_B$ is the Boltzmann constant and

$$ \Lambda=1 + \left( \frac{4 \lambda_D K_B T}{Z_1 Z_2 e^2} \right)^2$$

with $\lambda_D=(K_B T/(4 \pi n_ e e^2))^{1/2}$ (Debye-length) and $n_e$ the electron density.

Typically $D_s$ is very small (smaller by about 8 orders of magnitude) compared to $K/(c_p \rho)$ (the so-called thermal diffusivity) 
and the predicted semiconvective transport is very inefficient. 

The code KEPLER \cite{w:78, Heger00} 
implements a different diffusion coefficient for the semiconvective transport:

\begin{equation}
D_\mathrm{SC}= \frac{\alpha_{SC} K_T D_{\rm c}^{'}}{D_{\rm c}^{'}+ \alpha_{sc} K_T}
\end{equation}

where $D_{c}^{'}$ is the diffusion coefficient the layer would have in case the Schwarzschild criterion is used 
(fully efficient convection, $D^{'}_{\rm c}=\frac{1}{3} \alpha_{MLT} \ v_{c} \ H_P$),  
$\alpha_{SC}$ a free parameter usually fixed to 0.1 in KEPLER calculations \cite{sw:14}. 

Finally, results from recent 3D hydrodynamics simulations of layered semiconvective regions \cite{w:13} 
have been transposed into a diffusion coefficient $D_\mathrm{SC}$ implemented in the MESA code. The result is that in stellar conditions 
the mixing obtained with this coefficient is very fast and is essentially equivalent 
to calculations performed with instantaneous mixing in the semiconvective region \cite{moore:16}.

\subsection{Core He-burning phase in low-intermediate mass stars}

\begin{figure}[!h]
\centering\includegraphics[width=3.0in]{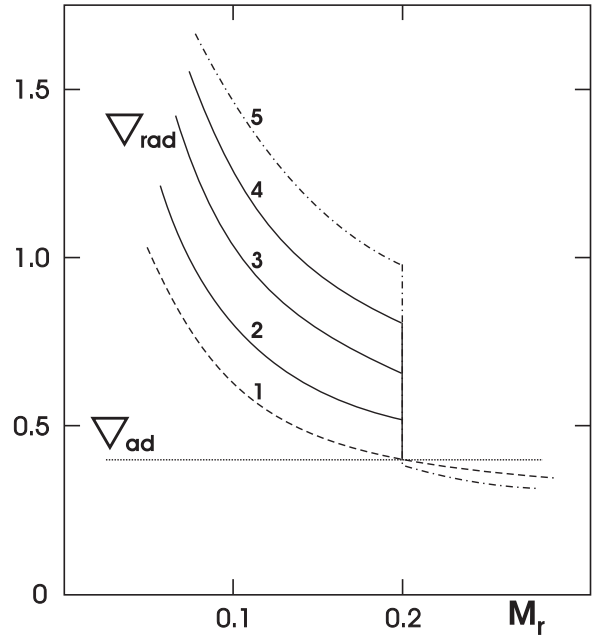}
\caption{Time evolution of $\nabla_{\rm rad}$ as a function of the mass enclosed within radius $r$  
inside a core He-burning model, if the discontinuity of $\nabla_{\rm rad}$ is maintained. 
Increasing numbers define a sequence of increasing time (see text for details).}
\label{semi1}
\end{figure}

Another important evolutionary stage where semiconvection plays a major role is the core He-burning phase of 
low- and intermediate-mass stars, and 
has been widely discussed in the literature (see \cite{cgr:71a, sg:74, rf:72, dr:93} and references therein).
If we consider a low-mass star (initial mass below $\sim$2$M_{\odot}$) after the core He-flash, He-burning is efficient 
in a convective core with central values $log(T_c)\sim8.07\pm0.01$, $log(\rho_c)\sim4.25\pm0.05$, 
and a chemical composition of almost pure He.
The opacity is dominated by electron scattering, but an important contribution (about 25\% of the total) comes also from 
{\sl free-free} absorption. The transformation of He to C due to nuclear burning increases the {\sl free-free} opacity, hence $\kappa$, and  
within the convective core $\nabla_{\rm rad}$, increases, developing a discontinuity of $\nabla_{\rm rad}$  at the inner 
convective boundary (see Fig.~\ref{semi1}).

This discontinuity of the radiative gradient is clearly unphysical. In fact, in convective regions far from the surface 
$\nabla_{\rm c}\sim \nabla_{\rm ad}$, 
and given that in the MLT $F_{\rm c}\sim (\nabla_{\rm rad}-\nabla_{\rm c})$ , there would be an increasing convective flux developing  
on the inner side of the convective boundary.
According to \cite{gn:14} the presence of this discontinuity is due to an 
incorrect application of the Schwarzschild criterion when 
a MLT picture of convection is considered, for a proper implementation requires always $\nabla_{\rm rad}=\nabla_{\rm ad}$ at the 
convective side of the boundary. Another interpretation is that even a very small amount of overshooting suffices to reach the neutral 
$\nabla_{\rm rad}=\nabla_{\rm ad}$ condition at the convective boundary, by means of a {\sl self-driving} mechanism. 

Let's consider an overshooting such that just one radiative layer is mixed. This layer will become 
fully convective on very short timescales, with locally $\nabla_{\rm rad}>\nabla_{\rm ad}$ because of the amount of carbon mixed from the layers below. 
This means that it is now 
the formal boundary of a new enlarged convective core. A small overshooting to mix the next radiative layer will have the same effect, and so on 
until finally $\nabla_{\rm rad}=\nabla_{\rm ad}$ at the convective side of the boundary. 
Another possibility is atomic diffusion \cite{michaud:07} (see Sect.~\ref{diffusion}), driven by the gradient in carbon abundance 
between the fully mixed convective core and the surrounding radiative layers. As carbon diffuses into the radiative He-rich layers, 
the local opacity increases, $\nabla_{\rm rad}$ becomes larger than $\nabla_{\rm ad}$ and these layers become convective. 

This diffusion 
is able to \lq{extend}\rq\ the convective core on short timescales (short compared to 
nuclear timescales), so that $\nabla_{\rm rad}=\nabla_{\rm ad}$ is satisfied at the 
convective boundary \cite{michaud:07}. 
Yet another possibility to extend the formally convective core to attain $\nabla_{\rm rad}=\nabla_{\rm ad}$ at the convective
side of the boundary is the shear instability \cite{chandra:61} (see also Sect.~\ref{rotation}).
At the formal boundary of the convective region (where the discontinuity of $\nabla_{\rm rad}$ appears) 
we have two fluids of mean molecular weight $\mu_1$ and $\mu_2$ respectively, whose surface of separation 
is perpendicular to the gravity field, and have a relative velocity $v_{tang}$ tangential to the separation surface. 
In this situation $v_{tang}\sim v_{\rm c}$, the velocity of the overturning convective elements when they reach the 
formal convective boundary, as obtained from the MLT\footnote{Even a difference in rotational velocity between the convective core  
and the overlying radiative layers can have a similar effect. In this case Eq.~\ref{kh_z} becomes 
$z= v_{tang}^2/(4g (1-\mu_1/\mu_2))$ where $v_{tang}$ is now difference in rotational velocity between the convective core (expected 
to rotate like a solid body) and the overlying radiative layers. Equation~\ref{kh_z2} becomes 
$v_p= (v_{tang}^2 \ v_{mix})/(4g\Lambda(1-(\mu_1/\mu_2)))$}.
In this case a mixed transition region should appear whose width is 

\begin{equation}
 z= \frac{v_{\rm c}^2}{4g (1-\mu_1/\mu_2)}
\label{kh_z}
\end{equation}

By denoting with 
$t_{mix}$ the typical time to mix this region of width $z$, and assuming that $t_{mix}$ is of the order of the characteristic time 
of convection at the boundary of the core (as suggested by \cite{cgr:71a}), i.e. $t_{mix}=\Lambda/v_{\rm c}$, with $\Lambda \sim H_P$, 
the velocity of advancement of the border of the mixed core is $v_p\approx z/t_{mix}\approx z \ v_{\rm c}/\Lambda$. Combining these relations  
with Eq.~\ref{kh_z} provides

\begin{equation}
 v_p= \frac{v_{\rm c}^3}{4g\Lambda(1-\mu_1/\mu_2)}
\label{kh_z2}
\end{equation}

It has been shown in \cite{cgr:71a} using a MLT approach that an overshooting length $z\approx 5$~cm guarantees an advancement of 
the fully mixed core with speed $v_p\approx 10^3 [1-(\nabla_{\rm ad}/\nabla_{\rm rad})]$~cm/yr. This is sufficient to enlarge the mass 
of the fully mixed core to the point where $\nabla_{\rm ad}=\nabla_{\rm rad})$ on timescales shorter than nuclear timescales. 
By adopting the values of $g$, $v_{\rm c}$, $\lambda$ and $\mu_1/\mu_2$ adopted by \cite{cgr:71a}, Eqs.~\ref{kh_z} and \ref{kh_z2} provide 
$z\sim 3$~cm, and $v_p\approx 10^3$~cm/yr, consistent with \cite{cgr:71a} results.
 
\begin{figure}[!h]
\centering\includegraphics[width=3.0in]{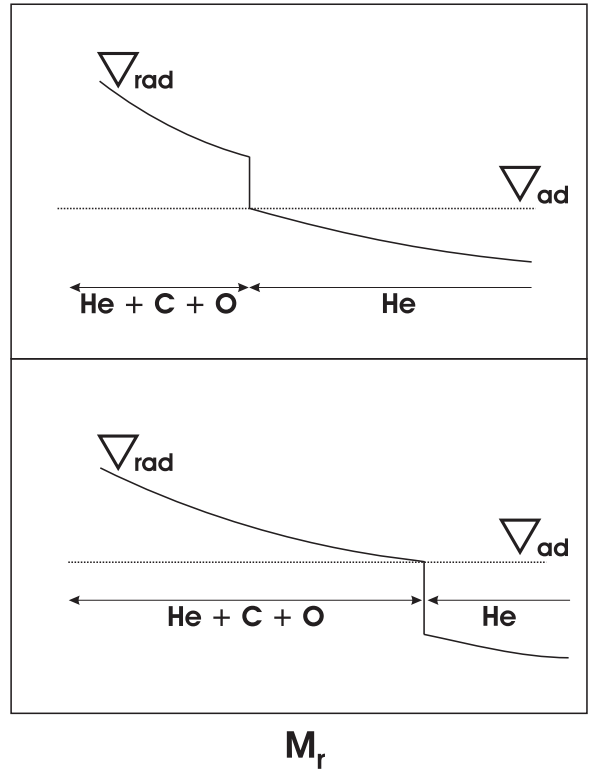}
\caption{Similar to Fig.~\ref{semi1}, qualitative sketch of the change in the fully mixed core  
when the discontinuity of $\nabla_{\rm rad}$ at its edge is maintained (top panel), and when the convective boundary is 
set where $\nabla_{\rm rad}=\nabla_{\rm ad}$ (see text for details).}
\label{driving}
\end{figure}

In practical terms, at every computational timestep, one can place the boundary of the fully mixed convective region at the layer where 
$\nabla_{\rm rad}=\nabla_{\rm ad}$ (see Fig.~\ref{driving}).

\begin{figure}[!h]
\centering\includegraphics[width=3.0in]{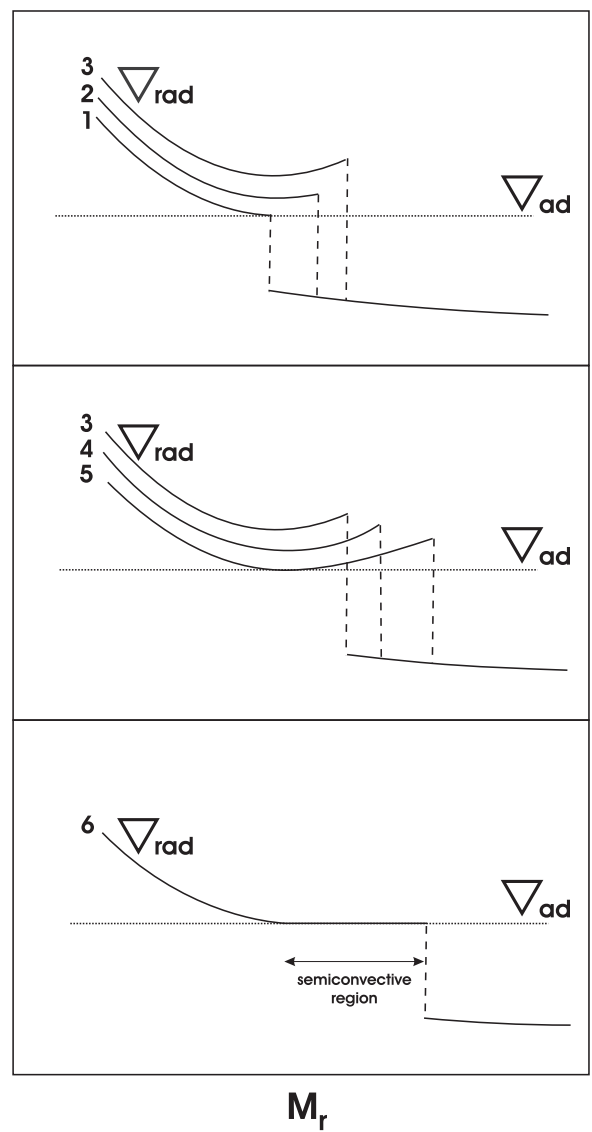}
\caption{Similar to Fig.~\ref{driving}, sketch of the time evolution of the radiative gradient profile near the
 boundary of the convective core during the central 
He-burning stage. The panels show the sequence of events which lead to the formation of the semiconvective zone 
(shown in the bottom panel). Increasing numbers denote a sequence of increasing time (see text for details).}
\label{semi2}
\end{figure}

After this early phase, when the extension of the convective core follows the increase of $\nabla_{\rm rad}$,  
the radiative gradient profile starts to show a minimum (profile~3 in Fig.~\ref{semi2}), as a consequence of 
the progressive outwards shift of the convective boundary. 
The presence of this minimum depends on the complex behaviour of the 
physical quantities involved in the definition of ${\rm\nabla_{rad}}$, such as opacity, pressure, temperature and local energy flux. 
In this situation, outward mixing to eliminate the $\nabla_{\rm rad}$ discontinuity will induce a 
general decrease of the radiative gradient in the whole convective core 
(due to an average increase of He and consequent decrease of C --see profile~4 in Fig.~\ref{semi2}). 
The radiative gradient will eventually decrease and become equal to $\nabla_{\rm ad}$ 
at the location of the minimum of $\nabla_{\rm ad}$
(profile~5 in Fig.~\ref{semi2}).

\begin{figure}[!h]
\centering\includegraphics[width=3.0in]{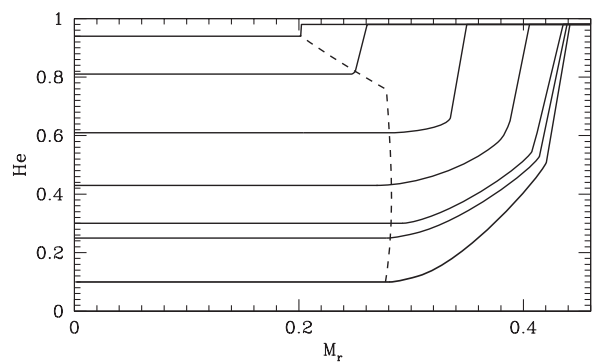}
\caption{Helium abundance profile inside the convective core and the semiconvective zone at various levels 
of central He depletion, for a low-mass core He-burning model. The dashed line marks the location of the fully mixed core 
boundary.}
\label{semichi}
\end{figure}

The semiconvective region is the region between the \lq{neutral}\rq\ $\nabla_{\rm rad}=\nabla_{\rm rad}$ point, and the 
overlying formally convective shell whose upper boundary still displays a discontinuity in $\nabla_{\rm rad}$.
In fact a full instantaneous mixing between the convective core located inside the minimum and the external convective 
shell, would have the consequence of decreasing the
radiative gradient in the whole mixed region. However, due to the presence of this minimum, $\nabla_{rad}$  in a portion of the 
mixed core would become lower than the $\nabla_{\rm ad}$, i.e. it would not be convective.
A solution for this inconsistency is to impose a partial mixing in the formally convective shell (see, e.g., 
\cite{cctp:85} for an example of implementation), such that 
the final chemical composition --shown in Fig.~\ref{semichi}-- 
satisfies the condition $\nabla_{\rm rad}=\nabla_{\rm ad}$ (profile 6 in Fig.~\ref{semi2}).
The mass location of the minimum of the radiative gradient moves outwards with time, because of the evolution of the chemical 
abundances caused by nuclear burning. As a final result, the C-enriched region increases its size outwards.

The effects of semiconvection on the evolution of these models are the following: The evolutionary tracks perform more 
extended loops in the HRD, and the central He-burning phase lasts longer because of a larger amount of fuel to burn.

When the central He-abundance has decreased to about $Y\sim0.10$, 
$\alpha$-captures by C nuclei tend to overcome C production by $3\alpha$ reactions, thus He-burning becomes 
mainly a ${\rm ^{12}C + \alpha}$ production of oxygen,
whose opacity is even larger than that of ${\rm ^{12}C}$. This causes an increase of the size of
the semiconvective region and, in turn, 
more fresh helium is transferred into the core, which is now nearly He-depleted. Even a small amount of He added to the mixed core 
enhances the rate of energy production, thus the luminosity increases, driving an increase of the radiative gradient.
As a consequence, a
phase of enlarged mixed zone starts, the so called {\sl breathing pulse}. After this breathing pulse, the star readjusts 
to burn steadily in the core the fresh He driven
there by semiconvection. Detailed calculations show that a few breathing pulses are expected before the complete exhaustion of He 
in the core. The evolutionary effects of the
breathing pulses are the following: The models perform a loop in the HRD at each pulse, the He-burning lifetime is slightly 
increased, and the mass of the CO-core at He exhaustion is increased. 

Empirical constraints --mainly the number ratio between horizontal branch (HB --core He-burning phase in low mass stars) 
and AGB stars in Galactic globular 
clusters (GGCs)-- suggest that the
efficiency of the breathing pulses phenomenon is very low, if any \cite{caputo:89, csi:03}. Therefore they are 
usually inhibited in stellar model computations 
by using \emph{ad hoc} numerical assumptions \cite{dr:93, cassisi:99}. 
Typically during the late stages of core He-burning one forces the extension of the mixed region not to lead to an
increase of the central He abundance from one model to the
next \cite{caputo:89}. Another option is to set to zero
the gravitational term in the energy generation equation for
the inner regions. In this way, the breathing pulses are also effectively
inhibited \cite{dr:93}.

This brief discussion highlights clearly the difficulty in modelling core mixing in HB stars. 
Notice that the standard treatment is usually instantaneous mixing, that may not be adequate in a semiconvective regime.
None of the diffusive semiconvective formalisms employed for semiconvection related to
H-burning is usually applied to this situation, although the diffusive mixing employed in the STARS code still leads
to the onset of breathing pulses during the core He-burning phase.
It is also worth recalling that with the inclusion of an extended overshooting region ($\sim$1 $H_P$) beyond the layer where the 
$\nabla_{\rm rad}$ discontinuity develops, the fully mixed core is always so large that the need to include semiconvective 
layers disappears \cite{sdi:03} (breathing pulses still seem to appear also with large overshooting).
An important consequence of including semiconvection, or large overshooting, and/or breathing pulses, is that the CO profile in  
the final CO core changes, with important consequences for the cooling times of the final WD stage\footnote{Even the 
current uncertainties in the ${\rm ^{12}C(\alpha,\gamma)^{16}O}$ reaction rate can affect the CO profiles, and the 
HB/AGB star count ratio}.

With increasing stellar mass the weight of the {\sl free-free} opacity decreases (because of higher 
core temperatures) and eventually all these problems disappear in the regime of intermediate mass stars (masses above a few 
solar masses).

\section{Thermohaline mixing}
\label{thermohaline}

\begin{figure}[!h]
\centering\includegraphics[width=2.5in]{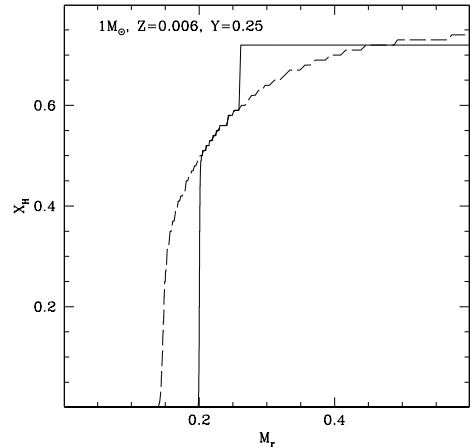}
\caption{H-abundance profile within a low-mass stellar model at the start (dashed line) and at the completion (solid line) 
of the first dredge-up, respectively. The horizontal coordinate displays 
the local value of the mass. At the completion of the dredge-up the He-core mass is equal to $\sim$0.2$M_{\odot}$ 
(it was $\sim$0.14$M_{\odot}$ at the start of the dredge-up) and the H abundance discontinuity 
left over by the fully mixed convective envelope at its maximum extension 
is located 0.25$M_{\odot}$ away from the centre. The varying H-abundance in the layers between the He-core and the discontinuity 
has been produced during core H-burning along the MS.
When the H-burning shell reaches the H-abundance discontinuity the RGB 
model experiences a temporary drop in luminosity (that in a stellar population 
produces a local increase in star counts, the so called \lq{RGB bump \rq}, 
before evolving again with increasing $L$ when the discontinuity is crossed.}
\label{Hprofile}
\end{figure}

In recent years the role played by thermohaline mixing in stellar evolution has been widely explored in connection with 
low-mass red giant branch (RGB) evolution\footnote{Thermohaline mixing due to the same reasons as along the RGB is also efficient 
during the HB and asymptotic giant branch evolution of low-mass stars, according to some calculations\cite{cl:10}.
An upper limit of 1.5$M_{\odot}$ \cite{cl:10} or 2.2$M_{\odot}$ \cite{clag:10} is found for RGB stars (the same 1.5$M_{\odot}$ 
limit is found by \cite{cl:10} for HB and AGB stars) to develop 
thermohaline mixing. The efficiency of the mixing in general tends to decrease with increasing mass. Thermohaline mixing 
is also important when carbon has ignited off-centre in the core of super-AGB stars, for it affects significantly the propagation of the flame 
\cite{siess09}.}.
When the convective envelope deepens after the TO, the surface chemical composition is altered due to the dredge up 
of H-burning processed matter --the so-called  
\emph{first dredge-up}-- that increases the abundance of N and He, decreases C and the ${\rm ^{12}C/^{13}C}$ ratio. 
The first dredge-up is completed when the convective region reaches 
its maximum extension, approaching closely (but not reaching) the H-burning shell around the inert (and electron degenerate) He-core.
From this moment on the receding convective boundary is not expected to modify further the surface abundances along the RGB.
Spectroscopic observations of metal poor Galactic halo stars provide however compelling evidence for an additional mixing process 
occurring when RGB stars reach the luminosity of the RGB bump (see Fig.~\ref{Hprofile}), that causes a sudden drop of 
the isotopic ratio ${\rm ^{12}C/^{13}C}$,  a decrease of Li and C, and an increase of N. 
This mixing affects $\sim95$\% of low-mass stars, regardless of whether 
they populate the halo field or clusters \cite{gratton:00, angelou:15}, and  
thermohaline mixing has been proposed to explain these observations\footnote{About 5\% of the 
post-bump RGB objects investigated do not show these abundance variations. A strong 
magnetic field can potentially inhibit thermohaline mixing \cite{cz:07b}.}.

In low-mass stars, the main H-burning mechanism during the MS is the \emph{p-p chain} 
that, due to its weak dependence on temperature, 
is efficient also in stellar layers far from the star centre. As a consequence, ${\rm ^3He}$ accumulates in a broad zone outside 
the main energy production region.  During the first dredge-up this ${\rm ^3He}$  is 
mixed within the convective envelope, with the consequence that during the following RGB evolution, 
the layers above the H-discontinuity  left over by the receding convective envelope at its maximum extension will have a uniform 
${\rm ^3He}$ abundance, larger than the initial one.

When the shell advances towards the surface during the RGB evolution, in the outer wing above 
the point of maximum burning efficiency there is a narrow region where ${\rm ^3He}$ is processed through the reaction  
${\rm ^3He(^3He,2p)^4He}$. In this nuclear reaction 2 nuclei transform into 3 and the mean 
mass per nucleus --the molecular weight $\mu$-- decreases.  
This leads to a small local decrease of $\mu$ when moving from the surface towards the centre of the star. 
As long as the H-burning shell advances through layers below the H-discontinuity --when 
the star evolves before the RGB bump, see Fig.~\ref{Hprofile}-- this effect is negligible because 
the shell is 
moving in a region with a large positive gradient, due to the H profile left over at the end of the MS. 
However, when the H-burning shell 
enters the region of uniform H-abundance above the discontinuity, 
the local inversion of the $\mu$  profile, of the order of one part in $10^4$, becomes important \cite{edl:06}.
This situation corresponds to the conditions for thermohaline mixing (see \cite{cz:07} and references therein)

Thermohaline mixing is usually included in stellar 
evolution codes as a diffusive process that works in the direction to erase the molecular weight inversion, 
with a diffusion coefficient derived from a linear analysis \cite{ulrich:72, kipp:80} 

\begin{equation}
D_ {\rm th} =  C_ {\rm th} \,  \frac{K}{c_p \rho}  \left(\phi \over \delta\right){\nabla_\mu \over (\nabla_{\rm rad} - \nabla_{\rm ad})} 
\label{ther}
\end{equation}

where $C_{\rm th}=(8/3) \pi^2 \alpha^2$, $\alpha$ being a free parameter related to the aspect ratio (length/width) of the mixing elements\footnote{The 
calculations by \cite{cl:10} and the MESA calculations by \cite{lattanzio:15} consider the constant $C_ {\rm th}$ as 
$C_ {\rm th} \equiv (3/2) \alpha_{\rm th}$, where $\alpha_{\rm th}$ is in this case the free parameter.}

\begin{figure}[!h]
\centering\includegraphics[width=2.5in]{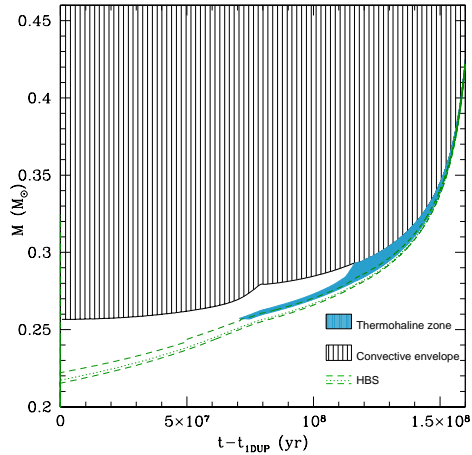}
\caption{Location of the mass boundaries as a function of time (zero point taken 
at the completion of the first dredge-up) of the convective envelope, H-burning shell, and the zone mixed by the 
thermohaline instability, for a low-mass RGB stellar model.}
\label{thermo2}
\end{figure}

As shown in Fig.~\ref{thermo2}, evolutionary calculations show that thermohaline mixing extends between the outer wing of 
the H-burning shell 
and the inner boundary of the convective envelope, merging with the outer convection in a 
short time ($\sim30$~Myr for a model with mass of the order of 
$1M_\odot$). Therefore, depending on the mixing efficiency (hence the choice of $C_{\rm th}$), 
a significant amount of nuclear processed matter in the hotter layers of the 
H-burning shell can be dredged up to the surface during the remaining RGB evolution, helping to explain the spectroscopic data. 

Comparisons with observations require $C_{\rm th}\sim 100-300$ from Li observations in Galactic globular clusters,  
and $C_{\rm th}\sim 1000$ for C in globulars, clearly mutually inconsistent. On the other hand models by \cite{cz:07, clag:10} 
are able to match the ${\rm ^{12}C/^{13}C}$  isotopic ratio, C and N abundances in halo RGB field stars with $C_{\rm th}\sim 1000$; the same 
choice of $C_{\rm th}$ allows also to match measurements of carbon isotopic ratio and [N/C] ratio in open clusters, and 
Li abundances in field disk stars. 
Results from 3D hydrodynamics simulations (rescaled to match stellar conditions) add an additional source of uncertainty, 
for they predict an efficiency equivalent to just $C_{\rm th} \approx10$ \cite{dm:11, traxler}.

A recent detailed numerical analysis \cite{lattanzio:15} has shown that 
the surface chemical abundances predicted by stellar models accounting for thermohaline mixing depend on 
numerical assumptions like spatial- and time resolution adopted in the computations. As a consequence, the 
predicted surface chemical abundances should be treated with caution until a firmer assessment on how to treat   
thermohaline mixing in model computations is achieved.

\section{Atomic diffusion}
\label{diffusion}

Microscopic effects related to collisions among the gas particles induce a 
{\sl slow} element transport within radiative regions. 
It is possible to show from first principles that 
individual ions are forced to move under the influence of 
pressure as well as temperature gradients, which both  
tend to displace the heavier elements toward the centre of 
the star, and of concentration gradients that oppose the above processes. 
Radiation --which doesn't have a major effect in the Sun \cite{turc}-- 
pushes the ions toward the surface, whenever the radiative 
acceleration imparted to an individual ion 
species is larger than the gravitational acceleration. 
The speed of the diffusive flow depends on 
the collisions with the surrounding 
particles, as they share the acquired momentum in a random way. 
It is the extent of these 'collision' effects that dictates the 
timescale of element diffusion within the stellar structure, once the  
physical and chemical profiles are specified. 

The most general treatment for the element transport in a multicomponent fluid associated with 
diffusion is provided by the Burgers equations \cite{burgers:69}.  
They are obtained assuming  
the gas particles have approximate Maxwellian velocity distributions,  
the temperatures are the same for all particle species, the mean 
thermal velocities are much larger than the diffusion velocities, 
magnetic fields are unimportant, and can be written as: 

\begin{eqnarray}\label{4difeq}
\frac{\textrm{d}p_i}{\textrm{d}r}+\rho_i(g-g_{{\rm rad}, i}) -n_i\bar{Z}_ieE=  \nonumber \\
\sum_{j\neq i}^N K_{ij}(w_j-w_i)+\sum_{j\neq i}^{N}K_{ij}z_{ij}\frac{m_jr_i-m_ir_j}{m_i+m_j},
\end{eqnarray}
including the heat flow equations,
\begin{eqnarray}\label{4heat}
\frac{5}{2}n_i K_B\nabla T =\frac{5}{2}\sum_{j\neq i}^Nz_{ij}\frac{m_j}{m_i+m_j}(w_j-w_i)-\frac{2}{5}K_{ii}z_{ii}''r_i \nonumber\\
-\sum_{j\neq i}^N\frac{K_{ij}}{(m_i+m_j)^2}(3m_i^2+m_j^2z_{ij}'+0.8m_im_jz_{ij}'')r_i\\
+\sum_{j\neq i}^N\frac{K_{ij}m_im_j}{(m_i+m_j)^2}(3+z_{ij}'-0.8z_{ij}'')r_j.\nonumber
\end{eqnarray}
In addition, there are the constraints of electric current neutrality,
\begin{equation}\label{4current}
\sum_i \bar{Z}_i n_i w_i = 0
\end{equation}
and local mass conservation,
\begin{equation}\label{4mass}
\sum_i m_i n_i w_i = 0.
\end{equation}

In the above $2N+2$ equations $p_i$, $\rho_i$, $n_i$, $\bar{Z}_i$ and $m_i$ denote the partial pressure, mass density, number density, 
mean charge and mass for species $i$, respectively. The total number of species (including electrons) is $N$. The $2N+2$ unknown variables are the $N$ 
diffusion velocities $w_i$, the $N$ heat fluxes $r_i$, the gravitational acceleration $g$ (the comparison of the derived $g$ with the 
known value from the integration of the stellar structure equations, provide an important check for the consistency of the results) 
and the electric field $E$. The coefficients $K_{ij}$, $z_{ij}$, $z_{ij}^\prime$ and $z_{ij}^{\prime\prime}$ have to be specified, together 
with the radiative accelerations $g_{\rm rad}$. 
Several stellar evolution codes --in one form or another-- use the routine by \cite{thoul:94} to solve the Burgers equations and calculate 
the velocities of the various chemical species. 
These diffusion velocities can then be inserted as an advection term in the equation for the time evolution of the mass fraction abundance $X_i$

\begin{equation}
\frac{\partial X_i}{\partial t}\bigg|_{M_r} = - \frac{1}{\rho r^2}\frac{\partial}{\partial r}(\rho r^2 X_i w_i)
\label{eqdiffu}
\end{equation}

where we show on the right-hand-side just the contribution of diffusion to the time evolution of the abundance $X_i$ of element $i$.

In Burgers' formalism the effect of collisions between ions is represented by the so-called resistance coefficients, i.e. 
the matrices $K$, $z$, $z^\prime$, 
$z^{\prime\prime}$, whose precise evaluation is essential to estimate correctly the diffusion timescales for the 
various elements. These resistance coefficients can be expressed in terms of the so-called 
reduced collision integrals 
${\Omega^{(l,s)}_{ij}}^\ast$
according to the following relationships: 
\begin{eqnarray} 
\frac{K_{ij}}{K_{ij}^0} & = & 4 
\frac{{T^\ast_{ij}}^2\,{\Omega^{(1,1)}_{ij}}^\ast}{\ln\left(\Lambda_{ij}^2 +1\right)}\mathrm{,} \label{kdef} \\ 
z_{ij} & = & 1 - 1.2 
\frac{{\Omega^{(1,2)}_{ij}}^\ast}{{\Omega^{(1,1)}_{ij}}^\ast}\mathrm{,} \\ 
z_{ij}^\prime & = & 2.5 - \frac{6\,{\Omega^{(1,2)}_{ij}}^\ast - 4.8\, 
{\Omega^{(1,3)}_{ij}}^\ast}{{\Omega^{(1,1)}_{ij}}^\ast}\\ 
z_{ij}^{\prime\prime} & = & 2 
\frac{{\Omega^{(2,2)}_{ij}}^\ast}{{\Omega^{(1,1)}_{ij}}^\ast}\mathrm{,} \label{zppdef} 
\end{eqnarray} 
where 
\begin{eqnarray} 
T^\ast_{ij} & = & K_B T\frac{\lambda_\mathrm{D}}{\left|Z_i Z_j 
e^2\right|} \label{Tdef} \\ 
K_{ij}^0 & = & \frac{2}{3}\sqrt{\frac{2\mu_{ij} \pi}{(K_B T)^3}} 
\left(Z_i Z_j e^2\right)^2 n_i n_j \ln\left(\Lambda_{ij}^2 +1\right)  
\end{eqnarray} 
with $\lambda_\mathrm{D}$ being the Debye-length, $\Lambda_{ij} = 
4 T_{ij}^\ast$ the plasma parameter (${\rm ln}(\Lambda_{ij})$ is called {\sl Coulomb logarithm}),
$\mu_{ij} = m_i m_j/(m_i+m_j)$ the reduced mass, and $Z_i$, $m_i$ and $n_i$ the 
charge number, mass and particle number density 
of species $i$, respectively.  

\begin{figure}[!h]
\centering\includegraphics[width=3.0in]{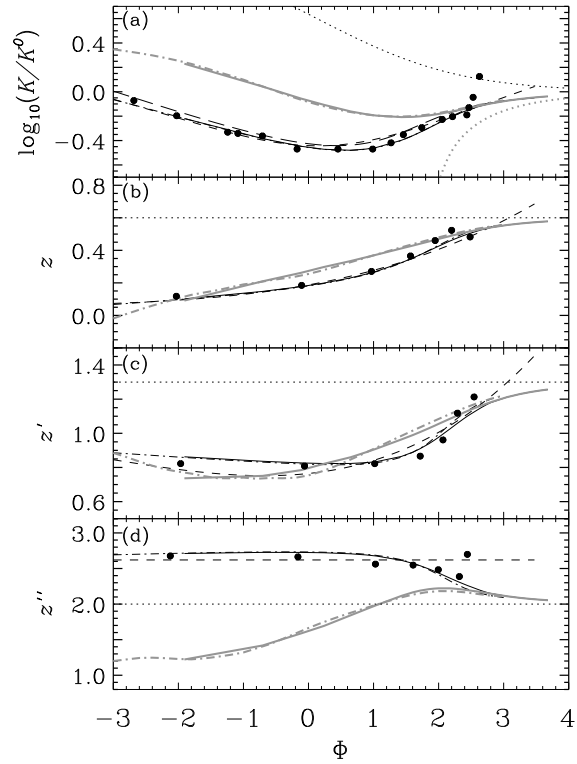}
\caption{Resistance coefficients $K$ (a), $z$ (b), $z^\prime$ (c) and 
$z^{\prime\prime}$ (d) as originally computed by Burgers \cite{burgers:69} 
(dotted lines), \cite{mason:67} (solid lines), \cite{muchmore:84} (short-dashed line), 
\cite{paquette:86} (dash-dotted lines) and \cite{iben:85} (long-dashed line) as a function of 
$\Phi_{ij} = \ln \left(\ln (1+\Lambda_{ij}^2)\right)$. 
Dark black lines denote the case 
of a repulsive potential, while the brighter ones denote attractive forces 
(not computed in \cite{muchmore:84}  and \cite{iben:85}). The dots represent the 
actual values computed by \cite{muchmore:84}, not his fitting formulae.}
\label{diff_1}
\end{figure}

The collisions between particles of species $i, j$ in the stellar plasma  
determine the values of the ${\Omega^{(l,s)}_{ij}}^\ast$ 
integrals; the physics of the collisions is specified by some form of 
the Coulomb interaction. 
This, as a first approximation,
can be described by a pure Coulomb potential with a long-range cut-off  
distance, typically equal to $\lambda_\mathrm{D}$. 
Using this truncated pure Coulomb potential the resistance coefficients originally computed by Burgers become \cite{burgers:69}  
\begin{eqnarray} 
\frac{K_{ij}}{K_{ij}^0} & = & 2\, \frac{\ln \Lambda_{ij} - C_\mathrm{E} 
\pm \frac{\pi^2}{4}}{\ln\left(\Lambda_{ij}^2 + 1\right)} \label{Kburg}{\rm,} \\  
z_{ij} & = & 0.6 \label{zb} \\ 
z^\prime_{ij}  & = & 1.3 \label{zpb}\\ 
z^{\prime\prime}_{ij} & \approx & 2 \label{zppb} 
\end{eqnarray} 
where $C_\mathrm{E}$ is Euler's constant and the $\pm$ signs denote   
repulsive ($+$) and attractive ($-$) pairs of particles, respectively.
More accurate collision integrals have been obtained by considering 
a Debye-H\"uckel type of potential 
\begin{equation} 
V_{ij}(r) =  \frac{Z_i Z_j e^2}{r}\; \mathrm{e}^{-r/\lambda_\mathrm{D}} \label{debhuck} 
\end{equation} 
where $r$ is the particle distance. The results are provided either in tabulated form \cite{mason:67} 
or as fitting formulae \cite{muchmore:84, iben:85, paquette:86, ss:03, zhang:16}, and some of them are compared in 
Fig.~\ref{diff_1}, as a function of $\Phi_{ij} = \ln \left(\ln(1+\Lambda_{ij}^2)\right)$.
Given that the calculations shown in the figure (apart from the Burgers' results) make all the same physical assumptions, 
results are similar, but they all differ significantly from Burgers' results when moving towards higher densities (lower values of $\Phi_{ij}$), 
where his approximations are no longer adequate. 

Some of these authors argued that whilst $\lambda_\mathrm{D}$ is a  
suitable screening distance at low densities, the Debye sphere 
loses its significance in denser plasmas, and that a more  
appropriate screening distance is in this case the mean interionic distance. Hence,  
they suggested to use the 
larger value between $\lambda_\mathrm{D}$ and the mean interionic distance 
(in the sun the former has always been larger than the latter). 
Whatever the choice for the actual screening distance, its
value has to be employed as $\lambda_\mathrm{D}$ in Eq.~\ref{Tdef} to
compute the appropriate $\Lambda_{ij}$ for determining the collision
integrals. 
The effect of quantum corrections on the resistance coefficients has been included in \cite{ss:03} whilst the calculations 
\cite{zhang:16} account for very recent developments in 
ionic transport properties in strongly coupled plasmas \cite{daligault:16}.

As for the radiative levitation of element $i$, the main physical interactions 
that drive the transfer of momentum from the radiation field to ions are {\sl bound-bound} and
{\sl bound-free} transitions, whilst it is practically ineffective for fully ionized elements. 
The radiative acceleration can be written as \cite{hu:11}
\begin{equation}\label{gradeq} 
g_{{\rm rad}, i} = \frac{\mu\kappa}{\mu_i c}\frac{l}{4\pi r^2}\gamma_i
\end{equation}
where $\mu$ is the mean atomic weight, $\mu_i$ is the atomic weight of element $i$, 
$l$ is the local luminosity and $r$ is the radius \cite{hu:11}. 
The dimensionless quantity $\gamma_i$ depends on the monochromatic opacity data 
\[
\gamma_i=\int \frac{(\sigma_i(u)[1-\exp(-u)]-a_i(u)){\rm d}u}{\sum_i f_i\sigma_i(u)}
\]
where $u=(h \nu)/(K_B T)$, $\sigma_i$ is the cross-section for absorption or scattering of radiation by element $i$, $a_i$ 
accounts for the fact that in  
{\sl bound-free} transitions only a fraction of the momentum of the ionizing photons is transferred to the ion, the rest being 
transferred to the electron lost by the ion, and $f_i$ is the number fraction of element $i$. 
The Opacity Project provides a set of codes called OPSERVER that enables the calculation of $g_{{\rm rad}, i}$ \cite{seaton:05}.

\begin{figure}[!h]
\centering\includegraphics[width=3.0in]{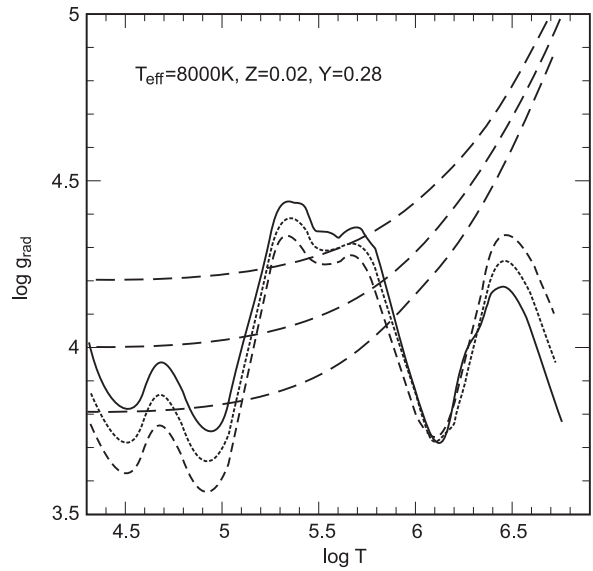}
\caption{Total radiative acceleration of Fe as a function of the temperature in a stellar envelope 
with the labelled $T_{eff}$ and chemical composition, for three values of the surface 
gravity log($g$), equal to 3.8 (short dashed line), 4.0 (dotted line) and 4.2 (solid line) respectively. The run of the local gravity with 
temperature is displayed by long dashed lines.}
\label{diff_2}
\end{figure}

Calculations of precise $g_{{\rm rad}, i}$ values involve carrying out the integration over about $10^4$ $u$ values for each atomic 
species \cite{leblanc:00}. 
Given that these calculations have to be repeated at each layer in the stellar model and at each time step during the computation of 
an evolutionary sequence, this explains why, 
to date, there are only few extended sets of stellar models that include also the effect of 
radiative levitation. 
One has also to note that the Rosseland mean opacity entering the stellar structure equations as well as Eq.~\ref{gradeq} has to be  
continuously recalculated at each mass layer, not just in the nuclear burning regions
--this is in principle true also for calculations including only the other diffusive processes listed above-- 
to be fully consistent with the composition changes. 

\begin{figure}[!h]
\centering\includegraphics[width=3.0in]{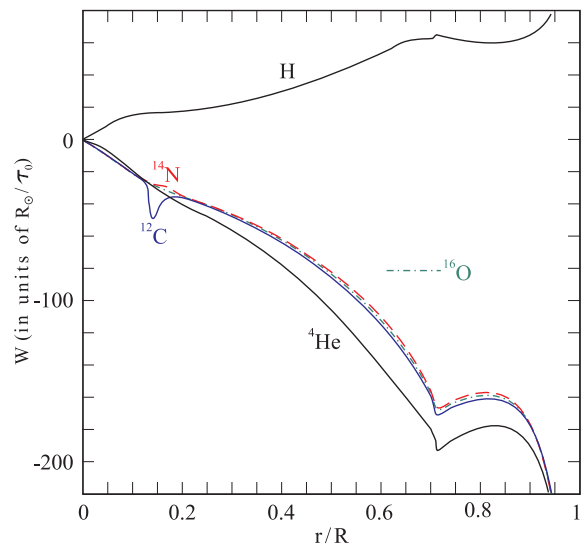}
\caption{Diffusion velocity of H, He, C, N, O 
(the lines displaying the diffusion velocities of C, N  and O almost overlap), 
as a function of the local fraction of the total radius, within a solar model. The velocity is in units of 
solar radius over the typical diffusion timescale for the Sun, that is of the order of $6 \times 10^{13}$~yr.
The base of the convective envelope is at $R/R_{\odot}\sim$0.71.}
\label{veldiff}
\end{figure}

As an example, Fig.~\ref{diff_2} displays the total radiative acceleration of iron in a stellar envelope. 
When the radiative acceleration is larger than gravity below a convective envelope -- if convection is present 
the effect is obviously negligible because of the much faster timescales of convective mixing--  the element 
can diffuse towards the surface.
Figure~\ref{veldiff} displays the velocity profiles of H (moving upwards), He and CNO (sinking) 
within a solar model, derived from the solution of the Burgers equations. 

The diffusion velocities are always dominated by the effect of pressure gradients. The radiative (upwards) acceleration of 
C, N, and O is only 
at most about 5\% of the local acceleration of gravity just below the convective envelope boundary \cite{turc}, and decreases fast 
moving towards the centre.
The local sharp increase of the settling velocity of C and the corresponding small 
decrease for N at $r/R\sim$0.15-0.20, are due to the effect 
of abundance gradients when these elements attain the equilibrium abundances of the CN cycle (C decreases whilst N increases).

\subsection{The effect of atomic diffusion on stellar models}
\label{ato}

Atomic diffusion (sometimes denoted as microscopic diffusion, and hereafter simply diffusion) has a major direct effect on the MS evolution, 
the chemical stratification of the external layers of hot horizontal branch stars\footnote{We discussed in Sect.~\ref{semiconvection} the effect of diffusion on 
the HB mixed cores.}  
and WDs, and the internal chemical stratification of cold WDs. However, 
some properties of other evolutionary phases are also indirectly affected, as discussed below 
\cite{cassisi:97, richard:02a, richard:02a, michaud:07, koester:09, michaud:11, cs:13}.

The impact of diffusion on stellar models will be discussed mainly in the context of low-mass stars, due to their evolutionary timescales 
comparable to the diffusion timescales during the MS phase. One has also to take into account that massive hot stars, where in principle 
radiative levitation can be extremely efficient, experience strong mass loss and rotational mixings (see later on in this section and 
Sect.~\ref{rotation}) that tend to limit or completely erase the effect of diffusion.

If we consider the evolution of a typical low-mass star with a convective envelope on the MS, 
the surface abundances of metals and He tend to decrease because of diffusion from the bottom boundary of 
the shrinking convective envelope  
(that maintains a uniform chemical profile due to the shorter convective timescales compared to diffusion) 
being replaced by hydrogen. However, if the radiative acceleration on some ion species is larger than the local gravitational acceleration below the 
convective envelope, 
these elements are slowly pushed into the convective zone and their surface abundance increases. In general the variation of the surface 
abundances during the MS phase is dictated by the interplay between radiative levitation and the sedimentation due mainly to 
pressure gradients (often denoted as {\sl gravitational settling}).
This variation of the surface abundances reaches a maximum around the TO. 

When, after the TO, the convective envelope starts to deepen, it will rehomogenize an increasingly larger fraction of the stellar mass.  
Once the star reaches the RGB and the first dredge-up is completed,  
the abundance changes previously developed are almost completely erased, apart for the very inner layers, where metals and He 
were sinking during the MS.
In general, the smaller (in mass) the convective envelope, the larger the change of surface abundances during the MS, because of the smaller 
diluting {\sl buffer} of matter with the initial chemical composition. 

\begin{figure}[!h]
\centering\includegraphics[width=3.0in]{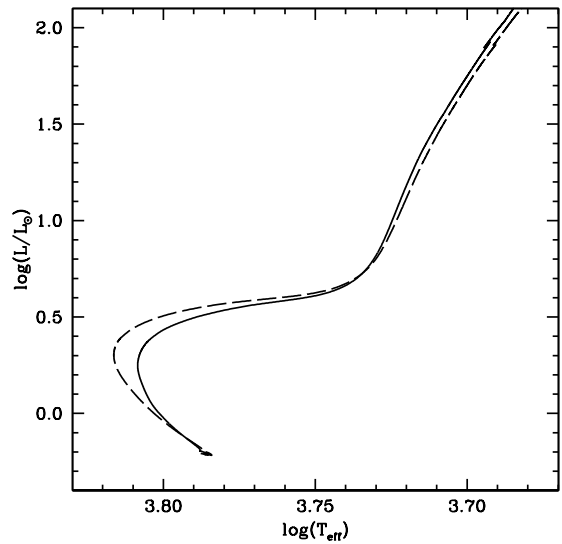}
\caption{Comparison of the HRD of two 0.82$M_{\odot}$, $Y$=0.25, $Z$=0.001 evolutionary track 
calculated with (solid line) and without (dashed line) atomic diffusion.}
\label{diff_3}
\end{figure}

Figure~\ref{diff_3} compares the HRD of low-mass, metal poor evolutionary tracks, 
calculated with and without the inclusion of atomic diffusion. The track with diffusion increasingly diverges from the 
no-diffusion one moving along the MS. Its TO is fainter and cooler, whilst the two tracks coverge again along the RGB, where 
the effect of diffusion is practically erased by the fully mixed deep convective zone.
The MS evolutionary times are also different, for the track with diffusion has a shorter lifetime 
because of the slow increase of He at the centre at the expenses of H during the MS 
phase\footnote{The surface abundance variations tend to increase the envelope opacities which, in turn, would tend to increase the MS lifetime. But the 
dominant process is the decrease of available fuel, and the net effect is a $\sim$10\% decrease of the MS lifetime.}.
These changes of TO luminosity and MS lifetime at a given stellar mass affect also theoretical isochrones,  
hence age determinations of old stellar populations based on the TO luminosity decrease by about 10\%.

\begin{figure}[!h]
\centering\includegraphics[width=3.0in]{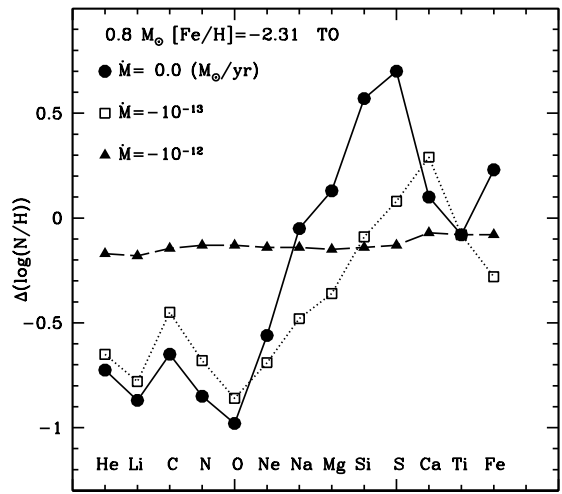}
\caption{Difference between the TO surface abundances of the most relevant elements, and their initial values, for a  
a low-mass metal poor model (see labels). The element abundances are given as log(N/H), e.g. as logarithms of 
their number fraction to hydrogen \cite{vick:13}. These abundance variations are shown for models calculated with no mass loss, 
and two different mass loss rates during the MS evolution.}
\label{diff_4}
\end{figure}

Figure~\ref{diff_4} shows the pattern of variations of TO surface abundances, for a typical stellar model at the TO of a metal poor 
globular cluster. These models have thin convective envelopes that maximize the effect of diffusion. 

It is straightforward to notice the selective effect of diffusion, due to radiative levitation. Elements like 
He, Li, C, N and O sink below the convective layers and their surface abundances are severely depleted up to a factor of about ten.
Other elements like Mg, Si, S and Fe are pushed into the convective envelope by the radiation pressure and their surface abundances 
increase during the MS. The size of these abundance variations decrease in models with larger convective envelopes. For example 
a 0.8$M_{\odot}$ model with initial [Fe/H]=$-$0.7 will show uniformly depleted abundances for all elements heavier than H 
at the TO, at level of just 0.1-0.15~dex.

After the first dredge up is completed, these abundance variations basically disappear. 
The He abundance post dredge-up is slightly lower 
in the model with diffusion ($\Delta Y\sim$0.01), the RGB bump magnitude decreases by $\sim$0.07~mag, 
the He-core mass at the He-flash increases by a few 0.001$M_{\odot}$. As a consequence the RGB tip luminosity increases 
at the level of just $\sim$0.01~mag. The abundances of metals (other than CNO) in the He-core are only a few 0.01~dex more abundant 
than originally, whilst they (and He) are less abundant than originally by about the same amount around the core.  

\begin{figure}[!h]
\centering\includegraphics[width=3.0in]{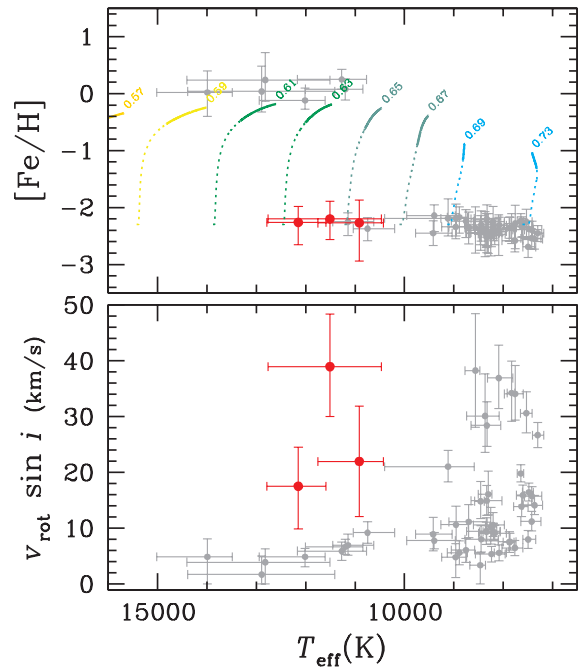}
\caption{The top panel displays the surface [Fe/H] values measured in sample of stars in the metal poor GGCs  
M15, M68 and M92, as a function of their $T_{eff}$. The evolution (dotted segments for an HB age between 0 and 10~Myr, 
solid segments for ages between 30 and 60~Myr) of the surface [Fe/H] 
for HB models with the labelled masses including diffusion is also displayed  
\cite{michaud:11}. The bottom panel displays the projected rotation velocities for the cluster stars in the top panel. The red 
points (with error bars) identify stars with a different behaviour regarding the correlation between $T_{eff}$, rotation velocity and surface [Fe/H] 
(see text for details).}
\label{diff_5}
\end{figure}

Moving to the following HB phase, the ZAHB luminosity decreases by $\sim$0.02~mag compared to the no-diffusion case, 
driven by the lower amount of He in the envelope, but the major effect is on the surface abundances during the following HB 
evolution, as shown in Fig.~\ref{diff_5}. 
At $T_{eff}$ above $\sim$6000~K the surface chemical composition of the models starts to be altered by diffusion; the selective 
effect increases during the HB evolution and is more pronounced for lower HB masses that evolve at higher $T_{eff}$. The abundance of He 
tends to decrease whilst overall the surface metal content increases. Figure~\ref{diff_5} displays, as an example, the predicted 
behaviour of the surface Fe abundance in metal poor HB models. One can notice an increase of surface Fe by a factor of $\sim$10 
within the first 10~Myr, and up to a factor $\sim$100 after 60~Myr. 

These surface abundance changes are again erased by the deepening envelope convection 
for all HB masses that evolve to the following AGB phase. The subsequent hot WD phase sees diffusion altering again in a major way the surface abundances.
Due to the high surface gravity, gravitational settling of metals (down to the edge of the electron degenerate CO core) 
in the doubly layered H and He (DA type) or He-dominated (non-DA type) 
envelopes is very fast. Timescales depend on the dominant element, the envelope thickness, $T_{eff}$ and gravity, but they are at most 
of the order of $10^6$ yr \cite{koester:09}.
Finally, in cool WDs, when the electron degenerate CO core is in the liquid phase, the slow gravitational settling of 
${\rm ^{22}Ne}$ lenghtens the cooling times 
of faint WDs from high metallicity progenitors by up to $\sim$1~Gyr, about 10\% or more of the total cooling time of the models  
\cite{bravo:92, deloye:02, althaus:10, garciaberro:10, camisassa:16}.
This element is the most abundant
{\sl impurity} present in the CO core, and originates from ${\rm ^{14}N}$ during the core He-burning phase. Due to the two additional neutrons 
hosted by the ${\rm ^{22}Ne}$ nucleus relative to ${\rm ^{12}C}$ and ${\rm ^{16}O}$ ions, this element experiences a net downward gravitational force and 
slow settling in the liquid region of the core (inhibited in the solid regions, due to the expected sudden increase of viscosity).
This regime is much denser than the case treated by the Burgers equations, and the ${\rm ^{22}Ne}$ diffusion velocity $w_{\rm ^{22}Ne}$ is determined from 
a diffusion coefficient $D_{\rm ^{22}Ne}$ estimated from molecular dynamics calculations for a strongly coupled plasma \cite{houghto:10} 
$$ w_{\rm ^{22}Ne}=\frac{2 m_p g D_{\rm ^{22}Ne}}{K_B T} $$
where $m_p$ is the proton mass and $g$ the local acceleration of gravity.

The most recent estimate of the diffusion coefficient $D_{\rm ^{22}Ne}$ in the liquid phase provides 

\begin{equation}
D_{\rm ^{22}Ne} \sim D_0 \ 0.53 \left[\frac{\overline{Z}}{Z_{^{22}Ne}}\right]^{2/3} \ (1+0.22 \Gamma) \ exp(-0.135 \Gamma^{0.62})
\label{difu2}
\end{equation}

In this equation ${\rm \overline{Z}}$ denotes the average atomic number of the degenerate core chemical composition, and 
$\Gamma$ the Coulomb parameter ($1  <\Gamma < 180$ in the liquid phase) defined as
 
\begin{equation}
\Gamma= \frac{\overline {Z^{5/3}} e^2}{a_e T}
\label{gammamix}
\end{equation} 

Here $\overline{Z^{5/3}}$ is an average over the ion charges, 
$T$ is the temperature, and the electron sphere radius $a_e$ is equal to $(3/4\pi n_e)^{1/3}$ with $n_e=\overline{Z} n$ the electron density, 
and $n$ the ion density.
 
In addition, 
$$ D_0=\frac{3 \omega_p a^2}{\Gamma^{4/3}} $$

with $\omega_p$ the plasma frequency 
$$ \omega_p = \left[\frac{4 \pi e^2 \overline{Z}^2 n}{\overline{M}}\right]^{1/2} $$

with $\overline{M}$ denoting the average mass of the ions.

The diffusion of ${\rm ^{22}Ne}$ increases the energy budget of the WD (hence increases the cooling times) 
through the contribution 
$(\partial U/\partial \mu)_{T,V} (\partial \mu/\partial t)$  
to the gravitational energy generation coefficient $\epsilon_g$ in the equations of stellar structure (see Eq.~\ref{epsg}).

Moving to stars with initial masses in the range $\sim$1.5-3.0$M_{\odot}$, that display vanishing convective surface layers 
during the MS, calculations with diffusion show the development of surface abundance variations already during the 
pre-MS \cite{vick:11}. During the MS phase they display large surface abundance variations (like He and Ca underabundances, 
iron-peak element overabundances); moreover, local enhancements of iron and nickel abundances below the thin surface convective layers 
lead to extra convective zones, that in some cases can trigger pulsations through the iron-induced $\kappa$-mechanism \cite{charpinet:97} 

\subsection{Inhibition of the efficiency of atomic diffusion}

As already mentioned, diffusion is an element transport mechanism that comes from basic physics principles, and  should be efficient 
in stars. Helioseismic observations tell us that diffusion is efficient in the Sun; its inclusion in stellar models improves 
the match of the inferred sound speed profile, as well as the present depth of the convective envelope and its He abundance 
\cite{bahcall:95, villante:14}.
The very efficient diffusion of metals from the WD envelopes is also confirmed by observations, whereby for the vast majority of 
WDs the chemical composition of the surface is either pure H or pure He \footnote{Metals observed in a minority of WD spectra are ascribed 
to the effect of accretion.}.
Also the diffusion of Ne in the CO liquid cores of WDs is indirectly favoured by the comparison of TO and WD ages for the old metal rich open cluster 
NGC6791. The effect of Ne diffusion increases the age obtained from the WD luminosity function, ensuring agreement with the TO age 
\cite{garciaberro:10}.

However, existing spectroscopic measurements of surface element abundances (e.g. Fe) in a sample of 
GGCs contradict the predictions of stellar 
models that include uninhibited
diffusion \cite{korn:07, lind:08, mucc:11, michaud:11, gruyters:16}. In a star cluster the initial 
mass of the stars evolving in post-MS stages is essentially constant, and measurements of chemical abundances at the TO and the base of the RGB 
should display a difference, due to the effect of diffusion on evolutionary tracks discussed before, 
that is maximized at the TO and essentially disappears along the RGB. 
For example, in case of NGC6397, with [Fe/H]$\sim -$2.1 along the RGB, the TO Fe abundance should be $\sim$0.3~dex lower, 
and the Ca abundance $\sim$0.1~dex larger than on the RGB according to models calculated with diffusion; 
observations show instead TO Fe and Ca abundances $\sim$0.15 and $\sim$ 0.1~dex 
lower than on the RGB respectively \cite{korn:06}. 

This points to a reduction of the effect of diffusion on the surface abundances by some competing mechanism. Nothing of course 
can be said about the efficiency of diffusion in the inner layers\footnote{If diffusion is inhibited only from the outer convective layers 
of low-mass stars, the effect on the interiors can still affect the age determination of clusters and field stars \cite{salaris:16}}. 
The same qualitative result is found in more metal rich clusters like 
NGC6752 and M4, and also on the HB of a number of clusters, as shown by Fig~\ref{diff_5}.  
In this figure one can clearly see that for HB stars hosted by three metal poor GGCs,  
the surface Fe abundance increases abruptly when $T_{eff}$ goes above $\sim$11000~K, whilst theory predicts Fe enhancements (compared to the 
initial value [Fe/H]=$-$2.3~dex typical of these clusters) at much lower temperatures.

Also comparisons of models in the mass range between $\sim$1.5 and $\sim$3.0$M_{\odot}$ calculated including diffusion with surface abundances 
measured in A and F stars, show that diffusion must be moderated by some competing mechanism \cite{richer:00, vick:11, deal:16}.
For example, in samples of A and F stars belonging to the Hyades, Pleiades and 
Coma Berenices Galactic clusters, observed abundances of elements like Na, Fe, Ni 
display a general pattern and star-to-star scatters that point to diffusion being moderated to different degrees by come competing process 
\cite{gebran:10}.

Finally, the observed constant surface Li abundance observed in field halo stars with $T_{eff}$ above $\sim$5500-6000~K and 
[Fe/H] below $\sim -$1.5~dex  --the so-called {\sl Spite plateau} \cite{spite:82}-- is also problematic for models including diffusion, because 
they predict for these stars a progressive decrease of surface Li with increasing $T_{eff}$, that is not observed 
\cite{salaris:01, richard:05}.

\subsubsection{Rotational mixing}

As we will see in Sect.~\ref{rotation}, rotational mixing can in principle counteract the effect of diffusion, and this 
has been invoked to explain the {\sl delayed} (in terms of $T_{eff}$) onset of diffusion in hot HB stars \cite{quievy:09}. 
This inference is related to the distribution of observed projected rotational velocities 
for the hot HB stars of Fig.~\ref{diff_5}, as shown in the lower panel of the same figure.  
All observed stars with $T_{eff}$ above $\sim$11000~K that show spectroscopically the signature of diffusion, are very slow rotators, 
whereas at lower $T_{eff}$ rotation rates are on average higher. Estimates (not based 
on fully consistent stellar evolution models) of the competing effects of meridional circulation and diffusion,  
show that moderation/inhibition of diffusion below $T_{eff} \sim$11000~K by rotational mixing is possible \cite{quievy:09}. 
Interestingly, in support of this idea Fig.~\ref{diff_5} shows three stars with $T_{eff} > $11000~K and large projected rotation velocities with  
no enhancement of surface Fe, at odds with the other slow-rotating objects at the same temperatures.
The link between (partial) inhibition of diffusion and rotation is however probably not justified for A and F stars, that often show 
slow rotation velocities and surface abundances clearly affected by a reduced efficiency of diffusion, compared to the model predictions.

\subsubsection{Mass loss}

Another process that can moderate diffusion in the stellar outer layers is mass loss 
\cite{swenson:95, vauclair:95, vick:13}. Assuming that mass loss (hereafter ML) is spherical and unseparated (chemical composition of the wind equal to 
the photospheric composition) its effect is simply to {\sl peel off} the outer layers of a model. Simple mass conservation 
constraints, coupled to the fact that the structure of the star is unaffected by the amount of mass lost between two consecutive computational timesteps, 
translates into the appearance of an outward flowing interior velocity due to the wind \cite{vick:10} given by

\begin{equation}
v_{w}(r)= -\frac{\dot{M}}{4\pi r^{2}\rho}\frac{m_r}{M_*}
\label{eqn:vwind}
\end{equation}
where $\rho$ is the local density at radial distance $r$ from the centre, $m_r$ is the mass interior to $r$, $M_*$ the total  
mass, and $\dot{M}$ (negative) the mass loss rate. 

Figure~\ref{diff_4} displays the effect of ML on the surface abundances of a typical model at the TO of a metal poor GGC. 
For elements affected mainly by gravitational settling the effect of ML is to increase their surface abundance compared to the 
pure diffusion case. The reason is that with the chosen mass loss rates, below the convective envelope 
$v_w$ counteracts the settling velocity, hence the degree of depletion is reduced. Larger negative values of $\dot{M}$ imply larger 
positive $v_w$ and less diffusion from the envelope. In case of elements supported by radiative levitation in some layers 
below the convective envelope, the effect of ML on the surface 
abundances is more complex, and it is due to the interplay between $v_w$ and the depth at which $g_{\rm rad}$ is larger than the local 
gravitational acceleration $g$. For example, let's assume a ML rate $\dot{M}=-10^{-13}$ and an age of 10~Gyr;   
if $v_w$ is larger than the settling velocity down to $\Delta M=10^{-3} M_*$ from the surface, 
and in those layers  $g_{{\rm rad}, i}> g$ for a generic element $i$, after 10~Gyr the 
wind will have advected to the surface matter at that location, and the surface abundance of element $i$ will be enhanced compared to the initial 
value.

This complex behaviour of the elements supported by radiative levitation can be seen in Fig.~\ref{diff_4}. For an element like Mg 
a ML rate $\dot{M}=- 10^{-13} M_{\odot}/yr$ transforms the surface overabundance of about 0.10~dex into an underabundance 
of about 0.4~dex. A further increase of  $\dot{M}$ reduces the underabundance to about just 0.1~dex. In case of Ca the 
overabundance cause by diffusion is instead increased by a ML rate $\dot{M}=- 10^{-13} M_{\odot}/yr$, and then transformed into an underabundance 
when $\dot{M}=- 10^{-12} M_{\odot}/yr$
 
ML rates of the order of $\approx 10^{-12} M_{\odot}/yr$ are required to explain the Spite-plateau and the behaviour of surface 
abundances in the metal poor GGC NGC6397 \cite{swenson:95, vick:13}. These rates are unlikely, being much larger than the 
solar current mass loss rate $\dot{M}=- 2 \times 10^{-14} M_{\odot}/yr$. 
 
Rates of the order of $\approx 10^{-13} M_{\odot}/yr$ are invoked to explain at least some of the surface abundance patterns seen in A and F stars 
\cite{vick:10}.

\subsubsection{Thermohaline mixing}

The modifications of the chemical abundance profiles caused by diffusion lead to variations of the 
local mean molecular weight. For example He always sinks, introducing a stabilizing contribution to the local $\mu$-gradient. 
On the other hand heavy elements supported by radiative levitation will lead to a destabilizing $\mu$-gradient (increasing outward). 
If this second effect prevails, these layers are subject to thermohaline mixing (see Sect.~\ref{thermohaline}), that 
will tend to erase the abundance changes, hence to moderate the effect of diffusion. 

Published calculations of models for A and F stars including both diffusion and thermohaline mixing,
show that thermohaline mixing (employing the diffusion coefficient derived by \cite{brown:13}) 
play an important role in moderating the effect of diffusion on the surface abundances, although the 
surface abundances of the models do not yet match observations \cite{deal:16}. Both processes have been included
also in calculations to explain the observed population of carbon enhanced metal poor stellar stars 
\cite{matrozis}.

\subsubsection{Generic turbulence}

A pragmatic solution adopted in several investigations is to add {\sl by hand} some turbulence able to counteract 
the efficiency of atomic diffusion in the outer layers. Turbulence here means simply an additional {\sl ad-hoc} term in  
the chemical evolution equations, that acts towards suppressing the development of chemical abundance gradients, not explicitly 
connected to a specific physical mechanism. 
Equation~\ref{eqdiffu}, that showed the contribution of diffusion to the time evolution of the abundance $X_i$ of element $i$, is therefore modified 
to include an additional term

\begin{equation}
\frac{\partial X_i}{\partial t}\bigg|_{M_r} = \frac{1}{\rho r^2}\frac{\partial}{\partial r}\Big(
D_{\rm turb}\rho r^2\frac{\partial X_i}{\partial r}\Big) - \frac{1}{\rho r^2}\frac{\partial}{\partial r}(\rho r^2 X_i w_i),
\label{turbdiff}
\end{equation}

where $D_{\rm turb}$ is the chosen turbulent diffusion coefficient.

The literature presents at least three different choices for $D_{\rm turb}$. To date, the only formulation employed 
to interpret several different sets of data is the following \cite{richer:00, richard:02a, richard:02b}

\begin{equation}\label{turbmodel}
D_{\rm turb, Rich}=f\,D({\rm He})_0\big( \frac{\rho}{\rho_0}\big)^n 
\end{equation}
where 
\[
D({\rm He})_0=\Big[\frac{3.3\times 10^{-15}T^{2.5}}{4\rho \ln(1+1.125\times10^{-16}T^3\rho)}\Big]_0
\]
is the He diffusion coefficient\footnote{The He settling velocity $w_{\rm He}$ can be related to a diffusion coefficient $D({\rm He})$ through
$$ D({\rm He}) \sim \frac{K_B \ T \ w_{\rm He}}{m_{He} \ g}$$
where $m_{\rm He}$ is the mass of the He nucleus.} 
at a certain reference depth, in the approximation of trace amount in an ionized hydrogen plasma. 
The subscript 0 indicates the reference depth 
in terms of either a reference density $\rho_0$ or, more often, a reference temperature $T_0$. In this latter case $\rho_0$ is the density 
at the layer where $T=T_0$. 
The turbulent diffusion coefficient $D_{\rm turb}$ is $f$ times the He diffusion coefficient at the reference depth 0, and 
varying as $\rho^{n}$. 

A choice of parameters $f=400$, ${\rm log}(T_0)=5.5$, $n=-3$ (denoted as T5.5D400-3) allows to match the 
abundances of A and F stars, while a T6.0D400-3 or T6.09D400-3 choice of parameters 
in Eq.~\ref{turbmodel} reproduce the flatness of the Li Spite-plateau \cite{richard:02a}.
Spectroscopy of stars in NGC6397, NGC6752 and M4, three GGCs spanning a range of about 1~dex in initial metallicity, suggest 
choices between T6.0D400-3 and T6.2D400-3 to match the observations \cite{korn:07, mucc:11, gruyters:14}. Interestingly,
for the $\sim$4~Gyr old, solar metallicity open cluster M67, predictions from stellar
models with unhinibited diffusion are generally consistent with observed abundances 
\cite{ogk:14}, although the predicted abundance variations are small.

Two additional forms for $D_{\rm turb}$ have been proposed, but not widely applied, to comparisons with
spectroscopic observations. 
The first one applied to low mass star models with convective envelopes is 

\begin{equation}
D_{\rm turb, VdB}= f \frac{(\rho_{BCZ}/ \rho)^3}{(1-(M_{BCZ}/M_{tot}))^n}
\end{equation}

with $f$=15 and $n$=1.5 \cite{vandenberg:12}, where $BCZ$ denote quantities at the lower boundary of the convective envelope and 
$M_{tot}$ is the total mass of the model.
Due to the steep power-law dependence of the density ratio, $D_{\rm turb}$ becomes negligible in the nuclear-burning 
regions, ensuring that the assumed turbulence will not affect the inner chemical profiles resulting from nucleosynthesis and gravitational settling.

The other proposed parametrization is proportional to the radiative viscosity \cite{morel:02}

\begin{equation}
D_{\rm turb, visc}=f \frac{4 a T^4}{15 c \kappa \rho^2}
\end{equation}

where $f$ a free parameter that is constrained in the range 
$f=1^{+2.0}_{-0.2}$ by helioseismic observations, and spectroscopy of Hyades stars and OB stars (in this latter case 
assuming no competing effect from ML).

\begin{figure}[!h]
\centering\includegraphics[width=3.0in]{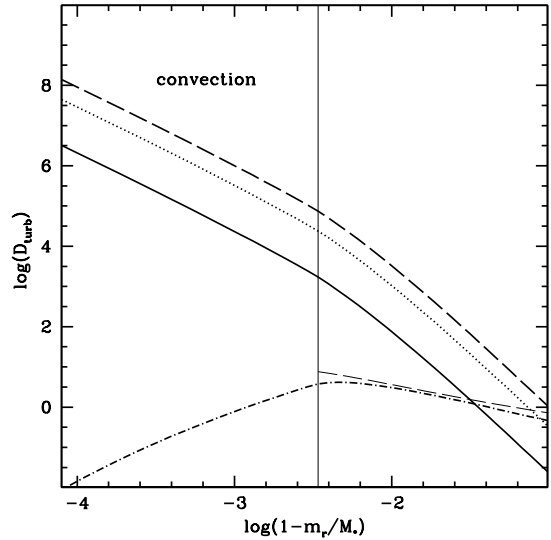}
\caption{Comparison of $D_{\rm turb, Rich}$ solid line for the T6.0D400-3 choice of free parameters, dotted line for T6.2D400-3), 
$D_{\rm turb, VdB}$ (dashed line) and $D_{\rm turb, visc}$ (dot-dashed line) as a function of the mass 
location within a 0.8$M_{\odot}$, [Fe/H]=$-$1.3 model, in the latter phase of its MS evolution. The vertical thin line 
marks the bottom of the convective envelope, whilst the thin dashed line displays the diffusion coeffcient of He (see text for details).}
\label{diff_6}
\end{figure}
 
Figure~\ref{diff_6} compares the different formulations for $D_{\rm turb}$ discussed above, for the outer layers 
of a 0.8$M_{\odot}$, [Fe/H]=$-$1.3 model, in the latter phase of its MS evolution. In case of $D_{\rm turb, Rich}$ we  
display both the T6.0D400-3 and T6.2D400-3 results, the ones that allow a match to results from 
spectroscopic observations, as discussed before. Notice how when moving the reference temperature 
from $log(T_0)=6.0$ to $log(T_0)=6.2$ the entire profile of $D_{\rm turb, Rich}$ shifts to deeper layers, thus counteracting the 
efficiency of diffusion in deeper stellar regions.   
Also, $D_{\rm turb, VdB}$ (displayed with the choice of parameters recommended in \cite{vandenberg:12}) appears to be very similar 
to  $D_{\rm turb, Rich}$ for the T6.2D400-3 choice. On the other hand, both absolute values and trends of $D_{\rm turb, visc}$ (for $f$=1) 
with mass depth are very different from the other cases.

For the sake of comparison we also show the diffusion coefficient of He (that is never supported 
by radiative levitation) at the edge of the convective envelope and below. All $D_{\rm turb}$ choices displayed in the figure  
are able to counteract the diffusion of He from the bottom of the convective zone. Even $D_{\rm turb, visc}$, that is the smallest 
turbulent diffusion coefficient in the figure, matches the one for He, hence is able to inhibit its settling below the convective envelope. 

\section{Phase separation in WDs}
\label{wds}

Besides convection in the non-degenerate envelopes and the diffusion of ${\rm ^{22}Ne}$ in the liquid core, 
another important process that redistributes the chemical abundances in cold WDs is the so-called 
\lq{chemical separation \rq} upon crystallization \cite{garbe:88}. In a nutshell, when a layer with given C and O abundances 
in the liquid phase crystallizes (we assume for the moment the core being made of just C and O, given that the sum C+O is
always more than $\sim$95\% by mass in any realistic WD model), the equilibrium composition in the solid phase --given by the 
phase diagram of the CO mixture-- is very likely to be different. 
This will cause a gradual change of the overall chemical profiles within the CO core, that impacts the energy budget and cooling times, as detailed 
below. 

\begin{figure}
  \resizebox{16pc}{!}{\includegraphics{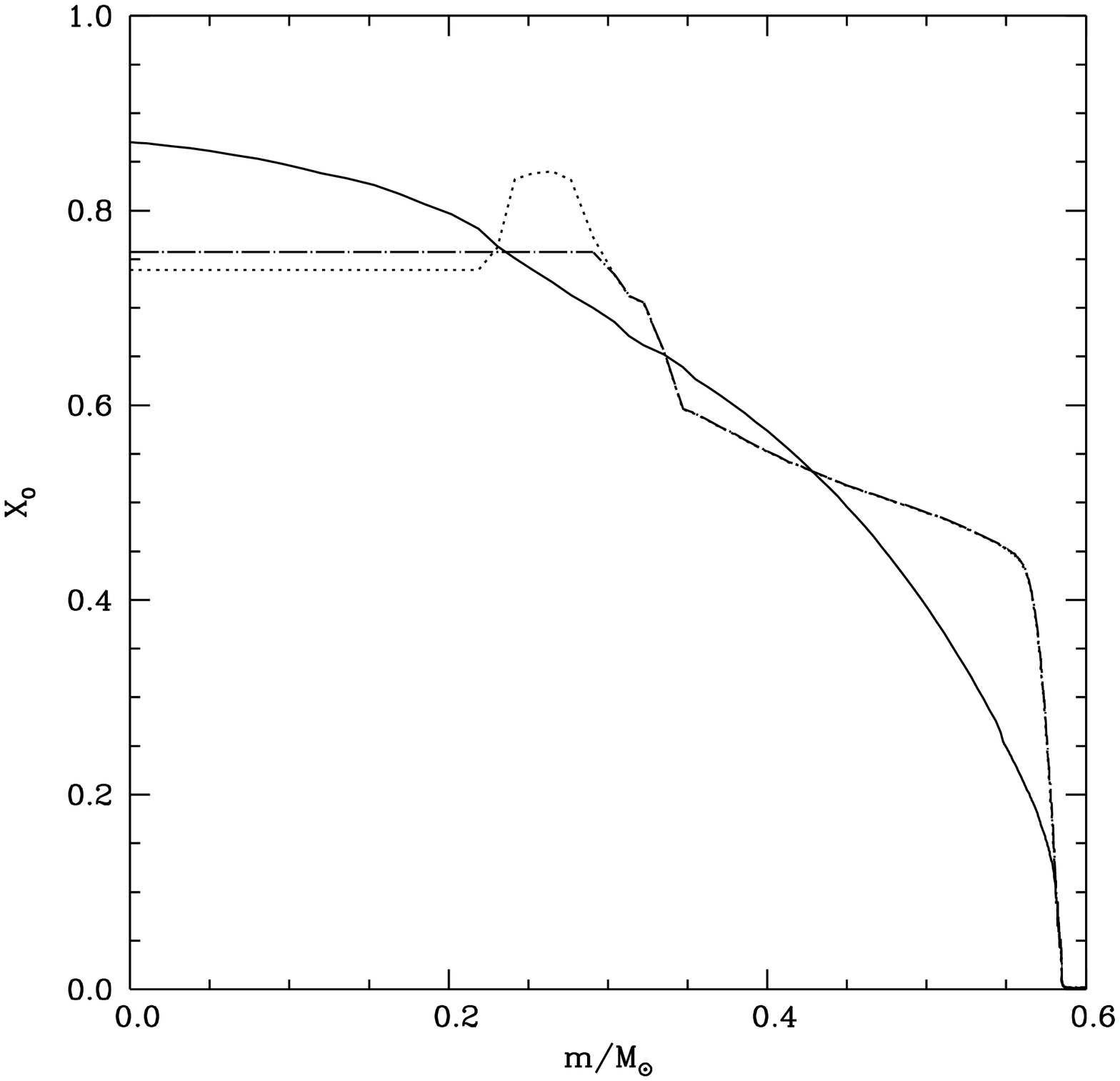}}
  \resizebox{16pc}{!}{\includegraphics{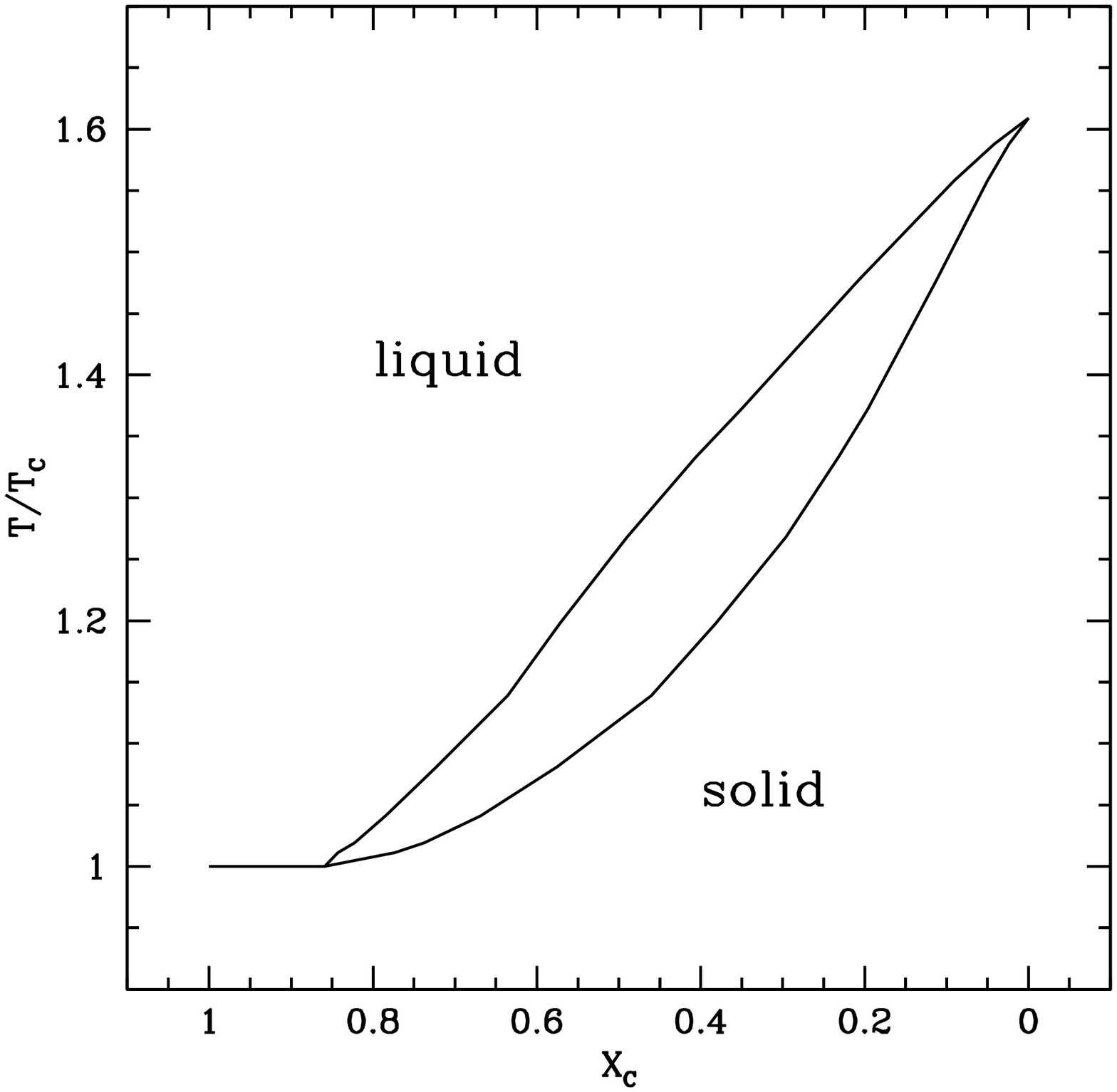}}
\caption{$Left:$ Oxygen profile (in mass fraction) of a 0.61 $M_{\odot}$ WD 
at the beginning of the thermal pulsing phase of the progenitor (dotted line), 
and the same after rehomogenization due to the $\mu$ inversion (dot-dashed line) and after 
complete crystallization (solid line) \cite{salaris:97}. 
$Right:$ Phase diagram of the CO mixture adopted to calculate the profiles in the left panel 
\cite{sc:93}. $T_{\rm C}$ denotes the crystallization temperature of a pure C composition, $X_{\rm C}$ the C mass fraction.}
\label{WD}
\end{figure}

The right-hand  panel of Fig~\ref{WD} displays a phase diagram for the CO mixture \cite{sc:93}, with the regions of liquid and solid phase labelled.
For heuristic purposes let's  
assume a C mass fraction $X_{\rm C}$=0.50 uniformly throughout the core (hence the oxygen mass fraction $X_{\rm O}$ is also $\sim$0.50). 
When the mixture starts to
crystallize at the centre (the value of the Coulomb parameter 
$\Gamma$=180 for the onset of crystallization is reached first in the centre, because of higher densities), the chemical 
composition in the solid phase is determined as follows. One needs to draw a vertical line in the phase diagram of Fig.~\ref{WD} 
with horizontal coordinate equal to 0.50, that runs through the
region belonging to the liquid phase until it intersects the
upper line describing the phase diagram. 
The vertical coordinate of the intersection
point gives the crystallization temperature of the WD
centre (corresponding to $T\sim 1.3 T_{\rm C}$, where $T_{\rm C}$ 
denotes the crystallization temperature of a pure C mixture). From this point one has to draw an horizontal line that will intersect 
the lower segment of the phase diagram in correspondence of a carbon abundance 
$X_{\rm C}\sim 0.30$. This is the equilibrium abundance of carbon in the solid phase 
(the corresponding oxygen abundance is $X_{\rm O}=1-X_{\rm C}$).

Given that $X_{\rm C}$ in the now crystallized centre is
lower than the initial value, conservation of mass requires that 
the carbon abundance in the liquid phase at the
crystallization boundary is increased (hence $X_{\rm O}$ is decreased) with respect to
the original value. This means that
right above the crystallized boundary the molecular weight is now 
lower than in the overlying layers still in the liquid phase
(where the ratio $X_{\rm O}/X_{\rm C}$ is higher). 
An increase of molecular weight with increasing distance from the
centre in the liquid phase causes an instability to develop, and the resulting fast-mixed region extends
outwards in mass as long as the new uniform average $X_{\rm C}$ value  
is higher than the abundance in the next, unperturbed layer (in this case it will reach the edge of the CO core) \cite{moch:83, imgh:97}. 
After mixing is completed, the liquid phase will have an overall
enhanced $X_C$ abundance compared to the value before the crystallization of the central layer.

\begin{figure}[!h]
\centering\includegraphics[width=3.0in]{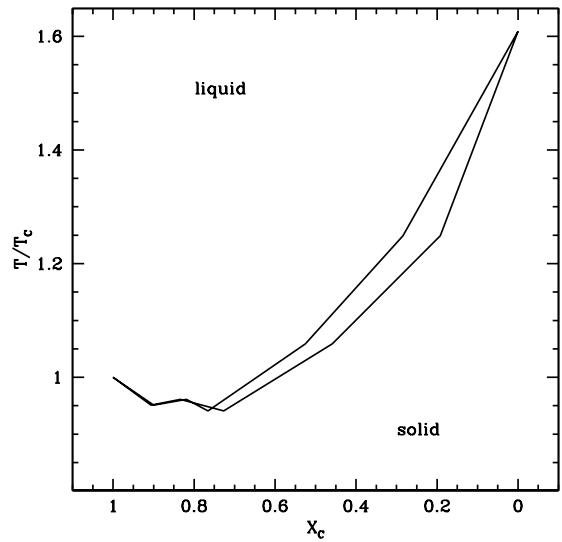}
\caption{The most recent reevaluation of the phase diagram for a CO mixture \cite{horowitz:10}.}
\label{phdiag}
\end{figure}

Let's now suppose that the new value of $X_{\rm C}$ at the boundary of the solid
core is equal to 0.55, when this layer crystallizes --at a lower
temperature than the core because of lower density. The abundance 
in the solid phase can be derived in the same way as before, and
it is now equal $X_{\rm C}\sim 0.35$. This
implies again that right outside the newly crystallized layer the abundance $X_C$ must be 
higher (and $X_{\rm O}$ lower) than in the overlying layer. The instability in the liquid
phase ensues again and the cycle is repeated (mixing in the liquid phase eventually 
stopping when the carbon abundance of the newly crystallized layer 
becomes lower than or equal to the overlying layers still in the liquid phase) 
until the whole degenerate core is crystallized.
The final profile of $X_{\rm C}$ and $X_{\rm O}$ after crystallization is completed 
is no longer homogeneous; $X_{\rm O}$ will display central values higher than in the liquid phase, and 
decrease from the centre outwards, while the opposite is true for $X_{\rm C}$. 

This variation of the $X_{\rm O}/X_{\rm C}$ profile (hence the local value of the molecular weight $\mu$) in the WD core  
during crystallization due to the CO phase diagram has an important impact on the 
WD cooling times, because of the term 
$(\partial U/\partial \mu)_{T, v} (\partial \mu/\partial t)$ in $\epsilon_g$ \cite{salaris:97}.
This means that more energy is available to be released and cooling times are longer. 
Depending on the thickness and chemical composition 
of the non-degenerate layers surrounding the CO core (that regulate the energy release), delays 
induced by chemical separation upon crystallization can reach $\sim$10\% for the coolest objects.

The left-hand panel of Fig.~\ref{WD} displays the evolution of the O-profile 
for a realistic 0.61$M_{\odot}$ WD model \cite{salaris:97}. The first stage of this temporal sequence is represented by 
the profile in the electron degenerate core at the beginning of the thermal pulse phase of the AGB progenitor, 
built during the core He-burning. 
One can notice the local maximum in $X_{\rm O}$ is about 0.25$M_{\odot}$ away from the centre. The associated 
local maximum of the molecular weight $\mu$ causes an instability that homogeneizes the internal layers 
during the liquid phase of the cooling \cite{salaris:97}, 
generating the profile that is maintained until 
the start of core crystallization. The final very different $X_{\rm O}$ profile is attained when the whole core is crystallized, and it 
is produced by the chemical separation upon crystallization. 

The most recent reevaluation of the phase diagram for a binary CO mixture is displayed in Fig.~\ref{phdiag} \cite{horowitz:10}. Compared 
to the widely used phase diagram of Fig.~\ref{WD}, it causes slightly smaller abundance variations for a given initial chemical profile 
in the liquid phase.

Within the CO core there are also small amounts of other metals -- so called minor species, 
with mass fractions of at most the order of the initial 
progenitor metallicity -- that have been either processed through the previous 
burning phases (i.e. Ne), or are unchanged since the formation of the progenitor (i.e. Fe). 
Due to the extreme complexity of calculating a multi-component phase diagram, 
a ternary CONe mixture is often assumed to behave as an effective binary mixture composed of
neon (or iron) plus an element of average charge $\langle
Z\rangle=7$ determined from the C and O abundances. 
Ne is important, given that  $^{22}$Ne is the most 
abundant species after C and O, with a mass fraction $X_{\rm ^{22}Ne}\sim Z$, where 
$Z$ is the initial metallicity of the progenitor ($Z\sim 0.02$ for the solar metallicity). 
A detailed phase diagram for a three component mixture 
CONe shows that the effect of Ne 
on the final WD cooling times is negligible, when the separation of carbon and oxygen is accounted for \cite{segr:96}. 

\section{Rotation and rotational element transport mechanisms}
\label{rotation}

Basic considerations about the angular momentum evolution in a contracting  
protostellar cloud, observations of the solar magnetic field, stellar spectroscopy (and now also asteroseismology) 
dictate/show that stars do rotate.
Observations of MS O- and B-type stars in the Galaxy and the Magellanic Clouds reveal 
average projected rotational velocities 
of the order of 150 km/s \cite{hunter:08, pennygies:09, braganca:12, huang:10}, with values up to 350-400 km/s. 
Average rotational velocities of the order of 150~km/s are observed also in MS 
A- and F-type stars \cite{royer:07}, fast decreasing when moving to later spectral types \cite{stauff:86}, 
whilst giant stars display typically slow projected rotational velocities below $\sim$10 Km/s \cite{carlberg:11}. 

All fast rotators among O-stars show surface He-enrichments not predicted by standard non-rotating models 
\cite{Herrero:92}. Also, high rotational velocities along the MS are associated to enhancements of surface N 
that cannot be reproduced by non-rotating stellar models \cite{hunter:09, martins:16}.

For a long time rotation has not been an ingredient of 
standard stellar models, due to the increase 
in complexity and uncertainty --free parameters-- related to the inclusion of  
rotation. Moreover, the basic principle that the explanation relying on the smallest
number of hypotheses is the one to be preferred, coupled to the many successes of non-rotating 
stellar models, means that rotation has been generally considered only a second order effect. 
It is however clear that it can have important effects 
on the structure and evolution of stars, mainly through its effect on the evolution of the chemical stratification.

\subsection{1D modelling of rotating stars}

The physical basis for the inclusion of rotation in detailed stellar evolution computations was laid down fifty years ago 
\cite{monaghan:65, ostriker:68} 
and more recently hydrodynamical treatments of the problem have appeared \cite{deupree:90, deupree:01, rieutord:13}.
Detailed stellar evolution calculations and comprehensive libraries of stellar models 
are however still (and for the foreseeable future) possible only with standard 1D modelling, and  
the basic assumptions to simulate the average mechanical and thermal distortions induced by rotation 
without basically altering the standard equations, are the following\footnote{It may seem surprising 
to be able to calculate 1D rotating models, given that the centrifugal forces reduce the effective gravity according 
to the latitude and introduce deviations from spherical simmetry.}:

\begin{enumerate}
\item{Roche approximation, i.e. the gravitational potential $\Psi$ is the same as if the total mass of the star were 
concentrated at the centre;}
\item{The angular velocity $\Omega$ and chemical composition are constant along isobars, e.g. surfaces of constant pressure ({\sl shellular} rotation). 
This follows the assumption that turbulence is anisotropic, with a stronger transport in the horizontal (tangential to an isobar) 
direction than in the vertical (perpendicular to an isobar) one \cite{z:92}. The angular velocity in this case depends very weakly on the colatitude, hence
$\Omega \equiv \Omega(r)$, meaning that it varies according only to the radial coordinate of the isobars.}
\item{The mean radius $r_P$ of an isobar is defined by $V_P=({4\pi}/{3})r_P^3$, where $V_P$ is the volume inside the isobar.}
\end{enumerate} 

With these assumptions, the set of equations of stellar structure for a rotating star can be written in 1D as \cite{mm:97, maederbook}

\begin{eqnarray}
\frac{\partial r_P}{\partial m_P} &=& \frac{1}{4 \pi r_P^2
  \bar{\rho}} \label{eq-controt2} \nonumber\\
\frac{\partial P}{\partial m_P}& = & - \frac{G m_P}{4 \pi r_P^4} f_p \nonumber \\
\frac{\partial L_P}{\partial m_P} & = & \epsilon_n -\epsilon_\nu + \epsilon_g \label{eqrotstruct}\\
\frac{\partial \bar{T}}{\partial m_P} & = & - \frac{G m_P T}{ 4 \pi
  r_P^{4} P} \nabla_P \label{eq-heatrot2} \nonumber
\end{eqnarray}
with $\nabla_P$ being the appropriate temperature gradient ${\rm d {\rm ln}}(\bar{T})/{\rm d {\rm ln}}(P)$. In case of radiative transport 
\begin{equation}
\nabla_{P, rad} = - \frac{3 \kappa}{ 16 \pi a c G} \frac{P}{T^4}
\frac{L_P}{m_P} \frac{f_T}{f_p}
\label{eqsrot}
\end{equation}

When comparing these equations to Eqs,~\ref{eqss1}, \ref{eqss2}, \ref{eqss3} and \ref{eqss4}, one can notice that 
they are formally identical, apart from the 
{\sl form factors} $f_P$ and $f_T$. The only difference is the interpretation of the variables.
In case of rotating models $r_{P}$ is the mean radius of an isobar defined by the value $P$ of the pressure, 
$\bar{\rho}$ and $\bar{T}$ are the volume-averaged density and temperature between two contiguous isobars (the difference 
with averages on the isobars is negligible if the mass grid of the models is dense enough), 
$L_P$ and $m_P$ are the mass and luminosity inside a given isobar.
Calculations of the energy generation coefficients, adiabatic gradient, opacity, make use of 
$\bar{\rho}$ and $\bar{T}$. The equation of state is expressed in terms of $P=P(\bar{\rho}, \bar{T})$ and the chemical composition 
on the isobar $P$.  

Analogous to the case of non-rotating models the system of equations is solved considering 
the \lq{Lagrangian}\rq\ independent variable $m_P$, with $r_{P}, L_{P}, P, \bar{T}$ as unknowns.

The form factors $f_P$ and $f_T$ are defined as 

\begin{equation}
f_P = \frac{4 \pi r_P^4}{G m_P S_P} \frac{1}{\langle g^{-1} \rangle}.
\end{equation}

\begin{equation}
f_T = \left( \frac{4 \pi r_P^2}{S_P}\right)^2 \frac{1}{\langle
    g \rangle \langle g^{-1} \rangle}.
\end{equation}

The quantities $\langle g \rangle$ and $\langle g^{-1} \rangle$
are average values of the gravity $g$ over an isobar $P$ with surface area $S_P$. For a generic variable $f$ this average is defined as 

\begin{equation}
<f>=\frac{1}{S_P}\int_{P=constant}f \ d\sigma
\label{average}
\end{equation}

where ${\rm d \sigma}$ is an infinitesimal element of the isobar surface $P$. 
At the non-rotation limit $f_P$ and $f_T$ converge to unity and the equations 
are reduced to their non-rotating form.

To solve equation \ref{eqsrotstruct} one needs to evaluate $f_P$ and $f_T$, and this requires the calculation of the surfaces of isobars, 
e.g. surfaces where $\Psi_P$ is constant, with
\begin{equation}
\Psi_P = -\Phi + \frac{1}{2}\Omega^2 r^2 \sin^2\theta
\end{equation}

where $\Phi$ is the Roche gravitational potential, $r$ the radial distance from the centre and $\theta$
the colatitude ($\theta$=0 at the poles;  see e.g. \cite{zeng:02} for an example of how to calculate integrals \ref{average} for a given
$\Psi_P$). It is through the calculation of $f_P$ and $f_T$ that the angular velocity profile enters 
the equation of stellar evolution.  
It is worth noticing that all this formalism works also in the case of a \lq{conservative}\rq potential (for example the simple case of 
solid body rotation), e.g. in case 
the centrifugal acceleration can be derived from a potential \cite{kt:70, endal:76}, that is not the case for shellular rotation. 
A potential of this type is used for example in \cite{kt:70}
\begin{equation}
\Psi = \Phi - \frac{1}{2}\Omega^2 r^2 \sin^2\theta
\end{equation}
In this case instead of isobaric surfaces one reads equipotential surfaces, and the equations 
(including $f_P$ and $f_T$) are the same as the shellular case.

\subsection{Chemical element and angular momentum transport}

A tricky issue for modelling rotating stars with 1D stellar models is how to 
describe the transport of angular momentum --that determines the evolution of $\Omega$-- and 
the chemical mixing associated with rotation. These are two facets of the same problem, for  
rotation triggers hydrodynamical instabilities and large scale motions of the 
gas in radiative regions, that result in transport of both angular momentum and chemical elements.
In depth discussions about rotation driven instabilities can be found in \cite{talon:07, palacios:13, maederbook}. 
Here we just give a brief overview, focussing on the actual implementation in stellar evolution calculations.

The Eddington-Sweet meridional circulation is one of the major instabilities caused by rotation. In simple terms, 
we can compare a non-rotating star in radiative equilibrium with its solid-body rotating counterpart.
The equipotential surfaces of the non-rotating star are spherical, whilst in case of a  
solid-body rotator they are 
rotational ellipsoids, and two contiguous equipotential surfaces will diverge in distance from each other at the equator. 
Given that the effective gravity $g$ is proportional to the gradient of the potential (that is normal to the 
equipotential surfaces), it will vary with latitude on an equipotential 
surface. As a consequence the temperature will be hotter at the poles and cooler at the equator --as demonstrated by von Zeipel 
almost a century ago, the energy flux is proportional to the local value of $g$ (the von Zeipel theorem)-- preventing the star 
from maintaining hydrostatic equilibrium. 

The solution to this paradox is to invoke large scale mass motions that transport energy, the so-called 
{\sl meridional circulation}, moving material inwards from the equator and upwards along the rotational axis towards the poles.
The timescale for this mixing process -- the \lq{Eddington-Sweet}\rq\ timescale -- was estimated to be

\begin{equation}
t_{ES} \approx \frac{G M^2}{LR} \frac{G M}{\Omega^2 R^3}
\end{equation}

where $L$, $M$ and $R$ are the stellar luminosity, mass and radius, $\Omega$ the angular velocity, and the first term is the Kelvin-Helmholtz thermal timescale. 
These early estimates of $t_{ES}$ were much shorter than the MS lifetime of stars, even for modest rotation rates, hence 
rotating stars should be fully mixed, contradicting observational data. 
The presence of chemical abundance stratifications ($\mu$-gradients) in the interior of rotating stars can however increase 
$t_{ES}$ considerably compared to the early estimates.
A widely used modelling of the meridional circulation included in stellar evolution codes (that will be used later when discussing the implementation of rotational mixing)
develops the circulation velocity vector {\bf U} into two components \cite{z:92}: 

\begin{equation}
{\bf U}= U_2(r) P_2(cos(\theta)) {\bf e}_r + V_2(r) \frac{dP_2(\theta)}{d \theta} {\bf e}_\theta 
\end{equation}

where $r$ is the radial coordinate, $\theta$ the colatitude, and $P_2$ the Legendre polynomial of order 2.
The radial components of these velocities, that will be used to treat chemical element transport, are  
related through:

\begin{equation}
\frac{1}{r} \frac{d}{dr} [\rho r^2 U_2(r)]-6 \rho V_2(r)) = 0 
\end{equation}

Another important effect of meridional circulation is that it advects also angular momentum. 
Local variations of angular velocity with time due to this angular momentum transport, plus the effect of contraction and expansion of the stellar layers 
(and eventually angular momentum loss 
from mass loss) will generate --starting for example from solid body rotation, usually assumed for pre-MS fully convective stars-- 
a variation of angular rotation velocity with depth in radiative regions.
Hence a \lq{shear}\rq\ develops between neighbouring layers, that leads to instabilities. This stems from the fact that the minimum energy 
state of a rotating fluid is solid body rotation, and if the star develops differential rotation, it is possible to extract energy by homogenizing the 
velocities through transport of material.

The strong thermal and molecular weight radial stratification in radiative zones tend to oppose the homogeneization of the 
rotational velocities, while it is reasonable to assume that 
no restoring forces oppose horizontal displacements. As a consequence, horizontal 
shear (along an isobar) is expected to generate a strong turbulence on short dynamical timescales, justifying the assumption of 
shellular rotation.
Eventually, thermal diffusion and horizontal shear can reduce the stabilizing effect of the vertical (radial) thermal and chemical 
stratification and induce element and angular momentum transport if the Richardson number $Ri$ satisfies this criterion: 
\begin{equation}
Ri \equiv \frac{N^2}{(d u/d z )^2} < Ri_{\rm crit} = \frac{1}{4}
\end{equation}
where $u$ is the velocity of the fluid elements and
$z$ designates the vertical direction\footnote{Hydrodynamical simulations 
show that shear mixing is already efficient for values of $Ri$ higher than 1/4 \cite{bruggen:01}.}. 

The Solberg-H\o iland instability, plus additional instabilities (i.e. Goldreich-Schubert-Fricke, ABCD instabilities) 
that can develop when equipotentials and isobars do not coincide  --so-called baroclinic instabilities \cite{maederbook}--  
do potentially contribute to the transport of angular momentum and chemicals\footnote{The Solberg-H\o iland criterion for stability 
is essentially the Ledoux criterion when rotation is present. For a moderate outward decrease of the angular velocity rotation  
favours stability compared to the Ledoux criterion.}.  
Some of these instabilities are implemented in calculations of rotating stellar models (see Section \ref{diffusive}), even though 
their treatment is considered to be much more uncertain than the case of meridional circulation and shear instabilities 
\cite{talon:07, palacios:13}. 

In case of radiatively driven winds the von Zeipel result has important consequences on the mass loss rates to be employed in the calculation 
of rotating stellar models, because in general it causes an increase of the mass loss efficiency compared to the case of a non-rotating counterpart.
Various prescriptions of the mass loss enhancement due to rotation can be found in the literature. As an example we report the prescription 
employed in the MESA code \cite{mesa13}:

\begin{equation}
  \frac{dM}{dt}(\Omega)=\frac{dM}{dt}(\Omega=0) \left(\frac{1}{(1-\Omega/\Omega_{crit})}\right)^{\zeta}
\end{equation} 

with $\zeta$=0.43, and $\Omega_{crit}^2=(1-L/L_{Edd})GM/R^3$ and $L_{Edd}=4 \pi c G M/\kappa$ averaged over a certain optical depth range.

\subsubsection{Advective/diffusive implementation}
\label{advect}

Meridional circulation and shear instability are considered as the main mechanisms for angular momentum transport and rotational mixing 
in a number of stellar evolution codes, e.g. STAREVOL \cite{palacios:06, decressin09}, 
CESTAM \cite{marques13}, the Geneva stellar evolution codes \cite{mm:97, genevacode} and the  
FRANEC code \cite{cl13}, that implement advective+diffusive transport of angular momentum and diffusive rotational chemical mixing.

The transport equation for a chemical element with mass fraction $X_i$ is written 
as\footnote{From now on, for simplifying the notation, 
we will denote with $r$ the radius of the isobar $r_P$.}:
\begin{eqnarray}
  {\partial X_i \over \partial t}\bigg|_{M_r} = {1 \over \rho r^2} \, {\partial \over \partial r}
  \left (\rho \, r^2 \, D_{\text{chem}} \, {\partial X_i \over \partial r}\right )  \
\label{difel}
\end{eqnarray}
where $D_{\text{chem}}$ is the sum of the 
vertical shear diffusion coefficient $D_{\rm shear}$ and the effective diffusion 
coefficient, $D_{\rm eff}$, which accounts for the combined  
effect of the strong horizontal shear diffusion, $D_{\rm h}$ and of the meridional currents:
\begin{equation}
  D_\text{eff} =  \frac{1}{30} \frac{\left| r\ U_2^2(r) \right|^2}{D_{\rm h}}
\end{equation}

The equation for the transport of angular momentum is written as \cite{mmproc:05}: 
\begin{eqnarray}
{\partial \over \partial t}(r^2 \Omega)\bigg|_{M_r}
={1 \over 5 \, \rho r^2}{\partial \over \partial r}(\rho \,  r^4 \Omega \,U_2(r))
+{1 \over \rho r^2}{\partial \over \partial r}
\left(\rho  \, D_{\text{ang}} \,  r^4 \,  {\partial \Omega \over \partial r} \right) \
\label{angmom}
\end{eqnarray}

here $D_{\text{ang}}$ denotes the diffusion coefficient for angular momentum.
The second term on the right-hand side of Eq.~\ref{angmom} is a diffusion term, similar in its form to Eq.~\ref{difel}, 
while the first term is an advective term, modelling
the transport by a velocity current. Equation~(\ref{difel}) does not 
contain an advective term, following \cite{Chaboyer1992} who showed that the combined effect of turbulence and 
circulation currents is equivalent to a diffusion process for the element transport.

There are various expressions in the literature for $U_2$. The complete description by \cite{mz98} in case of shellular rotation provides:
\begin{eqnarray}
U_2 (r)& = &\frac{P}{\bar{\rho} \bar{g} C_P \bar{T}}\frac{1}{\left( \nabla _{\rm ad} -
  \nabla + (\phi/\delta) \nabla _\mu \right)} \left[ \frac{L}{M} \left( E_\Omega + E_\mu \right) +  \frac {\bar{T}C_P}{\delta} 
\frac{\partial \Theta}{\partial t}\right] \nonumber\\
\end{eqnarray} 

The barred symbols mean averages over the isobar.
$E_{\Omega}$ and $E_{\mu}$ are complicated terms that depend on the angular velocity profile and mean molecular weight fluctuations 
on an isobar (see \cite{maederbook} for the complete formulas and the derivation), $\Theta$ describes the density fluctuation on an isobar,
$M$ here denotes the reduced mass
\begin{equation}
M=M_{r_P} \left(1-\frac{\Omega^2}{2 \pi G \rho_m}\right)
\end{equation} 
where $\rho_m$ is the mean density inside the considered isobar surface, and $L$ is the luminosity within the same surface.

In case of convective regions the transport of angular momentum is very uncertain, 
as the interaction between convection and meridional currents is not well understood.
Traditionally, one can choose between the following two limiting cases. 
If convection inhibits meridional currents, angular momentum is redistributed 
very efficiently by convection, as in case of chemical elements, and the result will be solid body rotation, as in the solar convective envelope. 
If meridional currents dominate -- one can hypothesize that this is the case applicable to large, rarefied RGB envelopes, 
where convective elements may collide 
elastically rather than inelastically as in the solid body case -- it is the specific angular momentum that is expected to be 
uniform. Hydrodynamical simulations have shown that the extended deep convective envelopes of red giant stars are likely to undergo 
radial differential rotation, with an angular velocity profile of the form $\Omega(r) \propto r^{-0.5}$ \cite{palacios:09}.
In general, solid body rotation is usually prescribed in convective regions.

Angular momentum losses due to stellar winds need also to be accounted for. In the assumption that the mass loss is spherically symmetric, 
the rate of angular momentum loss during a given timestep will be approximately equal to the average specific angular momentum at the surface of 
the star multiplied by the assumed mass loss rate.

\begin{figure}[!h]
\centering\includegraphics[width=3.0in]{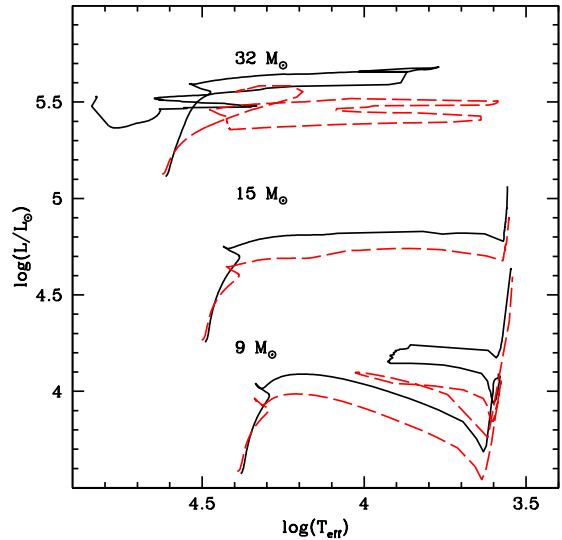}
\caption{Comparison of the HRD of 9, 15 and 32$M_{\odot}$ solar initial composition models, with 
(solid line) and without (dashed line) rotation, from the Geneva calculations \cite{ekstroem12}.}
\label{rot_1}
\end{figure}

To discuss the effect of rotation and rotational mixing on stellar evolution models, we consider as an example models from \cite{ekstroem12}, 
calculated with the Geneva code.
Figure~\ref{rot_1} compares the HRD of 9, 15 and 32$M_{\odot}$ solar initial chemical composition models with and without rotation, 
from the zero age main sequence (ZAMS) to the end of core C-burning. 
In these calculations $D_\text{h}$ has been taken as 

\begin{equation}
  D_\text{h} =  \frac{1}{c_\text{h}}\ r\ \left| 2\,V_2(r) - \alpha\,U_2(r) \right| 
\end{equation}
from \cite{z:92}, assuming $c_\text{h}$=1 and $\alpha = \frac{1}{2} \frac{\text{d} \ln (r^2 \Omega)}{\text{d} \ln r}$. $D_{\rm shear}$ is expressed as   
\begin{equation}
  D_\text{shear} =  f_\text{energ} \frac{H_P}{g\delta}\frac{K_T}{\left[\frac{\varphi}{\delta}\nabla_\mu 
        + \left( \nabla_\text{ad} - \nabla_\text{rad} \right)\right]} 
    \left( \frac{9\pi}{32}\ \Omega\ \frac{\text{d} \ln \Omega}{\text{d} \ln r} \right)^2
\end{equation}

from \cite{maed97}, where $f_\text{energ} = 1$ (the fraction of the excess energy in the shear that contributes to mixing). 

The models are initialized as solid bodies at the ZAMS, and then evolved according to the prescriptions 
for the angular momentum and chemical transport described above. The prescribed initial equatorial rotation velocity is 0.4 times the 
critical velocity (when the centrifugal acceleration in the equatorial plane exactly compensates the gravitational 
acceleration) of the corresponding stellar mass. This is equal to values between $\sim$250 and $\sim$300 km/s 
for the three displayed masses.

The effect of rotation on the tracks (that of course depends on initial velocities) is striking. Models calculated with rotation 
evolve to higher luminosities during the MS and stay more luminous also during the following evolutionary phases.
This is mainly due  
to the consequences of the chemical element transport associated to rotation. In general, rotational mixing works in the direction of erasing chemical 
gradients within the star, hence during the MS this causes a continuous slow ingestion of fresh hydrogen into the H-depleted convective core, as well 
as a slow transport towards the surface of elements whose abundances are increased by H-burning (e.g., He and ${\rm ^{14}N}$). 
The effect of the ingestion of H in the convective core is to increase the MS lifetime compared to the non-rotating counterpart, and 
to produce slightly larger He-cores at the end of core H-burning.

\begin{figure}[!h]
\centering\includegraphics[width=5.5in]{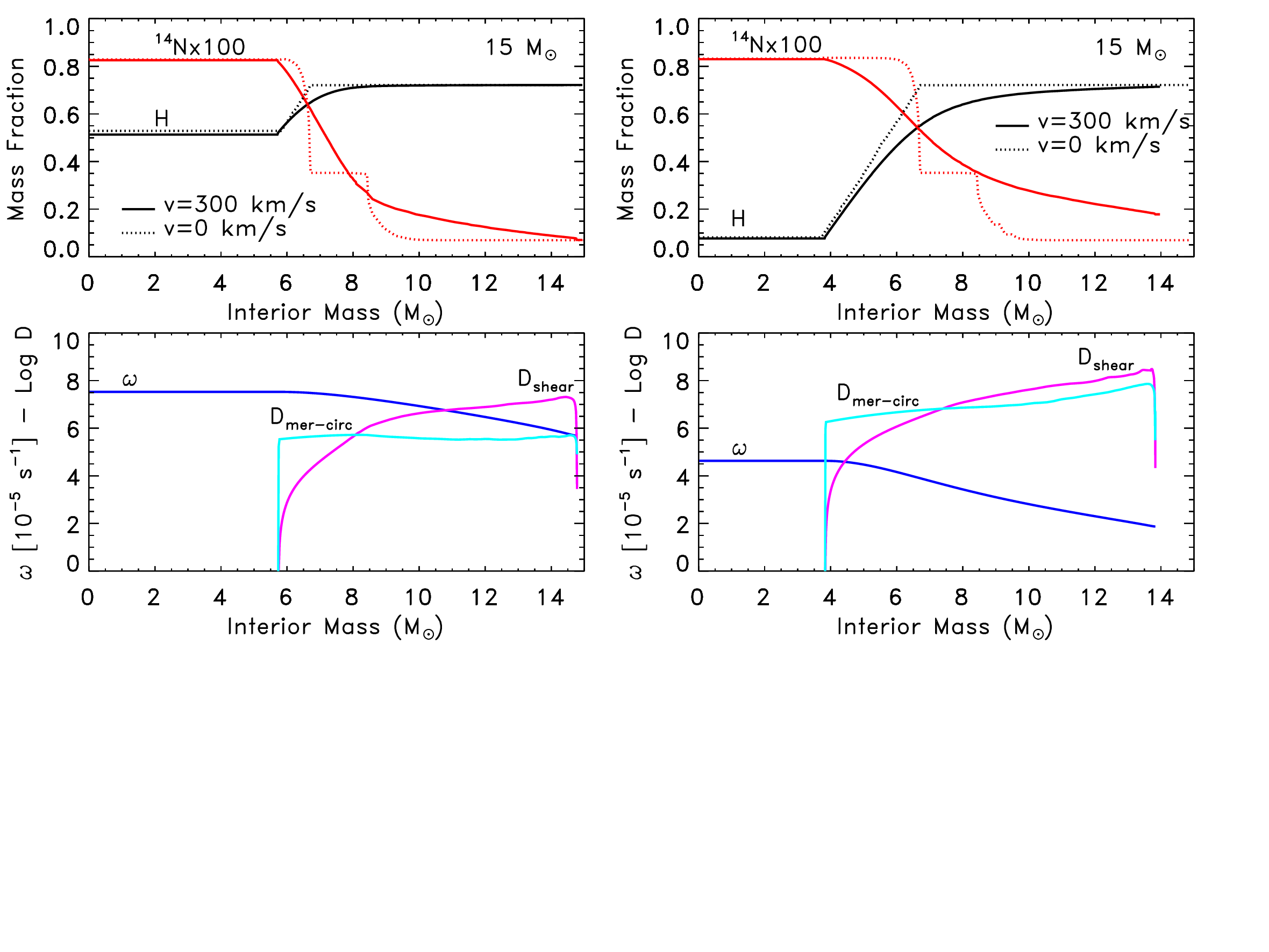}
\caption{Internal profiles of H and N abundances, 
angular velocity and chemical diffusion coefficients for shear and meridional circulation respectively, as a function 
of mass, within 15$M_{\odot}$ solar initial composition models from \cite{cl13}, with the labelled initial rotation rates. 
The different panels refer to two MS stages, when the central H mass fraction 
is equal to 50\% (left panel) and 10\% (right panel) respectively (courtesy of M. Limongi).}
\label{rot_2}
\end{figure}

To this purpose, Figure~\ref{rot_2} displays the internal profiles (taken during the MS 
when central H is reduced to --from left to right-- 50\% and 10\% by mass, respectively) of H and N abundances, 
angular velocity and chemical diffusion 
coefficients for shear and meridional circulation in solar initial composition, 
15$M_{\odot}$ rotating and non-rotating evolutionary models from \cite{cl13}, similar to the Geneva calculations.
One can see very clearly the effect of rotational mixing, that smooths out gradients in the N and H abundance profiles. Also, 
shear mixing dominates in the external layers, where the gradient of angular velocity gets progressively steeper,  
while meridional circulation is the more efficient chemical transport process close to 
the edge of the convective core. Notice also the flat angular velocity profile in the convective core, due to the solid body assumption.

The different paths in the HRD followed by rotating and non-rotating models together with the different lifetimes lead also 
to different mass loss histories, hence different total masses at the end of the MS and during the post-MS phases.
This impacts for example (for a fixed convective mixing scheme) the occurrence of \lq{loops}\rq in the HRD, for the blue or red 
location of a model during the giant phase depends on thickness of the H-rich envelope (thick enough envelopes keep the 
models at the red side of the HR, whilst when their mass decrease below a threshold value, the models move to the blue).
During the He-burning phase the slow ingestion of fresh He in the core lowers the final C/O ratio and increases the mass 
of the CO core \cite{cl13}, affecting the pre-supernova structure of the massive models.

\begin{figure}[!h]
\centering\includegraphics[width=2.5in]{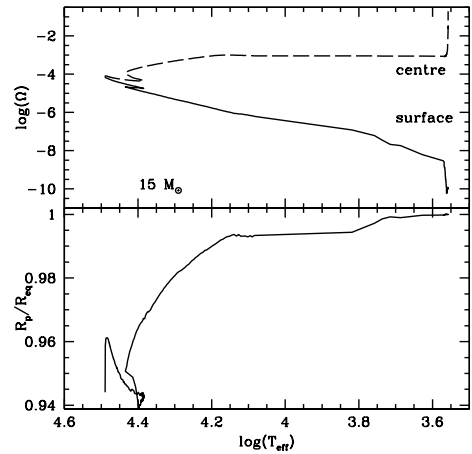}
\caption{The upper panel displays the evolution of the surface and central angular velocity  as a function of 
$T_{eff}$ for the 15$M_{\odot}$ models in Fig.~\ref{rot_1}. The lower panel 
displays the evolution of the ratio of the 
polar and equatorial radius.}
\label{rot_3}
\end{figure}


The upper panel of Fig.~\ref{rot_3} shows the evolution of the angular velocity $\Omega$ taken at the surface and at the centre of 
the 15$M_{\odot}$ models. 
After the ZAMS model with enforced solid body rotation is left to evolve, 
there is a readjustment of the rotational profile at the very beginning of the evolution, until  
the equilibrium rotational profile is reached, and the angular velocity starts to evolve under the action of the transport mechanisms. 
The general trend is to transfer angular momentum from the contracting core towards the external layers, but this is counterbalanced by the mass loss 
that removes angular momentum from the surface, and eventual expansions of the convective envelopes that slow down the surface.

For the model shown in Fig.~\ref{rot_3} the net effect is a constant decrease of the surface angular velocity along the whole evolution, 
whereas the centre of the star displays a moderate increase of angular velocity during the MS, followed by a plateau and a sharp increase during the 
giant phase.
The models are never very far from spherical symmetry, as shown by the lower panel of  Fig.~\ref{rot_3}, that displays the ratio of the 
polar to the equatorial radius along the whole evolution.

\begin{figure}[!h]
\centering\includegraphics[width=3.5in]{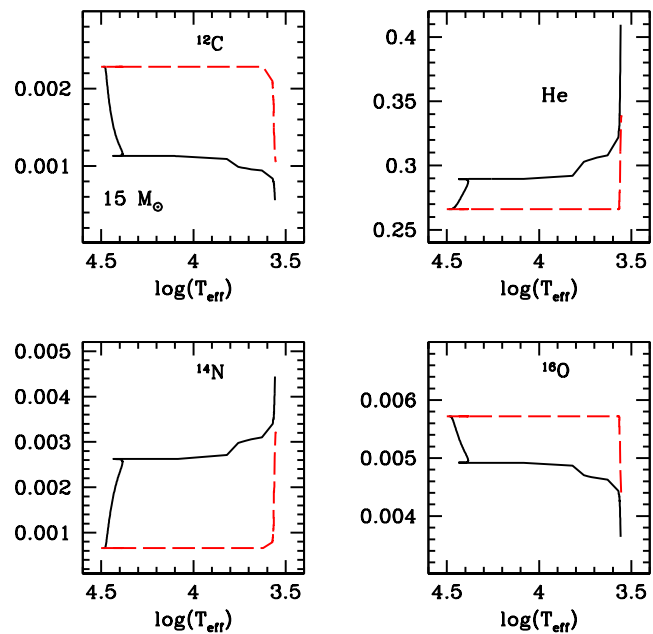}
\caption{Evolution with $T_{eff}$ of the surface abundances (in mass fractions) 
of the labelled elements, for the rotating (solid lines) and 
non rotating (dashed lines) models in Fig.~\ref{rot_1}.}
\label{rot_4}
\end{figure}


The effect of element transport on the surface abundances (in mass fractions) of some key elements (He, C, N, O) 
for the same 15$M_{\odot}$ evolution is displayed in Fig.~\ref{rot_4}. Notice that in the non-rotating models the 
abundances change only due to the dredge-up  
during the red giant phase.
Rotating models display instead abundance changes already during the MS, due to rotational mixing that tends to erase chemical gradients. 
This explains the increase of ${\rm ^{14}N}$ and He, whose abundance increase in the central regions due to CNO H-burning, and the depletion of 
${\rm ^{12}C}$ and ${\rm ^{16}N}$, caused by the decreased O and N equilibrium abundances, compared to the initial scaled solar values.  
The abundances remain almost constant during the fast transition to the giant phase, and then change again due to the dredge-up.

The detailed behaviour of the evolution of the angular momentum and chemical abundance profiles is however strongly dependent 
on the precise choice of the diffusion coefficients in Eqs.~\ref{difel} and \ref{angmom}.
We can see for example that the coefficient employed in these calculations contain parameters like $c_\text{h}$ 
and $f_\text{energ}$ that are set to fixed constant values not derived from first principles.

\begin{figure}[!h]
\centering\includegraphics[width=3.5in]{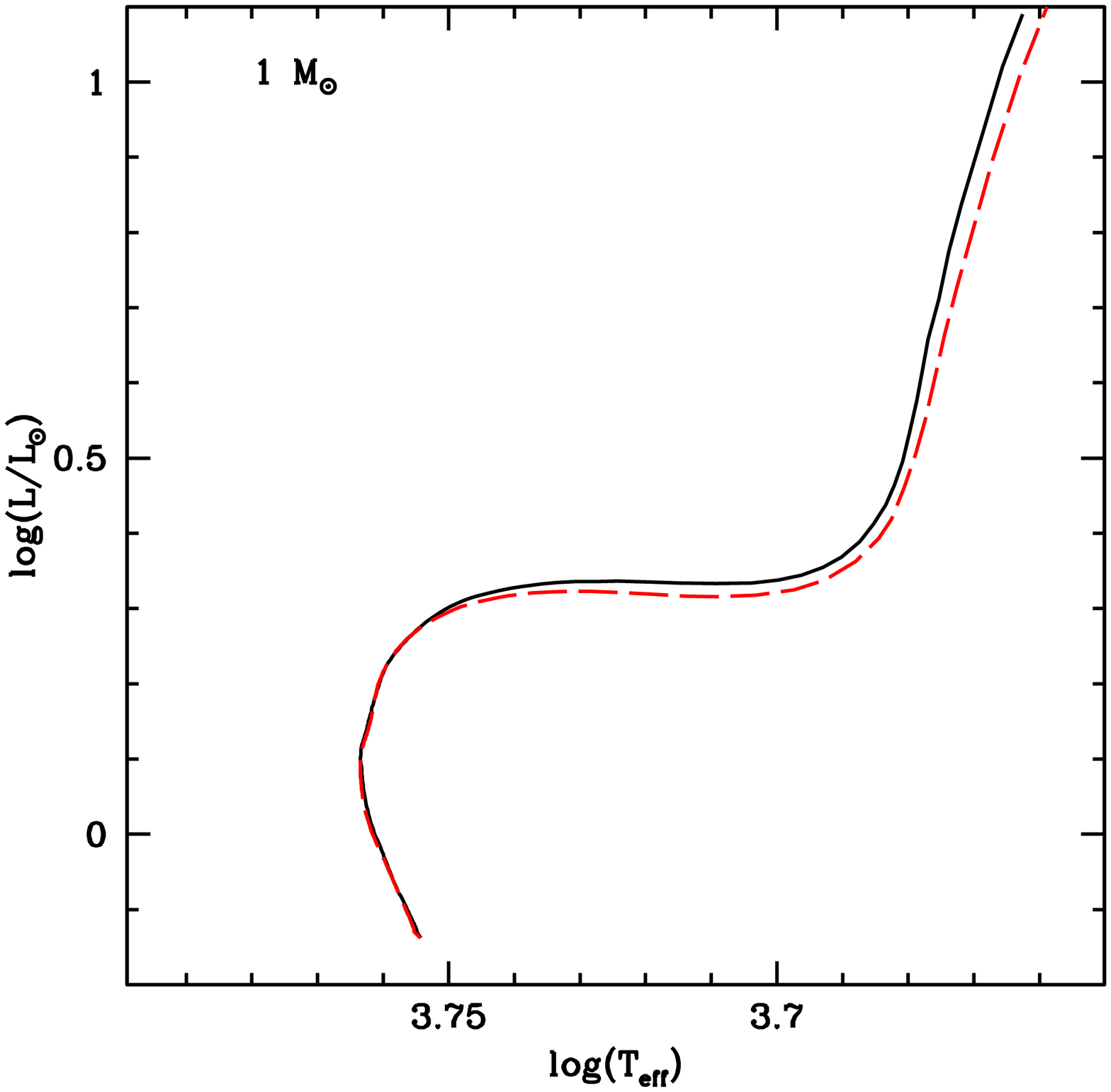}
\caption{As Fig~\ref{rot_1} but for 1$M_{\odot}$ models.}
\label{rot_5}
\end{figure}

It is extremely interesting to analyze also the case of the 1$M_{\odot}$ rotating (ZAMS equatorial velocity equal to 50 Km/s) 
and non-rotating models for the same initial chemical composition. Both   
calculations account also for atomic diffusion, but without including radiative levitation, whose effect is practically negligible at 
this metallicity. Figure~\ref{rot_5} compares their HRDs\footnote{The model with rotation is not calculated up to the tip of the RGB, but 
from independent calculations with rotation we know that He-core masses at He-ignition are larger in rotating models. This 
causes a brigther tip of the RGB and a brighter HB luminosity compared to the non-rotating case.}, 
that are virtually identical along the MS, whilst the rotating SGB is slightly brighter and the 
following RGB phase slightly hotter. The MS lifetime of the rotating model is just $\sim$4\% longer than the non-rotating case. 

\begin{figure}[!h]
\centering\includegraphics[width=3.5in]{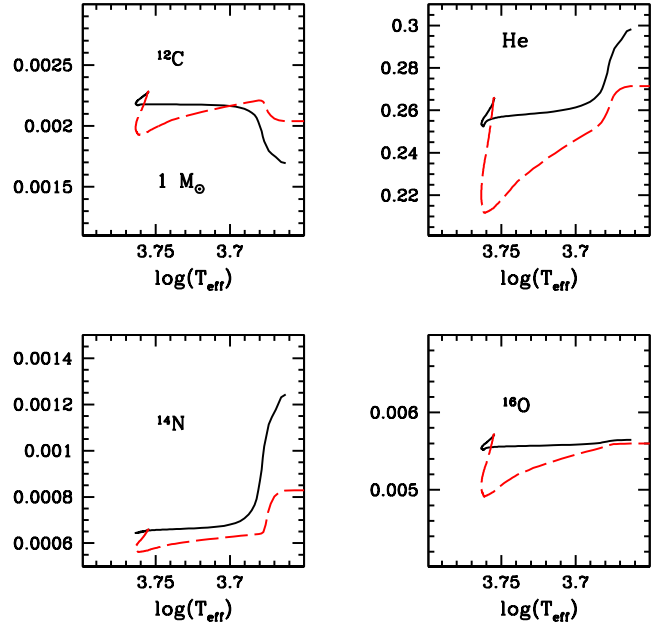}
\caption{as Fig.~\ref{rot_4} but for 1$M_{\odot}$ models.} 
\label{rot_6}
\end{figure}

Figure~\ref{rot_6} compares the evolution with $T_{eff}$ of the surface abundances of key elements affected by rotational mixing. 
In case of non-rotating models the signature of efficient diffusion during the MS is very clear (see Sect.~\ref{diffusion}). 
The abundances of all elements decrease, reaching a minimum around the TO, and then increase when convection deepens, before the signature of 
the first dredge-up can be seen for C, N and He.
When rotation is included the abundances stay constant along the MS, almost equal to the initial values, showing that rotational mixing (for the 
chosen initial rotational velocity at this metallicity) 
strongly inhibit the effect of diffusion from the convective envelopes. 
Along the RGB of the rotating models displays stronger enhancements of N and He, and larger a depletion of C compared to the non-rotating counterparts.

There are also alternative expressions for $D_{\rm shear}$ and $D_{\rm h}$, in addition to the ones employed in the calculations we are discussing, namely:

\begin{description}
\item[$D_{\rm shear}$ from \cite{talzahb97}]
  \begin{equation}
    D_\text{shear} =  f_\text{energ} \frac{H_P}{g\delta}\frac{\left(K_T+D_\text{h}\right)}{\left[\frac{\varphi}{\delta}\nabla_\mu\left(1+\frac{K}{D_\text{h}}\right) + \left( \nabla_\text{ad} - \nabla_\text{rad} \right)\right]} \left( \frac{9\pi}{32}\ \Omega\ \frac{\text{d} \ln \Omega}{\text{d} \ln r} \right)^2
  \end{equation}
  with $K$, $f_\text{energ}$, and $\varphi$ as in \cite{maed97}.
\end{description}

and 

\begin{description}
\item[$D_{\rm h}$ from \cite{maed03}]
  \begin{equation}
    D_\text{h} =  A\ r\ \left( r\Omega \ V_2(r)\ \left| 2V_2(r)-\alpha U_2(r) \right| \right)^{1/3} \\
  \end{equation}
  with $\alpha$ as in \cite{z:92}, and $A=0.002$.
\item[$D_{\rm h}$ from \cite{math:04}]
  \begin{equation}
    D_\text{h} =  \left( \frac{\beta}{10} \right)^{1/2} \left( r^2 \Omega \right)^{1/2} 
    \left( r \left| 2V_2(r)-\alpha U(r) \right| \right)^{1/2}
  \end{equation}
  with $\alpha$ as in \cite{z:92}, and $\beta=1.5\cdot10^{-6}$.
\end{description}

The effect of using these various combinations of $D_{\rm shear}$ and $D_{\rm h}$ in the Geneva code has been investigated recently 
\cite{meynetproc:13}\footnote{See also \cite{ptb:12} for a similar analysis}. They found that MS lifetimes, the evolution of surface velocities and the angular momentum of the core 
have a weak dependence of the choice of these diffusion coefficients. The shape of the evolutionary tracks, the surface enrichment (for a fixed 
initial rotation velocity), the blue-to-red evolution in the HRD and the extension of the blue loops are however significantly 
affected by the choice of $D_{\rm shear}$ and $D_{\rm h}$.


\subsubsection{Diffusive implementation}
\label{diffusive}

There is an alternative approach to include transport of chemicals and angular momentum, 
used in codes like the Yale evolutionary code \cite{pins89}, 
KEPLER, STERN and MESA. 
In this case the temporal evolution of angular momentum and chemical abundances 
due to rotation is described by a set of 
two diffusion equations, computationally easy to implement :
\begin{eqnarray}
\frac{\partial \Omega}{\partial t}\bigg|_{M_r} & = & \frac{1}{\rho r^4}
  \frac{\partial}{\partial r} \left( \rho r^4 \nu \frac{\partial
    \Omega}{\partial r}\right)\\
\frac{\partial X_i}{\partial t}\bigg|_{M_r} & = & \frac{1}{\rho r^2}
  \frac{\partial}{\partial r} \left( \rho r^2 D \frac{\partial
    X_i}{\partial r}\right)
\end{eqnarray}

where $\nu$  and $D$ are respectively the total turbulent viscosity and the total 
diffusion coefficient defined 
as a sum of all diffusion coefficients associated to
all the transport processes taken into account. Each of these diffusion
coefficients is built as the product of the velocity $v$ and the path
length $l$ of the redistribution currents with
\begin{equation}
l = min\left( r, \left| \frac{\partial r}{\partial \ln v} \right| \right).
\end{equation}
$\left| \frac{\partial r}{\partial \ln v} \right| $ is the velocity
scale height. The time-scale associated to the {\em redistribution}
over the path scale is simply $l^2/D = l/v$.\\
This description is usually employed to include within the same formalism also convection and semiconvection, treated 
as diffusive processes.
Appropriate diffusion coefficients for convective and rotational transports are adopted, often with free parameters to be calibrated 
against some sets of observations, given that these coefficients arise from order-of-magnitude considerations. 
Rotational mixing processes include at least meridional circulation, and shear, plus 
eventually additional rotational instabilities \cite{Heger00, mesa13} not usually included in the codes that 
employ the advective/diffusive implementation.

\begin{figure}[!h]
\centering\includegraphics[width=3.5in]{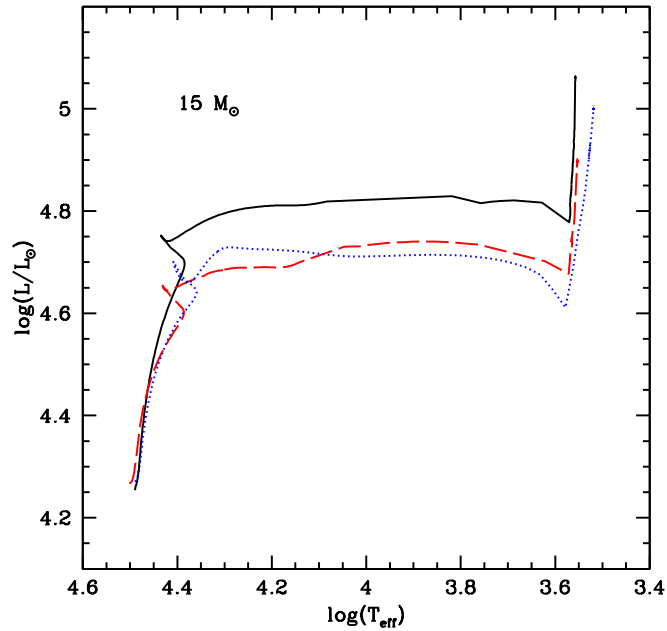}
\caption{Comparison of the HRD of the Geneva models in Fig~\ref{rot_1} (same line styles) with the 
rotating counterpart (dotted line) from MESA calculations \cite{mesa13}.}
\label{rot_7}
\end{figure}

\vspace*{-0pt}

The viscosity $\nu$ is prescribed by 
\begin{equation}
\nu =  D_{\rm conv} + D_{\rm semiconv} +  \sum_{i, rot inst} D_{i} 
\end{equation}
 and the diffusion coefficient is usually written as the following sum \cite{Heger00}
\begin{equation}
D =  D_{\rm conv} + D_{\rm semiconv} + f_c \times\left(\sum_{i, rot inst} D_{i} \right)
\end{equation}
$f_c$ is one of the efficiency parameters entering the diffusive 
formalism and it is calibrated on observations like the solar lithium abundance ($f_c$=0.046 in the models  
\cite{pins89}), or the main trend of the observed nitrogen surface abundances with the projected rotational velocity 
for the nitrogen enriched fast rotators in the LMC ($f_c$=0.0228 in the models \cite{Brott11a}). 

An example of these implementations is given in the following, for the meridional circulation \cite{pins89, Heger00}.
The characteristic velocity of these currents in a radiative region is assumed to be 

\begin{equation}
v_{ES_0}=\frac{\nabla_{\rm ad}}{(\delta (\nabla_{\rm ad}-\nabla)} \frac{\Omega^2 r^3 L}{(GM)^2} 
\left( \frac{2 \epsilon r^2}{L}- \frac{2 r^2}{M}- \frac{3}{4 \pi \rho r}\right)
\end{equation}

with $\epsilon$ denoting the energy generation rate per unit mass.

The counteracting effect of $\mu$-gradients is accounted for according to
\begin{equation}
v_{\mu}=f_{\mu} \frac{\psi H_P}{\delta (\nabla_{\rm ad}-\nabla) \tau_{KH}}\frac{\left|\nabla_{\mu}\right|}{\mu}
\end{equation}

$\psi=(d {\rm ln}(\rho)/d {\rm ln}(\mu))_{P, T}$, $\tau_{KH}$ 
is the local thermal timescale, and $f_{\mu}$ a free parameter that allows to vary the sensitivity 
of meridional circulation to $\mu$ gradients.

The effective value of the meridional circulation velocity used to determine the associated diffusion coefficient 
is finally calculated as $v_{ES}=\left| v_{ES_0} \right|-\left| v_{\mu} \right|$ if $v_{ES_0} > v_{\mu}$, or $v_{ES}$=0 otherwise.

\begin{figure}[!h]
\centering\includegraphics[width=3.5in]{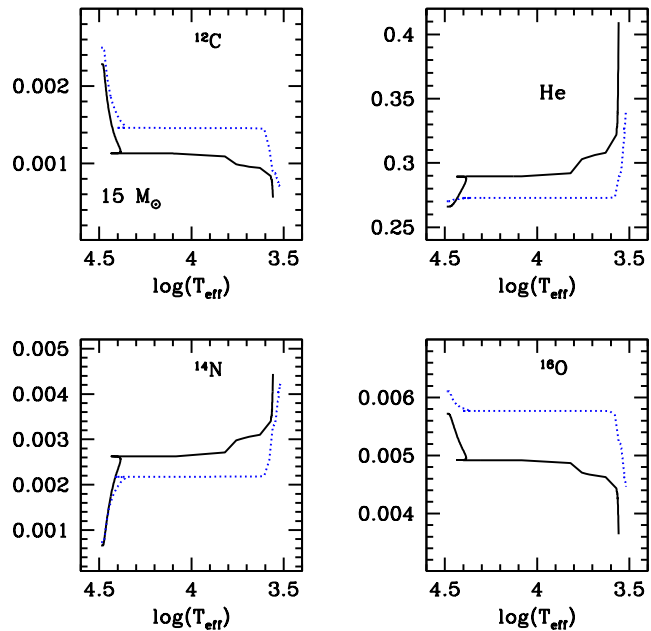}
\caption{Comparison of the evolution of the surface abundances 
(in mass fractions) as a function of $T_{eff}$ of the labelled elements, for the two 
sets of rotating models in Fig.~\ref{rot_7}. Solid lines refer to the Geneva calculations, dotted lines to the MESA models.}
\label{rot_8}
\end{figure}

Figures~\ref{rot_7} and \ref{rot_8} compare results for the initial solar composition 15$M_{\odot}$ models  
discussed in Figs.~\ref{rot_1}, \ref{rot_3} and \ref{rot_4}, with results of a MESA calculation 
(that employs this diffusive implementation for the transport of  angular momentum and chemicals) 
for the same mass, solar initial chemical composition and approximately the same initial rotational velocity \cite{choi:16}.
The HRD shows that these rotating models are almost equivalent to the non-rotating Geneva calculations. 
However, the evolution of the surface chemical elements shown in 
Fig.~\ref{rot_8} displays clear signatures of rotational mixing, although the quantitative effect is very different 
from the Geneva results. The enhancement of the surface abundances during the MS is more moderate 
in the MESA calculations (even accounting for the slightly different initial abundances of some elements), 
and He is almost unchanged. Even the effect of the dredge-up seem to differ between the calculations.

It is difficult to determine the exact cause of these differences, given that also other elements of the 
model input physics (apart from the implementation of rotational transports) differ between the two sets of calculations. 
Very importantly, MESA calculations include also the effect of atomic diffusion (and radiative levitation) however moderated 
by some additional turbulence \cite{choi:16}.
But taken at face value, this comparison may give an idea of the uncertainties in the outputs of the current generation of 
rotating models.

\subsection{Additional effects}

There are at least two additional processes that may affect substantially the evolution of rotating stars, for they 
modify the transport of angular momentum and, directly or indirectly, also the transport of chemical elements. Their implementation 
in stellar model calculations is still uncertain and they are not generally included in the current generation of stellar models. 

\subsubsection{Gravity waves}

Inside a star the density changes monotonically with the depth and the molecular weight typically increases in the direction of increasing gravity. 
When a gas element in this stable stratification is perturbed, the competition between buoyancy and gravity gives 
rise to an oscillatory motion around the equilibrium position and creates a so-called 
\lq{internal gravity wave}\rq\ (IGW).

These IGWs are expected to be generated 
by the injection of kinetic energy from a turbulent region into an adjacent stable region, 
as observed both in 2D and 3D simulations of convective mixing, e.g. by 
convective overshooting in a stable region and bulk excitation or excitation by Reynolds stresses inside the convection zone \cite{zahn:97,talon:02, talon:05}. 
The IGWs penetrate into the radiation zone, transporting angular momentum that is deposited where they are dissipated through 
heat diffusion by photon exchange, which produces an 
\lq{attenuation factor}\rq\ ($\tau$) proportional to the thermal diffusivity and inversely proportional to the IGW frequency $\nu$ and amplitude. 
It is by shaping the internal rotation profile that IGWs contribute indirectly to the mixing of chemical elements.

Angular momentum transport by gravity waves is seldom included in evolutionary stellar model calculations.  
The current approximate treatment implemented in some stellar evolution calculations \cite{talon:08} 
expresses the solution of the equations describing the propagation of IGWs in a rotating star in terms of Legendre polynomials.   
At each point within a radiative region, the total angular momentum \lq{luminosity}\rq\footnote{The total angular momentum luminosity is defined 
as the average angular momentum flux transported by IGWs through a surface of radius $r$.} associated to the IGWs propagation can be written as 

\begin{equation}
{\cal L}_J(r) = \sum_{\rm waves} {\cal L}_{J, \nu,\ell,m}(r_c)\,  \exp \left[ -\tau(r,\sigma,\ell)\right]
\end{equation}

where $r_c$ denotes the radiation/convection interface, $\ell$ and $m$ represent, respectively, 
the spherical order and the azimuthal number of the Legendre polynomial, $\nu$ the frequency of the IGW when launched from the convective zone, 
$\sigma=\nu-m(\Omega(r)-\Omega(c))$ is the local wave frequency measured in the co-rotating frame 
with angular velocity $\Omega(r)$ --$\Omega(c)$ being the 
angular velocity of the solid body rotating convective zone-- and 

\begin{equation}
{\cal L}_{J, \nu, \ell, m} = 4 \pi r^2 {\cal F}_J (\nu, \ell, m) = 4 \pi r^2 \frac{2m}{\nu} {\cal F}_E (\ell, \omega)  
\label{eq-monochromLJ}
\end{equation}

${\cal F}_J (\nu, \ell, m) $ is the mean flux of angular momentum carried by a monochromatic wave 
with emission frequency $\nu$. The amplitude of the waves is assumed to be \cite{golmurrkum:94}

\begin{eqnarray}
{\cal F}_E \left( \nu, \ell \right) &=& \frac{\nu^2}{4\pi} \int dr \frac{\rho^2}{r^2}
   \left[\left(\frac{\partial \xi_r}{\partial r}\right)^2 +
   \ell(\ell+1)\left(\frac{\partial \xi_h}{\partial r}\right)^2 \right]  \nonumber \\
 && \times  \exp\left[ -h_\nu^2 \ell(\ell+1)/2r^2\right] \frac{v_c^3 \Lambda^4 }{1
  + (\nu \tau_\Lambda)^{15/2}}
\label{eq-IGWflux}
\end{eqnarray}

with $\xi_r$ and $[\ell(\ell+1)]^{1/2}\xi_h$ being the vertical and horizontal displacement wave functions normalized to unit energy flux at the 
edge of the considered convection zone, $v_c$ the convective velocity, $\Lambda=\alpha_{\rm MLT} H_P$ the radial size of a convective element,
$\tau_\Lambda \sim \Lambda/v_c$ the convective timescale, $h_\nu = \lambda \min\{1, (2\nu\tau_\Lambda)^{-3/2}\}$. 
The radial ($k_r$) and horizontal ($k_h$) wave numbers are related by

\begin{equation}
k_r^2 = \left( \frac{N^2}{\omega^2} -1 \right) k_h^2 = 
\left( \frac{N^2}{\omega^2} -1 \right) \frac{\ell \left( \ell +1 \right)}{r^2} \label{kradial}
\end{equation}

The local damping rate $\tau(r,\sigma,\ell)$ can be written as

\begin{equation}
\tau(r, \sigma, \ell) = \left[ \ell(\ell+1) \right] ^{3/2} \int_r^{r_c} 
\left( K_T + \nu_v \right) \; {N {N_T}^2 \over
\sigma^4}  \left({N^2 \over N^2 - \sigma^2}\right)^{1/2} \frac{dr}{r^3} 
\label{eq-tauIGW}
\end{equation}

where $N^2_T$ is the thermal part of the Brunt-V\"ais\"al\"a frequency 
and $\nu_v$ is the vertical shear turbulent viscosity

\begin{equation}
\nu_v = \frac{8}{5} \frac{Ri_{crit} \left( r \frac{d\Omega}{d r}
  \right)^2}{\frac{N^2_T}{K_T+\nu_h} + \frac{N^2_\mu}{\nu_h}}.
\label{eq-Dv}
\end{equation}

with $N^2_\mu$ denoting the molecular weight stratification part of the Brunt-V\"ais\"al\"a frequency, 
and $\nu_h$ the horizontal shear turbulent viscosity, that can be set to $D_{\rm h}$, and $Ri_{crit}$=1/4.

The deposition of angular momentum is then given by the radial derivative of ${\cal L}_J$. Given that, as a first approximation, 
only the radial dependency of IGW transport is considered, 
all quantities required are evaluated from horizontal (on isobars) averages. The angular momentum evolution due only 
to IGW transport is given by 

\begin{equation}
\rho \frac{\rm d}{\rm dt} \left[ r^2 {\Omega} \right]  = \pm \frac{3}{8\pi} \frac{1}{r^2} \frac{\partial{\cal L}_J(r)}{\partial r}
\label{amevol}
\end{equation}

The $+$ or $-$ sign in front of the angular  momentum luminosity corresponds to prograde ($m>$0) or retrograde ($m <$0) waves.

\begin{figure}[!h]
\centering\includegraphics[width=2.5in]{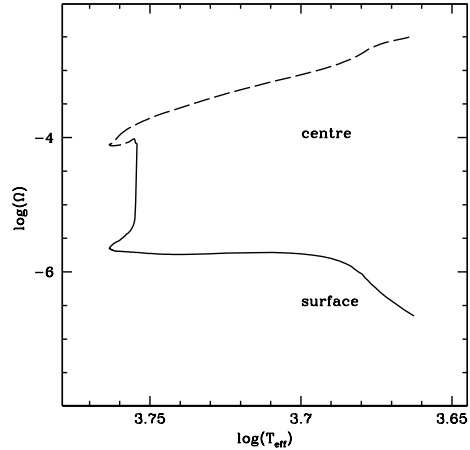}
\caption{Evolution with $T_{eff}$ of the central and surface angular velocity, 
for 1$M_{\odot}$ initial solar chemical composition rotating models, calculated with the Geneva code \cite{ekstroem12}.}
\label{rot_9}
\end{figure}


It is clear that in the absence of differential damping for inward- and outward-travelling waves, 
there will be no angular momentum transport. This is the case for example of solid body rotation. However, 
differential rotation naturally filters the IGWs that propagate within the star. 
Considering for example the realistic case of a low-mass star with interior layers 
-- see i.e. Fig.~\ref{rot_9} for the 1$M_{\odot}$ solar chemical composition Geneva rotating model discussed before-- 
that rotate faster than the convective envelope. In the inner radiative regions we have 
$\Omega(r) > \Omega(c)$ and for prograde $m>0$ waves $\sigma=\nu-m(\Omega(r)-\Omega(c))$ is shifted to lower frequencies 
when travelling from the edge of the convective envelope towards the interior layers. 
The radial damping length of the waves given approximately by  

\begin{equation}
l_d=\frac{2 r^3 \sigma^4}{[l(l+1)]^{3/2}N N_T^2 K_T}
\end{equation}

progressively decreases for these prograde waves. The retrograde $m<0$ waves are instead 
boosted to higher frequencies, and their damping length increases, 
propagating further within the star compared to the prograde waves. As a consequence,  
prograde waves are absorbed before they can propagate far into more rapidly rotating layers of a star, whilst 
retrograde waves pass through, and upon dissipation they deposit their 
negative angular momentum (they contribute with a negative sign in Eq.~\ref{amevol}), spinning down the rapidly rotating layers. 
The star then tends to evolve towards solid body rotation.

The angular momentum transport associated to IGWs is considered to be a possibility to explain the solar rotation profile, 
that is much flatter than generally predicted by rotating models that do not include IGWs \cite{ct:05}.
Also, recent asteroseismic observations \cite{beck:12, deheuvels:12} show that the core of low-mass RGB stars do not rotate 
faster than the surface as much as predicted by current rotating stellar models. The angular momentum redistribution by IGWs may 
provide one way to reconcile theory with observations. Recent multidimensional detailed 
hydrodynamical simulations of generation and propagation of IGWs within stars, confirm their ability to transport efficiently angular momentum 
\cite{rogers:13, alvan:14}.

\subsubsection{Magnetic fields}

Another way to favour the redistribution of angular momentum within a rotating star is to consider the effect of 
internal magnetic fields. They are generally implemented following the dynamo mechanism presented in \cite{spruit:02}, that envisages the 
creation of magnetic fields in the radiative regions of differentially rotating stars at the expenses of the shear, due to the so-called 
{\sl Tayler-Spruit} instability\footnote{The efficiency of this mechanism is however questioned by recent numerical simulations \cite{zahn:07}.}

The STERN code, that uses a diffusive implementation of element and angular momentum transports, 
accounts for the effect of magnetic fields as follows \cite{petrovic:05}.

Denoting with $q=\mathrm{d}\ln \Omega/\mathrm{d}\ln r$ the shear, 
the effective radial viscosity produced by the magnetic field can be written as
\begin{equation}
\nu_{re}=\frac{\nu_{e0}\nu_{e1}}{\nu_{e0}+\nu_{e1}}f(q),
\end{equation}
where
\begin{equation}
\nu_{e0}=r^2\Omega q^2 \left (\frac{\Omega}{N_{\mu}} \right )^4,
\end{equation}

\begin{equation}
\nu_{e1}=r^2\Omega \ {\rm max} \left [\left (\frac{\Omega}{N_T} \right )^{1/2} \left (\frac{\kappa}{r^2N_T}
\right )^{1/2},q^2\left (\frac{\Omega}{N_T} \right )^4 \right ],
\end{equation}

Denoting with $q_{\rm min}$ the minimum rotational gradient necessary for the dynamo
to operate
\begin{equation}
f(q)=1-q_{\rm min}/q,  (q>q_{\rm min}),
\end{equation}
and
\begin{equation}
f(q)=0, (q\le q_{\rm min}).
\end{equation}  

The effective diffusivity for transport of chemical elements is given as:
\begin{equation}
D_e=\frac{D_{e0}D_{e1}}{D_{e0}+D_{e1}}f(q),
\end{equation}  

where 

\begin{equation}
D_{e0}=r^2\Omega q^4 \left (\frac{\Omega}{N_{\mu}} \right )^6,
\end{equation}

\begin{equation}
D_{e1}=r^2\Omega \ {\rm max} \left [ \left (\frac{\Omega}{N_T} \right )^{3/4} \left (\frac{K_T}{r^2N_T}
\right )^{3/4},q^2 \left (\frac{\Omega}{N_T} \right )^6 \right]. 
\end{equation}

According to \cite{spruit:02} $q_{\rm min}$ is given by

\begin{eqnarray}
q_{\rm min}=q_0+q_1 \\
q_0=\left(\frac{N_\mu}{\Omega} \right)^{7/4} \left(\frac{\eta}{r^2 N_\mu} \right)^{1/4}\\
q_1=\left(\frac{N_{T}}{\Omega} \right)^{7/4} \left(\frac{\eta}{r^2 N_{T}} \right)^{1/4} 
\left(\frac{\eta}{K_T}\right)^{3/4}
\end{eqnarray}

with $\eta=  7 \times 10^{11} {\rm ln} (\Lambda) T^{-3/2}$ denoting the Spitzer magnetic diffusivity, and 
${\rm ln}(\Lambda)$ the Coulomb logarithm (see Sect.~\ref{diffusion}). 

These viscosities and diffusivities have been included as additional terms to the diffusion coefficients of, respectively,  
angular momentum and chemicals, in both the STERN \cite{petrovic:05, Brott11a} and KEPLER \cite{Heger:05}
codes, that use a diffusive implementation of element and angular momentum transports. In this latter code modifications 
are applied to the viscosities and diffusivities derived before in case of semiconvective regions, or where the thermohaline mixing is efficient, 
that means in regions where the timescale of mixing is much longer than in convective layers.
In case of semiconvection a dynamo effective viscosity is defined as
$\nu_{re}=\nu_{e0} f(q)$
and the final expression for the viscosity in the semiconvective region is set to be
$\nu_{e}=\sqrt{\nu_{re} \nu_{SC}}$, 
where $\nu_{SC}=(1/3) H_p v_{conv}$ and $v_{conv}$ is the convective velocity derived from the MLT. 
The effective diffusion coefficient for the transport of elements is set to $(D_e+D_{SC})$, with $D_{SC}$ being the diffusion coefficient 
due to semiconvection.
In the thermohaline mixing region $D_e$ is se to $D_{e1}$, $\nu_e$ is se to $\nu_{e1}$, and $q_{min}$ to $q_1$. 

\begin{figure}[!h]
\centering\includegraphics[width=3.5in]{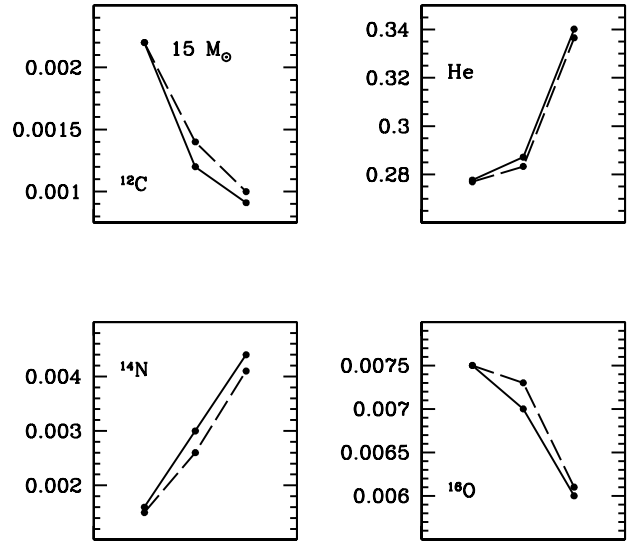}
\caption{In each panel, from left to right, the abundances (mass fractions --filled circles) correspond to the MS stage 
when the central H mass fraction is 0.35, to central He-burning with a 0.50 mass fraction of He, and at the pre-supernova stage. Solid lines 
correspond to rotating models without magnetic fields, dashed lines to rotating models with magnetic fields (see text for details).}
\label{rot_10}
\end{figure}

The inclusion of internal magnetic fields decreases the rotation velocity contrast between core and envelope 
during the MS, approaching near solid body rotation. This affects both the shape of the evolutionary tracks as well as the abundance 
profiles within the models.
Figure~\ref{rot_10} displays, as an example, the effect on the surface chemical abundances at selected 
evolutionary stages in the calculations with the code KEPLER \cite{Heger:05}, for 15$M_{\odot}$ solar metallicity models with a ZAMS 
equatorial rotation velocity of 200 Km/s.

Internal magnetic fields generated by the Tayler-Spruit instability have been also included in the 
Geneva code \cite{maedmeyn:05} --that implements the advective/diffusive formalism for rotational transports-- in a slightly different form. 
By denoting with $x$ the ratio 
$$x=(\omega_A/\Omega)^2$$
where $\omega_A$ is  the  Alfv\'en frequency of a magnetic field of intensity $B$, the solution of this equation 

\begin{equation}
\frac{r^2 \Omega}{q^2 K_T} \left(N_{\mathrm{T}}^2 + N_{\mu}^2 \right)  x^4-
\frac{r^2 \Omega^3}{K_T} x^3 + 2 N_{\mu}^2 \; x - 2 \Omega^2 q^2 = 0 \; .
\label{equx}
\end{equation} 

provides the unknown quantity $\omega_A$. The diffusion coefficient for the vertical transport of angular momentum is then given by
 \begin{equation}
 \nu_{re} = \frac{\Omega \; r^2}{q} \;
 \left( \frac{\omega_{\mathrm{A}}}{\Omega}\right)^3 \; 
\left(\frac{\Omega}{N}\right) \; 
 \end{equation}

and the diffusion coefficient for the transport of chemicals by

\begin{equation}
D_e \;= \; \frac{r^2 \; \Omega}{q^2} \; \left( \frac{\omega_{\mathrm{A}}}
{\Omega}\right)^6  
\end{equation}

Once  $\omega_A$ is calculated, the condition on the minimum shear for the dynamo to work is tested as

\begin{equation}
q > \left(\frac{N}{\Omega} \right)^{7/4} \left(\frac{D_e}{r^2 N} \right)^{1/4}
\end{equation}

If this condition is not realized, a formalism for the treatment of low rotation rates --expanding upon \cite{spruit:02} 
work--  is implemented.
We notice that in this treatment the Brunt-V\"ais\"al\"a frequency is rewritten as 

\begin{equation}
N^2=\frac{D_e/K_T}{(D_e/K_T)+2} (N_T^2+N_{\mu}^2)
\end{equation}

to account for the presence of a magnetic diffusivity $D_e$.

A set of 15$M_{\odot}$ solar metallicity 
models calculated with this implementation \cite{maedmeyn:05} also predict an almost solid body rotation during the MS, 
with obviously a negligible element transport by shear mixing. The transport of elements due to the magnetic fields is also 
negligible, whilst the meridional circulation is strongly enhanced by the flat rotational profile. The net effect during the MS 
is an enhancement of the surface abundance variations compared to rotating models without internal magnetic fields.
This is different from the result of Fig.~\ref{rot_10} for a KEPLER model with the same mass, initial chemical composition 
and very similar initial rotation rate. In this latter case the variation of the surface abundances during the MS is almost negligible 
compared to the case without magnetic fields.

Finally, we mention very briefly the effect of magnetic braking due to magnetized winds\footnote{The 
complex angular momentum evolution 
during the pre-MS phase due to influence of the surrounding disk, is generally not modelled by stellar evolution calculations.}. 
Winds with magnetic fields --in case of surface magnetic fields, whatever their origin is-- 
exert a braking torque that is significantly larger
than for non-magnetic cases \cite{bouvier:97, matt:12}. The reason is that the magnetic field connects the mass lost from the surface 
of the star to the envelope, and when the wind finally decouples from the magnetic field, it has the same angular velocity as  
the surface but a much larger moment of inertia. This increases the amount of angular momentum lost, compared to the case 
with no magnetic fields, and it is accepted as the reason for the slow rotation rate of low-mass MS stars.
The effect on the chemical element transport is indirect, through the variation of the star rotational profile.
Surface magnetic fields are observed mainly in low mass stars with convective envelopes, where they are expected to be generated 
by a dynamo mechanism, and this type of magnetic braking 
is implemented mainly in models of low-mass stars. There are various prescriptions in the literature 
that require the calibration of free parameters on observations, given the lack of knowledge about the surface magnetic fields \cite{amard:16}.
For example, the rate of angular momentum loss due to magnetized winds employed in \cite{amard:16} is a 
variation of \cite{kawaler:88, chaboyer:95}:

\begin{equation}
\frac{dJ}{dt} = -K_W\Omega\Omega_{sat}^2\left(\frac{R}{R_{\odot}}\right)^{1/2} \left(\frac{M}{M_{\odot}}\right)^{-1/2} \ {\rm for} \ \; \Omega \geq \Omega_{sat}
\end{equation}
\begin{equation}
\frac{dJ}{dt} = -K_W\Omega^{3}\left(\frac{R}{R_{\odot}}\right)^{1/2} \left(\frac{M}{M_{\odot}}\right)^{-1/2} \ {\rm for} \ \; \Omega < \Omega_{sat}
\end{equation}

where $K_W = 2.10^{48}$ in cgs units to reproduce the solar case and $\Omega_{sat}$ is a free parameter.

\begin{figure}[!h]
\centering\includegraphics[width=3.0in]{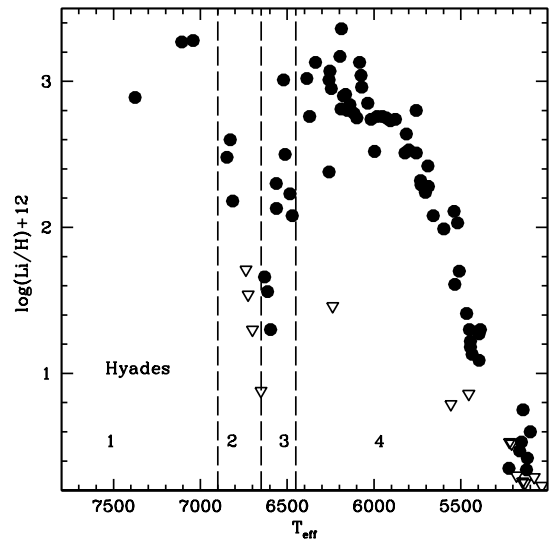}
\caption{Li abundances measured in a sample of Hyades MS stars, as a function of their $T_{eff}$ \cite{boesgaard:86a, thorburn:93}. 
Inverted open triangles denote upper limits. The four different regions marked in the diagram are discussed in the text.}
\label{rot_11}
\end{figure}

\section{An example of possible synergy amongst several element transport processes}
\label{synergy}

After the description of all major element transport mechanisms included in modern stellar evolution calculations, we show just an example 
of how their synergy might explain some puzzling observations of chemical abundances in star clusters. Figure~\ref{rot_11} displays the trend 
of the surface Li abundance as a function of $T_{eff}$ for a sample of MS stars in the $\sim$600~Myr old Hyades open cluster \cite{boesgaard:86a, thorburn:93}.
It is easy to notice the so-called {\sl Li-dip} around 6600~K, a sharp local minimum of the Li abundance, that cannot be explained 
by standard stellar evolution models including only convective mixing (the Li-dip is observed also in other open clusters 
of different ages, at similar temperature). 

Lithium is a very fragile element that is burned at temperatures $\sim 2.5 \times 10^6$ K. 
In solar metallicity MS models calculated with just convective mixing, and $T_{eff}$ between $\sim$6400 and 7500~K, the Li burning temperature is well 
below the base of the convective envelope, hence the Li abundance is constant down to the radiative layers where the temperature reaches  
$\sim 2.5 \times 10^6$ K. This implies that these models --that also are not expected to experience pre-MS Li depletion-- predict a constant surface Li 
with $T_{eff}$ in this temperature range, contrary to observations.

To explain these observations of Li abundances, we outline below the scenario proposed by \cite{talon:03}, that involves the combination of several 
of the element transport mechanisms discussed above\footnote{Other scenarios to explain the Li-dip are described in \cite{michaud:91}.}. 
To this purpose we have divided the temperature range covered by the data of Fig.~\ref{rot_11} 
into four contiguous ranges:

\begin{enumerate}

\item{{\sl Region:1}-- These stars have shallow convective envelopes, not efficient at generating a surface magnetic field via a dynamo process, hence the surface is 
not slowed down by magnetic winds. As a result, rotational mixing is expected to be just 
about sufficient to counteract atomic diffusion of Li below the convective envelope (e.g. counteracts the creation of a Li gradient right below the 
convective boundary where Li diffused from the convective region accumulates).}

\item{{\sl Region:2}-- Moving towards lower temperatures the convective envelopes deepen, and dynamo generated weak surface magnetic 
fields are expected now to start slowing down the outer layers. The increased shear strenghtens the rotational mixing and Li depletion increases with decreasing $T_{eff}$. 
This happens because the more vigorous mixing is trying now to erase the gradient between the Li burning region (no Li) and Li-rich convective layers.}

\item{{\sl Region:3}-- Stars on the cool side of the Li-dip have even deeper convective envelopes, that sustain a very efficient dynamo and slow down 
even more the external layers. At the same time, these convective layers are now very efficient at generating IGWs, that redistribute 
momentum (driving the star towards solid body rotation) and reduce the efficiency of rotational mixing, inducing an increase of surface Li with decreasing $T_{eff}$.
Calculations based on the approximations describe before \cite{talon:03}, show that the expected efficiency of IGW induced angular momentum transport has the 
required dependence on stellar mass ($T_{eff}$),}

\item{{\sl Region:4}-- At these low $T_{eff}$ the convective envelope is deep enough to reach Li burning temperatures, causing an increasingly larger 
depletion with decreasing $T_{eff}$.}

\end{enumerate}

Measurements of projected rotation velocities in this cluster display a decreasing average velocity with decreasing $T_{eff}$, consistent with an 
increased efficiency 
of magnetic braking when moving towards lower stellar masses \cite{talon:98}.

\section{Conclusions}
\label{conclusions}

It is clear that the description of complex physical processes like turbulent convection, semiconvection, thermohaline mixing, rotation,  
implemented in stellar evolution calculations is necessarily simplified, with a predictive power 
in some cases hampered by the use of several free parameters of uncertain calibration. Also, the effect of 
interactions amongst the various instabilities in rotating stars --which usually are considered as independent-- has to be fully explored  
\cite{mmlc:13}.

It is also fair to say that in many cases the approaches used in stellar evolution models are probably reaching their limits, and further developments of
multi-dimensional hydrodynamical simulations are crucial to progress in the field. 
As we have discussed, there are already some constraints that current hydrodynamical simulations pose to the element transport 
processes efficient in stars. 
Even though there is clearly still a long way to go until complete stellar interior modelling with numerical hydrodynamics, 
physical insights provided by computer simulations are invaluable in improving our understanding 
of element transport processes \cite{viallet:13, am:16}.
Just to give two examples, recent results from hydrodynamical simulations provide new 
indications about the way to go to replace the MLT in stellar modelling 
\cite{am:15}, as well as how to implement a more consistent description of overshooting in different regimes \cite{viallet:15}.
 
At the same time, the booming field of asteroseismology is starting to provide new information on the efficiency of element transport processes, that 
can at the very least be used to add further constraints to our current recipes. Two perfect examples are recent works on mixing in low-mass  
core He-burning stars \cite{bossini:15,consta:15}, and angular momentum transport in RGB stars
\cite{elm:16, elm:16b}.

The hope is that a synergy between clues from hydrodynamical simulations and the powerful constraints coming from asteroseismic analyses 
will help to improve our description of element transport in stellar model computations and improve their predicting power.

\bigskip

\noindent
{\sl Data accessibility.} There is no supporting data accompanying the manuscript. 

\noindent
{\sl Authors' contributions.} M.S. has designed the outline of the review. Both authors have agreed on the content, shared the writing, 
and gave final approval for the publication.

\noindent
{\sl Competing interests.} The authors declare no competing interests.

\noindent
{\sl Funding.} S.C. acknowledges support from PRIN-INAF2014 (PI: S. Cassisi), and from grants: AYA2013-42781-P and AYA2016-77237-C3-1-P 
from the Spanish Ministry of Economy and Competitiveness (MINECO).

\noindent
{\sl Acknowledgments.} The authors are grateful to Marco Limongi, Victor Silva-Aguirre and David Hyder for preparing some of the figures.


\bibliographystyle{vancouver.bst}

\begin{thebibliography}{100}

\bibitem{eddington}
{Eddington} AS.
\newblock {The Internal Constitution of the Stars}; 1926.

\bibitem{sc:05}
{Salaris} M, {Cassisi} S.
\newblock {Evolution of Stars and Stellar Populations}; 2005.

\bibitem{am:16}
{Arnett} WD, {Meakin} C.
\newblock {Key issues review: numerical studies of turbulence in stars}.
\newblock Reports on Progress in Physics. 2016 Oct;79(10):102901.

\bibitem{henyey}
{Henyey} LG, {Forbes} JE, {Gould} NL.
\newblock {A New Method of Automatic Computation of Stellar Evolution.}
\newblock \apj. 1964 Jan;139:306.

\bibitem{kww}
{Kippenhahn} R, {Weigert} A, {Weiss} A.
\newblock {Stellar Structure and Evolution}; 2012.

\bibitem{smartt}
{Smartt} SJ.
\newblock {Observational Constraints on the Progenitors of Core-Collapse
  Supernovae: The Case for Missing High-Mass Stars}.
\newblock \pasa. 2015 Apr;32:e016.

\bibitem{kato:66}
{Kato} S.
\newblock {Overstable Convection in a Medium Stratified in Mean Molecular
  Weight}.
\newblock \pasj. 1966;18:374.

\bibitem{merryfield:95}
{Merryfield} WJ.
\newblock {Hydrodynamics of semiconvection}.
\newblock \apj. 1995 May;444:318--337.

\bibitem{w:13}
{Wood} TS, {Garaud} P, {Stellmach} S.
\newblock {A New Model for Mixing by Double-diffusive Convection
  (Semi-convection). II. The Transport of Heat and Composition through Layers}.
\newblock \apj. 2013 May;768:157.

\bibitem{langer:83}
{Langer} N, {Fricke} KJ, {Sugimoto} D.
\newblock {Semiconvective diffusion and energy transport}.
\newblock \aap. 1983 Sep;126:207.

\bibitem{sh:58}
{Schwarzschild} M, {H{\"a}rm} R.
\newblock {Evolution of Very Massive Stars.}
\newblock \apj. 1958 Sep;128:348.

\bibitem{cw:71}
{Cameron} AGW, {Fowler} WA.
\newblock {Lithium and the s-PROCESS in Red-Giant Stars}.
\newblock \apj. 1971 Feb;164:111.

\bibitem{bie:32}
{Biermann} L.
\newblock {Untersuchungen {\"u}ber den inneren Aufbau der Sterne. IV.
  Konvektionszonen im Innern der Sterne. (Ver{\"o}ffentlichungen der
  Universit{\"a}ts-Sternwarte G{\"o}ttingen, Nr. 27. ) Mit 5 Abbildungen.}
\newblock \zap. 1932;5:117.

\bibitem{bv:58}
{B{\"o}hm-Vitense} E.
\newblock {{\"U}ber die Wasserstoffkonvektionszone in Sternen verschiedener
  Effektivtemperaturen und Leuchtkr{\"a}fte. Mit 5 Textabbildungen}.
\newblock \zap. 1958;46:108.

\bibitem{tas:90}
{Tassoul} M, {Fontaine} G, {Winget} DE.
\newblock {Evolutionary models for pulsation studies of white dwarfs}.
\newblock \apjs. 1990 Feb;72:335--386.

\bibitem{sc:08}
{Salaris} M, {Cassisi} S.
\newblock {Stellar models with the ML2 theory of convection}.
\newblock \aap. 2008 Sep;487:1075--1080.

\bibitem{ce:76}
{Cloutman} LD, {Eoll} JG.
\newblock {Comments on the diffusion model of turbulent mixing}.
\newblock \apj. 1976 Jun;206:548--554.

\bibitem{weaver:78}
{Weaver} TA, {Zimmerman} GB, {Woosley} SE.
\newblock {Presupernova evolution of massive stars}.
\newblock \apj. 1978 Nov;225:1021--1029.

\bibitem{langer:85}
{Langer} N, {El Eid} MF, {Fricke} KJ.
\newblock {Evolution of massive stars with semiconvective diffusion}.
\newblock \aap. 1985 Apr;145:179--191.

\bibitem{vc:05}
{Ventura} P, {Castellani} M.
\newblock {Time dependent mixing in He-burning cores: The case of NGC 1866}.
\newblock \aap. 2005 Feb;430:1035--1047.

\bibitem{bonaca}
{Bonaca} A, {Tanner} JD, {Basu} S, {Chaplin} WJ, {Metcalfe} TS, {Monteiro}
  MJPFG, et~al.
\newblock {Calibrating Convective Properties of Solar-like Stars in the Kepler
  Field of View}.
\newblock \apjl. 2012 Aug;755:L12.

\bibitem{tr:14}
{Trampedach} R, {Stein} RF, {Christensen-Dalsgaard} J, {Nordlund} {\AA},
  {Asplund} M.
\newblock {Improvements to stellar structure models, based on a grid of 3D
  convection simulations - II. Calibrating the mixing-length formulation}.
\newblock \mnras. 2014 Dec;445:4366--4384.

\bibitem{sc:15}
{Salaris} M, {Cassisi} S.
\newblock {Stellar models with mixing length and T({$\tau$}) relations
  calibrated on 3D convection simulations}.
\newblock \aap. 2015 May;577:A60.

\bibitem{mwa:15}
{Magic} Z, {Weiss} A, {Asplund} M.
\newblock {The Stagger-grid: A grid of 3D stellar atmosphere models. III. The
  relation to mixing length convection theory}.
\newblock \aap. 2015 Jan;573:A89.

\bibitem{cd:02}
{Christensen-Dalsgaard} J.
\newblock {Helioseismology}.
\newblock Reviews of Modern Physics. 2002 Nov;74:1073--1129.

\bibitem{pate:93}
{Paterno} L, {Ventura} R, {Canuto} VM, {Mazzitelli} I.
\newblock {Helioseismological test of a new model for stellar convection}.
\newblock \apj. 1993 Jan;402:733--740.

\bibitem{vdm:08}
{Ventura} P, {D'Antona} F, {Mazzitelli} I.
\newblock {The ATON 3.1 stellar evolutionary code. A version for
  asteroseismology}.
\newblock \apss. 2008 Aug;316:93--98.

\bibitem{cm:91}
{Canuto} VM, {Mazzitelli} I.
\newblock {Stellar Turbulent Convection: A New Model and Applications}.
\newblock \apj. 1991 Mar;370:295--311.

\bibitem{canuto:92}
{Canuto} VM.
\newblock {Turbulent convection with overshooting - Reynolds stress approach}.
\newblock \apj. 1992 Jun;392:218--232.

\bibitem{canuto:93}
{Canuto} VM.
\newblock {Turbulent Convection with Overshooting: Reynolds Stress Approach.
  II.}
\newblock \apj. 1993 Oct;416:331.

\bibitem{xiong:97}
{Xiong} DR, {Cheng} QL, {Deng} L.
\newblock {Nonlocal Time-dependent Convection Theory}.
\newblock \apjs. 1997 Feb;108:529--544.

\bibitem{ly:07}
{Li} Y, {Yang} JY.
\newblock {Testing turbulent convection theory in solar models - I. Structure
  of the solar convection zone}.
\newblock \mnras. 2007 Feb;375:388--402.

\bibitem{yl:07}
{Yang} JY, {Li} Y.
\newblock {Testing turbulent convection theory in solar models - II. Solar
  p-mode oscillations}.
\newblock \mnras. 2007 Feb;375:403--414.

\bibitem{canuto:11}
{Canuto} VM.
\newblock {Stellar mixing. I. Formalism}.
\newblock \aap. 2011 Apr;528:A76.

\bibitem{as:13}
{Andr{\'a}ssy} R, {Spruit} HC.
\newblock {Overshooting by convective settling}.
\newblock \aap. 2013 Nov;559:A122.

\bibitem{as:15}
{Andr{\'a}ssy} R, {Spruit} HC.
\newblock {Overshooting by differential heating}.
\newblock \aap. 2015 Jun;578:A106.

\bibitem{viallet:15}
{Viallet} M, {Meakin} C, {Prat} V, {Arnett} D.
\newblock {Toward a consistent use of overshooting parametrizations in 1D
  stellar evolution codes}.
\newblock \aap. 2015 Aug;580:A61.

\bibitem{basti}
{Pietrinferni} A, {Cassisi} S, {Salaris} M, {Castelli} F.
\newblock {A Large Stellar Evolution Database for Population Synthesis Studies.
  I. Scaled Solar Models and Isochrones}.
\newblock \apj. 2004 Sep;612:168--190.

\bibitem{dsep}
{Dotter} A, {Chaboyer} B, {Jevremovi{\'c}} D, {Kostov} V, {Baron} E, {Ferguson}
  JW.
\newblock {The Dartmouth Stellar Evolution Database}.
\newblock \apjs. 2008 Sep;178:89--101.

\bibitem{yy}
{Demarque} P, {Woo} JH, {Kim} YC, {Yi} SK.
\newblock {Y$^{2}$ Isochrones with an Improved Core Overshoot Treatment}.
\newblock \apjs. 2004 Dec;155:667--674.

\bibitem{vandenberg:06}
{VandenBerg} DA, {Bergbusch} PA, {Dowler} PD.
\newblock {The Victoria-Regina Stellar Models: Evolutionary Tracks and
  Isochrones for a Wide Range in Mass and Metallicity that Allow for
  Empirically Constrained Amounts of Convective Core Overshooting}.
\newblock \apjs. 2006 Feb;162:375--387.

\bibitem{roxb:89}
{Roxburgh} IW.
\newblock {Integral constraints on convective overshooting}.
\newblock \aap. 1989 Mar;211:361--364.

\bibitem{padua}
{Girardi} L, {Bressan} A, {Bertelli} G, {Chiosi} C.
\newblock {Evolutionary tracks and isochrones for low- and intermediate-mass
  stars: From 0.15 to 7 Msun, and from Z=0.0004 to 0.03}.
\newblock \aaps. 2000 Feb;141:371--383.

\bibitem{parsec}
{Bressan} A, {Marigo} P, {Girardi} L, {Salasnich} B, {Dal Cero} C, {Rubele} S,
  et~al.
\newblock {PARSEC: stellar tracks and isochrones with the PAdova and TRieste
  Stellar Evolution Code}.
\newblock \mnras. 2012 Nov;427:127--145.

\bibitem{zahn:91}
{Zahn} JP.
\newblock {Convective penetration in stellar interiors}.
\newblock \aap. 1991 Dec;252:179--188.

\bibitem{bressan:81}
{Bressan} AG, {Chiosi} C, {Bertelli} G.
\newblock {Mass loss and overshooting in massive stars}.
\newblock \aap. 1981 Sep;102:25--30.

\bibitem{roxb:92}
{Roxburgh} IW.
\newblock {Limits on convective penetration from stellar cores}.
\newblock \aap. 1992 Dec;266:291--293.

\bibitem{wd:01}
{Woo} JH, {Demarque} P.
\newblock {Empirical Constraints on Convective Core Overshoot}.
\newblock \aj. 2001 Sep;122:1602--1606.

\bibitem{schroder:97}
{Schroder} KP, {Pols} OR, {Eggleton} PP.
\newblock {A critical test of stellar evolution and convective core
  `overshooting' by means of zeta Aurigae systems}.
\newblock \mnras. 1997 Mar;285:696--710.

\bibitem{stancliffe:15}
{Stancliffe} RJ, {Fossati} L, {Passy} JC, {Schneider} FRN.
\newblock {Confronting uncertainties in stellar physics: calibrating convective
  overshooting with eclipsing binaries}.
\newblock \aap. 2015 Mar;575:A117.

\bibitem{herwig:00}
{Herwig} F.
\newblock {The evolution of AGB stars with convective overshoot}.
\newblock \aap. 2000 Aug;360:952--968.

\bibitem{mesa13}
{Paxton} B, {Cantiello} M, {Arras} P, {Bildsten} L, {Brown} EF, {Dotter} A,
  et~al.
\newblock {Modules for Experiments in Stellar Astrophysics (MESA): Planets,
  Oscillations, Rotation, and Massive Stars}.
\newblock \apjs. 2013 Sep;208:4.

\bibitem{garstec}
{Weiss} A, {Schlattl} H.
\newblock {GARSTEC -- the Garching Stellar Evolution Code. The direct
  descendant of the legendary Kippenhahn code}.
\newblock \apss. 2008 Aug;316:99--106.

\bibitem{freytag:96}
{Freytag} B, {Ludwig} HG, {Steffen} M.
\newblock {Hydrodynamical models of stellar convection. The role of overshoot
  in DA white dwarfs, A-type stars, and the Sun.}
\newblock \aap. 1996 Sep;313:497--516.

\bibitem{meakin:07}
{Meakin} CA, {Arnett} D.
\newblock {Turbulent Convection in Stellar Interiors. I. Hydrodynamic
  Simulation}.
\newblock \apj. 2007 Sep;667:448--475.

\bibitem{staritsin:13}
{Staritsin} EI.
\newblock {Turbulent entrainment at the boundaries of the convective cores of
  main-sequence stars}.
\newblock Astronomy Reports. 2013 May;57:380--390.

\bibitem{spp:15}
{Salaris} M, {Pietrinferni} A, {Piersimoni} AM, {Cassisi} S.
\newblock {Post first dredge-up [C/N] ratio as age indicator. Theoretical
  calibration}.
\newblock \aap. 2015 Nov;583:A87.

\bibitem{whw:02}
{Woosley} SE, {Heger} A, {Weaver} TA.
\newblock {The evolution and explosion of massive stars}.
\newblock Reviews of Modern Physics. 2002 Nov;74:1015--1071.

\bibitem{chiosi:70}
{Chiosi} C, {Summa} C.
\newblock {On the Evolution of OB Stars from the Main Sequence to the Helium
  Exhaustion Phase}.
\newblock \apss. 1970 Sep;8:478--496.

\bibitem{stothers:70}
{Stothers} R.
\newblock {Internal structure of upper main-sequence stars}.
\newblock \mnras. 1970;151:65.

\bibitem{simpson:71}
{Simpson} EE.
\newblock {Evolutionary Models of Stars of 15 and 30 M\_$\{$sun$\}$}.
\newblock \apj. 1971 May;165:295.

\bibitem{sag:11}
{Silva Aguirre} V, {Ballot} J, {Serenelli} AM, {Weiss} A.
\newblock {Constraining mixing processes in stellar cores using
  asteroseismology. Impact of semiconvection in low-mass stars}.
\newblock \aap. 2011 May;529:A63.

\bibitem{sc:75}
{Stothers} R, {Chin} CW.
\newblock {Stellar evolution at high mass with semiconvective mixing according
  to the Ledoux criterion}.
\newblock \apj. 1975 Jun;198:407--417.

\bibitem{sc:76}
{Stothers} R, {Chin} CW.
\newblock {Stellar evolution at high mass with semiconvective mixing according
  to the Schwarzschild criterion}.
\newblock \apj. 1976 Mar;204:472--480.

\bibitem{langer:95}
{Langer} N, {Maeder} A.
\newblock {The problem of the blue-to-red supergiant ratio in galaxies.}
\newblock \aap. 1995 Mar;295:685.

\bibitem{yl05}
{Yoon} SC, {Langer} N.
\newblock {Evolution of rapidly rotating metal-poor massive stars towards
  gamma-ray bursts}.
\newblock \aap. 2005 Nov;443:643--648.

\bibitem{spruit:92}
{Spruit} HC.
\newblock {The rate of mixing in semiconvective zones}.
\newblock \aap. 1992 Jan;253:131--138.

\bibitem{w:78}
{Weaver} TA, {Zimmerman} GB, {Woosley} SE.
\newblock {Presupernova evolution of massive stars}.
\newblock \apj. 1978 Nov;225:1021--1029.

\bibitem{Heger00}
{Heger} A, {Langer} N, {Woosley} SE.
\newblock {Presupernova Evolution of Rotating Massive Stars. I. Numerical
  Method and Evolution of the Internal Stellar Structure}.
\newblock \apj. 2000 Jan;528:368--396.

\bibitem{sw:14}
{Sukhbold} T, {Woosley} SE.
\newblock {The Compactness of Presupernova Stellar Cores}.
\newblock \apj. 2014 Mar;783:10.

\bibitem{moore:16}
{Moore} K, {Garaud} P.
\newblock {Main Sequence Evolution with Layered Semiconvection}.
\newblock \apj. 2016 Jan;817:54.

\bibitem{cgr:71a}
{Castellani} V, {Giannone} P, {Renzini} A.
\newblock {Overshooting of Convective Cores in Helium-Burning Horizontal-Branch
  Stars}.
\newblock \apss. 1971 Feb;10:340--349.

\bibitem{sg:74}
{Sweigart} AV, {Gross} PG.
\newblock {Horizontal-Branch Evolution with Semiconvection. I. Interior
  Evolution}.
\newblock \apj. 1974 May;190:101--108.

\bibitem{rf:72}
{Robertson} JW, {Faulkner} DJ.
\newblock {Semiconvection in the Core-Helium Phase of Stellar Evolution}.
\newblock \apj. 1972 Feb;171:309.

\bibitem{dr:93}
{Dorman} B, {Rood} RT.
\newblock {On partial mixing zones in horizontal-branch stellar cores}.
\newblock \apj. 1993 May;409:387--403.

\bibitem{gn:14}
{Gabriel} M, {Noels} A, {Montalb{\'a}n} J, {Miglio} A.
\newblock {Proper use of Schwarzschild Ledoux criteria in stellar evolution
  computations}.
\newblock \aap. 2014 Sep;569:A63.

\bibitem{michaud:07}
{Michaud} G, {Richer} J, {Richard} O.
\newblock {Horizontal Branch Evolution and Atomic Diffusion}.
\newblock \apj. 2007 Dec;670:1178--1187.

\bibitem{chandra:61}
{Chandrasekhar} S.
\newblock {Hydrodynamic and hydromagnetic stability}; 1961.

\bibitem{cctp:85}
{Castellani} V, {Chieffi} A, {Tornambe} A, {Pulone} L.
\newblock {Helium-burning evolutionary phases in population II stars. I
  Breathing pulses in horizontal branch stars}.
\newblock \apj. 1985 Sep;296:204--212.

\bibitem{caputo:89}
{Caputo} F, {Chieffi} A, {Tornambe} A, {Castellani} V, {Pulone} L.
\newblock {The 'Red Giant Clock' as an indicator for the efficiency of central
  mixing in horizontal-branch stars}.
\newblock \apj. 1989 May;340:241--248.

\bibitem{csi:03}
{Cassisi} S, {Salaris} M, {Irwin} AW.
\newblock {The Initial Helium Content of Galactic Globular Cluster Stars from
  the R-Parameter: Comparison with the Cosmic Microwave Background Constraint}.
\newblock \apj. 2003 May;588:862--870.

\bibitem{cassisi:99}
{Cassisi} S, {Castellani} V, {degl'Innocenti} S, {Salaris} M, {Weiss} A.
\newblock {Galactic globular cluster stars: From theory to observation}.
\newblock \aaps. 1999 Jan;134:103--113.

\bibitem{sdi:03}
{Straniero} O, {Dom{\'{\i}}nguez} I, {Imbriani} G, {Piersanti} L.
\newblock {The Chemical Composition of White Dwarfs as a Test of Convective
  Efficiency during Core Helium Burning}.
\newblock \apj. 2003 Feb;583:878--884.

\bibitem{cl:10}
{Cantiello} M, {Langer} N.
\newblock {Thermohaline mixing in evolved low-mass stars}.
\newblock \aap. 2010 Oct;521:A9.

\bibitem{clag:10}
{Charbonnel} C, {Lagarde} N.
\newblock {Thermohaline instability and rotation-induced mixing. I. Low- and
  intermediate-mass solar metallicity stars up to the end of the AGB}.
\newblock \aap. 2010 Nov;522:A10.

\bibitem{gratton:00}
{Gratton} RG, {Sneden} C, {Carretta} E, {Bragaglia} A.
\newblock {Mixing along the red giant branch in metal-poor field stars}.
\newblock \aap. 2000 Feb;354:169--187.

\bibitem{angelou:15}
{Angelou} GC, {D'Orazi} V, {Constantino} TN, {Church} RP, {Stancliffe} RJ,
  {Lattanzio} JC.
\newblock {Diagnostics of stellar modelling from spectroscopy and photometry of
  globular clusters}.
\newblock MNRAS. 2015 Jul;450:2423--2440.

\bibitem{cz:07b}
{Charbonnel} C, {Zahn} JP.
\newblock {Inhibition of thermohaline mixing by a magnetic field in Ap star
  descendants: implications for the Galactic evolution of $^{3}$He}.
\newblock \aap. 2007 Dec;476:L29--L32.

\bibitem{edl:06}
{Eggleton} PP, {Dearborn} DSP, {Lattanzio} JC.
\newblock {Deep Mixing of $^{3}$He: Reconciling Big Bang and Stellar
  Nucleosynthesis}.
\newblock Science. 2006 Dec;314:1580.

\bibitem{cz:07}
{Charbonnel} C, {Zahn} JP.
\newblock {Thermohaline mixing: a physical mechanism governing the photospheric
  composition of low-mass giants}.
\newblock A\&A. 2007 May;467:L15--L18.

\bibitem{ulrich:72}
{Ulrich} RK.
\newblock {Thermohaline Convection in Stellar Interiors.}
\newblock \apj. 1972 Feb;172:165.

\bibitem{kipp:80}
{Kippenhahn} R, {Ruschenplatt} G, {Thomas} HC.
\newblock {The time scale of thermohaline mixing in stars}.
\newblock A\&A. 1980 Nov;91:175--180.

\bibitem{lattanzio:15}
{Lattanzio} JC, {Siess} L, {Church} RP, {Angelou} G, {Stancliffe} RJ, {Doherty}
  CL, et~al.
\newblock {On the numerical treatment and dependence of thermohaline mixing in
  red giants}.
\newblock MNRAS. 2015 Jan;446:2673--2688.

\bibitem{dm:11}
{Denissenkov} PA, {Merryfield} WJ.
\newblock {Thermohaline Mixing: Does it Really Govern the Atmospheric Chemical
  Composition of Low-mass Red Giants?}
\newblock \apjl. 2011 Jan;727:L8.

\bibitem{traxler}
{Traxler} A, {Garaud} P, {Stellmach} S.
\newblock {Numerically Determined Transport Laws for Fingering
  (''Thermohaline'') Convection in Astrophysics}.
\newblock ApJL. 2011 Feb;728:L29.

\bibitem{turc}
{Turcotte} S, {Richer} J, {Michaud} G, {Iglesias} CA, {Rogers} FJ.
\newblock {Consistent Solar Evolution Model Including Diffusion and Radiative
  Acceleration Effects}.
\newblock \apj. 1998 Sep;504:539--558.

\bibitem{burgers:69}
{Burgers} JM.
\newblock {Flow Equations for Composite Gases}; 1969.

\bibitem{thoul:94}
{Thoul} AA, {Bahcall} JN, {Loeb} A.
\newblock {Element diffusion in the solar interior}.
\newblock \apj. 1994 Feb;421:828--842.

\bibitem{mason:67}
{Mason} EA, {Munn} RJ, {Smith} FJ.
\newblock {Transport Coefficients of Ionized Gases}.
\newblock Phys\ Fluids. 1967 Aug;10:1827--1832.

\bibitem{muchmore:84}
{Muchmore} D.
\newblock {Diffusion in white dwarf stars}.
\newblock \apj. 1984 Mar;278:769--783.

\bibitem{iben:85}
{Iben} I Jr, {MacDonald} J.
\newblock {The effects of diffusion due to gravity and due to composition
  gradients on the rate of hydrogen burning in a cooling degenerate dwarf. I -
  The case of a thick helium buffer layer}.
\newblock \apj. 1985 Sep;296:540--553.

\bibitem{paquette:86}
{Paquette} C, {Pelletier} C, {Fontaine} G, {Michaud} G.
\newblock {Diffusion coefficients for stellar plasmas}.
\newblock \apjs. 1986 May;61:177--195.

\bibitem{ss:03}
{Schlattl} H, {Salaris} M.
\newblock {Quantum corrections to microscopic diffusion constants}.
\newblock \aap. 2003 Apr;402:29--35.

\bibitem{zhang:16}
{Zhang} QS.
\newblock {Numerical integral of resistance coefficients in diffusion}.
\newblock ArXiv e-prints. 2016 Nov;.

\bibitem{daligault:16}
{Daligault} J, {Baalrud} SD, {Starrett} CE, {Saumon} D, {Sjostrom} T.
\newblock {Ionic Transport Coefficients of Dense Plasmas without Molecular
  Dynamics}.
\newblock Physical Review Letters. 2016 Feb;116(7):075002.

\bibitem{hu:11}
{Hu} H, {Tout} CA, {Glebbeek} E, {Dupret} MA.
\newblock {Slowing down atomic diffusion in subdwarf B stars: mass loss or
  turbulence?}
\newblock \mnras. 2011 Nov;418:195--205.

\bibitem{seaton:05}
{Seaton} MJ.
\newblock {Opacity Project data on CD for mean opacities and radiative
  accelerations}.
\newblock \mnras. 2005 Sep;362:L1--L3.

\bibitem{leblanc:00}
{LeBlanc} F, {Michaud} G, {Richer} J.
\newblock {Opacity Sampling in Radiative Acceleration Calculations}.
\newblock \apj. 2000 Aug;538:876--884.

\bibitem{cassisi:97}
{Cassisi} S, {degl'Innocenti} S, {Salaris} M.
\newblock {The effect of diffusion on the red giant luminosity function
  `bump'}.
\newblock \mnras. 1997 Sep;290:515--520.

\bibitem{richard:02a}
{Richard} O, {Michaud} G, {Richer} J, {Turcotte} S, {Turck-Chi{\`e}ze} S,
  {VandenBerg} DA.
\newblock {Models of Metal-poor Stars with Gravitational Settling and Radiative
  Accelerations. I. Evolution and Abundance Anomalies}.
\newblock \apj. 2002 Apr;568:979--997.

\bibitem{koester:09}
{Koester} D.
\newblock {Accretion and diffusion in white dwarfs. New diffusion timescales
  and applications to GD 362 and G 29-38}.
\newblock \aap. 2009 May;498:517--525.

\bibitem{michaud:11}
{Michaud} G, {Richer} J, {Richard} O.
\newblock {Horizontal branch evolution, metallicity, and sdB stars}.
\newblock \aap. 2011 May;529:A60.

\bibitem{cs:13}
{Cassisi} S, {Salaris} M.
\newblock {Old Stellar Populations: How to Study the Fossil Record of Galaxy
  Formation}; 2013.

\bibitem{vick:13}
{Vick} M, {Michaud} G, {Richer} J, {Richard} O.
\newblock {Population II stars and the Spite plateau. Stellar evolution models
  with mass loss}.
\newblock \aap. 2013 Apr;552:A131.

\bibitem{bravo:92}
{Bravo} E, {Isern} J, {Canal} R, {Labay} J.
\newblock {On the contribution of Ne-22 to the synthesis of Fe-54 and Ni-58 in
  thermonuclear supernovae}.
\newblock \aap. 1992 Apr;257:534--538.

\bibitem{deloye:02}
{Deloye} CJ, {Bildsten} L.
\newblock {Gravitational Settling of $^{22}$Ne in Liquid White Dwarf Interiors:
  Cooling and Seismological Effects}.
\newblock \apj. 2002 Dec;580:1077--1090.

\bibitem{althaus:10}
{Althaus} LG, {Garc{\'{\i}}a-Berro} E, {Renedo} I, {Isern} J, {C{\'o}rsico} AH,
  {Rohrmann} RD.
\newblock {Evolution of White Dwarf Stars with High-metallicity Progenitors:
  The Role of $^{22}$Ne Diffusion}.
\newblock \apj. 2010 Aug;719:612--621.

\bibitem{garciaberro:10}
{Garc{\'{\i}}a-Berro} E, {Torres} S, {Althaus} LG, {Renedo} I,
  {Lor{\'e}n-Aguilar} P, {C{\'o}rsico} AH, et~al.
\newblock {A white dwarf cooling age of 8Gyr for NGC 6791 from physical
  separation processes}.
\newblock \nat. 2010 May;465:194--196.

\bibitem{camisassa:16}
{Camisassa} ME, {Althaus} LG, {C{\'o}rsico} AH, {Vinyoles} N, {Serenelli} AM,
  {Isern} J, et~al.
\newblock {The Effect of $^{22}$NE Diffusion in the Evolution and Pulsational
  Properties of White Dwarfs with Solar Metallicity Progenitors}.
\newblock \apj. 2016 Jun;823:158.

\bibitem{houghto:10}
{Hughto} J, {Schneider} AS, {Horowitz} CJ, {Berry} DK.
\newblock {Diffusion of neon in white dwarf stars}.
\newblock \pre. 2010 Dec;82(6):066401.

\bibitem{vick:11}
{Vick} M, {Michaud} G, {Richer} J, {Richard} O.
\newblock {Abundance anomalies in pre-main-sequence stars. Stellar evolution
  models with mass loss}.
\newblock \aap. 2011 Feb;526:A37.

\bibitem{charpinet:97}
{Charpinet} S, {Fontaine} G, {Brassard} P, {Chayer} P, {Rogers} FJ, {Iglesias}
  CA, et~al.
\newblock {A Driving Mechanism for the Newly Discovered Class of Pulsating
  Subdwarf B Stars}.
\newblock \apjl. 1997 Jul;483:L123--L126.

\bibitem{bahcall:95}
{Bahcall} JN, {Pinsonneault} MH, {Wasserburg} GJ.
\newblock {Solar models with helium and heavy-element diffusion}.
\newblock Reviews of Modern Physics. 1995 Oct;67:781--808.

\bibitem{villante:14}
{Villante} FL, {Serenelli} AM, {Delahaye} F, {Pinsonneault} MH.
\newblock {The Chemical Composition of the Sun from Helioseismic and Solar
  Neutrino Data}.
\newblock \apj. 2014 May;787:13.

\bibitem{korn:07}
{Korn} AJ, {Grundahl} F, {Richard} O, {Mashonkina} L, {Barklem} PS, {Collet} R,
  et~al.
\newblock {Atomic Diffusion and Mixing in Old Stars. I. Very Large Telescope
  FLAMES-UVES Observations of Stars in NGC 6397}.
\newblock \apj. 2007 Dec;671:402--419.

\bibitem{lind:08}
{Lind} K, {Korn} AJ, {Barklem} PS, {Grundahl} F.
\newblock {Atomic diffusion and mixing in old stars. II. Observations of stars
  in the globular cluster NGC 6397 with VLT/FLAMES-GIRAFFE}.
\newblock \aap. 2008 Nov;490:777--786.

\bibitem{mucc:11}
{Mucciarelli} A, {Salaris} M, {Lovisi} L, {Ferraro} FR, {Lanzoni} B,
  {Lucatello} S, et~al.
\newblock {Lithium abundance in the globular cluster M4: from the turn-off to
  the red giant branch bump}.
\newblock \mnras. 2011 Mar;412:81--94.

\bibitem{gruyters:16}
{Gruyters} P, {Lind} K, {Richard} O, {Grundahl} F, {Asplund} M, {Casagrande} L,
  et~al.
\newblock {Atomic diffusion and mixing in old stars. VI. The lithium content of
  M30}.
\newblock \aap. 2016 May;589:A61.

\bibitem{korn:06}
{Korn} A, {Grundahl} F, {Richard} O, {Barklem} P, {Mashonkina} L, {Collet} R,
  et~al.
\newblock {New Abundances for Old Stars - Atomic Diffusion at Work in NGC
  6397}.
\newblock The Messenger. 2006 Sep;125.

\bibitem{salaris:16}
{Salaris} M.
\newblock {Age dating of old field stars: Challenges from the uncertain
  efficiency of atomic diffusion}.
\newblock Astronomische Nachrichten. 2016 Sep;337:805.

\bibitem{richer:00}
{Richer} J, {Michaud} G, {Turcotte} S.
\newblock {The Evolution of AMFM Stars, Abundance Anomalies, and Turbulent
  Transport}.
\newblock \apj. 2000 Jan;529:338--356.

\bibitem{deal:16}
{Deal} M, {Richard} O, {Vauclair} S.
\newblock {Hydrodynamical instabilities induced by atomic diffusion in A stars
  and their consequences}.
\newblock \aap. 2016 May;589:A140.

\bibitem{gebran:10}
{Gebran} M, {Vick} M, {Monier} R, {Fossati} L.
\newblock {Chemical composition of A and F dwarfs members of the Hyades open
  cluster}.
\newblock \aap. 2010 Nov;523:A71.

\bibitem{spite:82}
{Spite} F, {Spite} M.
\newblock {Abundance of lithium in unevolved halo stars and old disk stars -
  Interpretation and consequences}.
\newblock \aap. 1982 Nov;115:357--366.

\bibitem{salaris:01}
{Salaris} M, {Weiss} A.
\newblock {Atomic diffusion in metal-poor stars. II. Predictions for the Spite
  plateau}.
\newblock \aap. 2001 Sep;376:955--965.

\bibitem{richard:05}
{Richard} O, {Michaud} G, {Richer} J.
\newblock {Implications of WMAP Observations on Li Abundance and Stellar
  Evolution Models}.
\newblock \apj. 2005 Jan;619:538--548.

\bibitem{quievy:09}
{Quievy} D, {Charbonneau} P, {Michaud} G, {Richer} J.
\newblock {Abundances anomalies and meridional circulation in horizontal branch
  stars}.
\newblock \aap. 2009 Jun;500:1163--1171.

\bibitem{swenson:95}
{Swenson} FJ.
\newblock {Lithium in halo dwarfs: The undoing of diffusion by mass loss}.
\newblock \apjl. 1995 Jan;438:L87--L90.

\bibitem{vauclair:95}
{Vauclair} S, {Charbonnel} C.
\newblock {Influence of a stellar wind on the lithium depletion in halo stars:
  a new step towards the lithium primordial abundance.}
\newblock \aap. 1995 Mar;295:715.

\bibitem{vick:10}
{Vick} M, {Michaud} G, {Richer} J, {Richard} O.
\newblock {AmFm and lithium gap stars. Stellar evolution models with mass
  loss}.
\newblock \aap. 2010 Oct;521:A62.

\bibitem{brown:13}
{Brown} JM, {Garaud} P, {Stellmach} S.
\newblock {Chemical Transport and Spontaneous Layer Formation in Fingering
  Convection in Astrophysics}.
\newblock \apj. 2013 May;768:34.

\bibitem{richard:02b}
{Richard} O, {Michaud} G, {Richer} J.
\newblock {Models of Metal-poor Stars with Gravitational Settling and Radiative
  Accelerations. III. Metallicity Dependence}.
\newblock \apj. 2002 Dec;580:1100--1117.

\bibitem{gruyters:14}
{Gruyters} P, {Nordlander} T, {Korn} AJ.
\newblock {Atomic diffusion and mixing in old stars. V. A deeper look into the
  globular cluster NGC 6752}.
\newblock \aap. 2014 Jul;567:A72.

\bibitem{vandenberg:12}
{VandenBerg} DA, {Bergbusch} PA, {Dotter} A, {Ferguson} JW, {Michaud} G,
  {Richer} J, et~al.
\newblock {Models for Metal-poor Stars with Enhanced Abundances of C, N, O, Ne,
  Na, Mg, Si, S, Ca, and Ti, in Turn, at Constant Helium and Iron Abundances}.
\newblock \apj. 2012 Aug;755:15.

\bibitem{morel:02}
{Morel} P, {Th{\'e}venin} F.
\newblock {Atomic diffusion in star models of type earlier than G}.
\newblock \aap. 2002 Aug;390:611--620.

\bibitem{garbe:88}
{Garcia-Berro} E, {Hernanz} M, {Mochkovitch} R, {Isern} J.
\newblock {Theoretical white-dwarf luminosity functions for two phase diagrams
  of the carbon-oxygen dense plasma}.
\newblock \aap. 1988 Mar;193:141--147.

\bibitem{sc:93}
{Segretain} L, {Chabrier} G.
\newblock {Crystallization of binary ionic mixtures in dense stellar plasmas}.
\newblock \aap. 1993 Apr;271:L13.

\bibitem{salaris:97}
{Salaris} M, {Dom{\'{\i}}nguez} I, {Garc{\'{\i}}a-Berro} E, {Hernanz} M,
  {Isern} J, {Mochkovitch} R.
\newblock {The Cooling of CO White Dwarfs: Influence of the Internal Chemical
  Distribution}.
\newblock \apj. 1997 Sep;486:413--419.

\bibitem{moch:83}
{Mochkovitch} R.
\newblock {Freezing of a carbon-oxygen white dwarf}.
\newblock \aap. 1983 Jun;122:212--218.

\bibitem{imgh:97}
{Isern} J, {Mochkovitch} R, {Garc{\'{\i}}a-Berro} E, {Hernanz} M.
\newblock {The Physics of Crystallizing White Dwarfs}.
\newblock \apj. 1997 Aug;485:308--312.

\bibitem{horowitz:10}
{Horowitz} CJ, {Schneider} AS, {Berry} DK.
\newblock {Crystallization of Carbon-Oxygen Mixtures in White Dwarf Stars}.
\newblock Physical Review Letters. 2010 Jun;104(23):231101.

\bibitem{segr:96}
{Segretain} L.
\newblock {Three-body crystallization diagrams and the cooling of white
  dwarfs.}
\newblock \aap. 1996 Jun;310:485--488.

\bibitem{hunter:08}
{Hunter} I, {Lennon} DJ, {Dufton} PL, {Trundle} C, {Sim{\'o}n-D{\'{\i}}az} S,
  {Smartt} SJ, et~al.
\newblock {The VLT-FLAMES survey of massive stars: atmospheric parameters and
  rotational velocity distributions for B-type stars in the Magellanic Clouds}.
\newblock \aap. 2008 Feb;479:541--555.

\bibitem{pennygies:09}
{Penny} LR, {Gies} DR.
\newblock {A FUSE Survey of the Rotation Rates of Very Massive Stars in the
  Small and Large Magellanic Clouds}.
\newblock \apj. 2009 Jul;700:844--858.

\bibitem{braganca:12}
{Bragan{\c c}a} GA, {Daflon} S, {Cunha} K, {Bensby} T, {Oey} MS, {Walth} G.
\newblock {Projected Rotational Velocities and Stellar Characterization of 350
  B Stars in the Nearby Galactic Disk}.
\newblock \aj. 2012 Nov;144:130.

\bibitem{huang:10}
{Huang} W, {Gies} DR, {McSwain} MV.
\newblock {A Stellar Rotation Census of B Stars: From ZAMS to TAMS}.
\newblock \apj. 2010 Oct;722:605--619.

\bibitem{royer:07}
{Royer} F, {Zorec} J, {G{\'o}mez} AE.
\newblock {Rotational velocities of A-type stars. III. Velocity distributions}.
\newblock \aap. 2007 Feb;463:671--682.

\bibitem{stauff:86}
{Stauffer} JB, {Hartmann} LW.
\newblock {The rotational velocities of low-mass stars}.
\newblock \pasp. 1986 Dec;98:1233--1251.

\bibitem{carlberg:11}
{Carlberg} JK, {Majewski} SR, {Patterson} RJ, {Bizyaev} D, {Smith} VV, {Cunha}
  K.
\newblock {The Frequency of Rapid Rotation Among K Giant Stars}.
\newblock \apj. 2011 May;732:39.

\bibitem{Herrero:92}
{Herrero} A, {Kudritzki} RP, {Vilchez} JM, {Kunze} D, {Butler} K, {Haser} S.
\newblock {Intrinsic parameters of galactic luminous OB stars}.
\newblock \aap. 1992 Jul;261:209--234.

\bibitem{hunter:09}
{Hunter} I, {Brott} I, {Langer} N, {Lennon} DJ, {Dufton} PL, {Howarth} ID,
  et~al.
\newblock {The VLT-FLAMES survey of massive stars: constraints on stellar
  evolution from the chemical compositions of rapidly rotating Galactic and
  Magellanic Cloud B-type stars}.
\newblock \aap. 2009 Mar;496:841--853.

\bibitem{martins:16}
{Martins} F, {Simon-Diaz} S, {Barba} RH, {Gamen} RC, {Ekstroem} S.
\newblock {A study of the effect of rotational mixing on massive stars
  evolution: surface abundances of Galactic O7-8 giant stars}.
\newblock ArXiv e-prints. 2016 Nov;.

\bibitem{monaghan:65}
{Monaghan} FF, {Roxburgh} IW.
\newblock {The structure of rapidly rotating polytropes}.
\newblock \mnras. 1965;131:13.

\bibitem{ostriker:68}
{Ostriker} JP, {Mark} JWK.
\newblock {Rapidly rotating stars. I. The self-consistent-field method}.
\newblock \apj. 1968 Mar;151:1075--1088.

\bibitem{deupree:90}
{Deupree} RG.
\newblock {Stellar evolution with arbitrary rotation laws. I - Mathematical
  techniques and test cases}.
\newblock \apj. 1990 Jul;357:175--187.

\bibitem{deupree:01}
{Deupree} RG.
\newblock {Stellar Evolution with Arbitrary Rotation Laws. IV. Survey of
  Zero-Age Main-Sequence Models}.
\newblock \apj. 2001 May;552:268--277.

\bibitem{rieutord:13}
{Espinosa Lara} F, {Rieutord} M.
\newblock {Self-consistent 2D models of fast-rotating early-type stars}.
\newblock \aap. 2013 Apr;552:A35.

\bibitem{z:92}
{Zahn} JP.
\newblock {Circulation and turbulence in rotating stars}.
\newblock \aap. 1992 Nov;265:115--132.

\bibitem{mm:97}
{Meynet} G, {Maeder} A.
\newblock {Stellar evolution with rotation. I. The computational method and the
  inhibiting effect of the {$\mu$}-gradient.}
\newblock \aap. 1997 May;321:465--476.

\bibitem{maederbook}
{Maeder} A.
\newblock {Physics, Formation and Evolution of Rotating Stars}; 2009.

\bibitem{zeng:02}
{Zeng} YR.
\newblock {A more powerful evolution model for rotating stars}.
\newblock \aap. 2002 Nov;394:965--969.

\bibitem{kt:70}
{Kippenhahn} R, {Thomas} HC.
\newblock {A Simple Method for the Solution of the Stellar Structure Equations
  Including Rotation and Tidal Forces}.
\newblock In: {Slettebak} A, editor. IAU Colloq. 4: Stellar Rotation; 1970.
  p.~20.

\bibitem{endal:76}
{Endal} AS, {Sofia} S.
\newblock {The evolution of rotating stars. I - Method and exploratory
  calculations for a 7-solar-mass star}.
\newblock \apj. 1976 Nov;210:184--198.

\bibitem{talon:07}
{Talon} S.
\newblock {Transport Processes in Stars: Diffusion, Rotation, Magnetic fields
  and Internal Waves}.
\newblock In: {Charbonnel} C, {Zahn} JP, editors. EAS Publications Series.
  vol.~32 of EAS Publications Series; 2008. p. 81--130.

\bibitem{palacios:13}
{Palacios} A.
\newblock {Influence of Rotation on Stellar Evolution}.
\newblock In: {Hennebelle} P, {Charbonnel} C, editors. EAS Publications Series.
  vol.~62 of EAS Publications Series; 2013. p. 227--287.

\bibitem{bruggen:01}
{Br{\"u}ggen} M, {Hillebrandt} W.
\newblock {Mixing through shear instabilities}.
\newblock \mnras. 2001 Jan;320:73--82.

\bibitem{palacios:06}
{Palacios} A, {Charbonnel} C, {Talon} S, {Siess} L.
\newblock {Rotational mixing in low-mass stars. II. Self-consistent models of
  Pop II RGB stars}.
\newblock \aap. 2006 Jul;453:261--278.

\bibitem{decressin09}
{Decressin} T, {Mathis} S, {Palacios} A, {Siess} L, {Talon} S, {Charbonnel} C,
  et~al.
\newblock {Diagnoses to unravel secular hydrodynamical processes in rotating
  main sequence stars}.
\newblock \aap. 2009 Feb;495:271--286.

\bibitem{marques13}
{Marques} JP, {Goupil} MJ.
\newblock {The Influence of Initial Conditions on Stellar Rotation History}.
\newblock In: {Goupil} M, {Belkacem} K, {Neiner} C, {Ligni{\`e}res} F, {Green}
  JJ, editors. Lecture Notes in Physics, Berlin Springer Verlag. vol. 865 of
  Lecture Notes in Physics, Berlin Springer Verlag; 2013. p.~75.

\bibitem{genevacode}
{Eggenberger} P, {Meynet} G, {Maeder} A, {Hirschi} R, {Charbonnel} C, {Talon}
  S, et~al.
\newblock {The Geneva stellar evolution code}.
\newblock \apss. 2008 Aug;316:43--54.

\bibitem{cl13}
{Chieffi} A, {Limongi} M.
\newblock {Pre-supernova Evolution of Rotating Solar Metallicity Stars in the
  Mass Range 13-120 M solar masses and their Explosive Yields}.
\newblock \apj. 2013 Feb;764:21.

\bibitem{mmproc:05}
{Meynet} G, {Maeder} A.
\newblock {Rotation and Mixing in Massive Stars: Principles and Uncertainties}.
\newblock In: {Ignace} R, {Gayley} KG, editors. The Nature and Evolution of
  Disks Around Hot Stars. vol. 337 of Astronomical Society of the Pacific
  Conference Series; 2005. p.~15.

\bibitem{Chaboyer1992}
{Chaboyer} B, {Zahn} JP.
\newblock {Effect of horizontal turbulent diffusion on transport by meridional
  circulation}.
\newblock \aap. 1992 Jan;253:173--177.

\bibitem{mz98}
{Maeder} A, {Zahn} JP.
\newblock {Stellar evolution with rotation. III. Meridional circulation with MU
  -gradients and non-stationarity}.
\newblock \aap. 1998 Jun;334:1000--1006.

\bibitem{palacios:09}
{Brun} AS, {Palacios} A.
\newblock {Numerical Simulations of a Rotating Red Giant Star. I.
  Three-dimensional Models of Turbulent Convection and Associated Mean Flows}.
\newblock \apj. 2009 Sep;702:1078--1097.

\bibitem{ekstroem12}
{Ekstr{\"o}m} S, {Georgy} C, {Eggenberger} P, {Meynet} G, {Mowlavi} N,
  {Wyttenbach} A, et~al.
\newblock {Grids of stellar models with rotation. I. Models from 0.8 to 120
  solar masses at solar metallicity (Z = 0.014)}.
\newblock \aap. 2012 Jan;537:A146.

\bibitem{maed97}
{Maeder} A.
\newblock {Stellar evolution with rotation. II. A new approach for shear
  mixing.}
\newblock \aap. 1997 May;321:134--144.

\bibitem{talzahb97}
{Talon} S, {Zahn} JP.
\newblock {Anisotropic diffusion and shear instabilities.}
\newblock \aap. 1997 Feb;317:749--751.

\bibitem{maed03}
{Maeder} A.
\newblock {Stellar rotation: Evidence for a large horizontal turbulence and its
  effects on evolution}.
\newblock \aap. 2003 Feb;399:263--269.

\bibitem{math:04}
{Mathis} S, {Palacios} A, {Zahn} JP.
\newblock {On shear-induced turbulence in rotating stars}.
\newblock \aap. 2004 Oct;425:243--247.

\bibitem{meynetproc:13}
{Meynet} G, {Ekstrom} S, {Maeder} A, {Eggenberger} P, {Saio} H, {Chomienne} V,
  et~al.
\newblock {Models of Rotating Massive Stars: Impacts of Various Prescriptions}.
\newblock In: {Goupil} M, {Belkacem} K, {Neiner} C, {Ligni{\`e}res} F, {Green}
  JJ, editors. Lecture Notes in Physics, Berlin Springer Verlag. vol. 865 of
  Lecture Notes in Physics, Berlin Springer Verlag; 2013. p.~3.

\bibitem{pins89}
{Pinsonneault} MH, {Kawaler} SD, {Sofia} S, {Demarque} P.
\newblock {Evolutionary models of the rotating sun}.
\newblock \apj. 1989 Mar;338:424--452.

\bibitem{Brott11a}
{Brott} I, {de Mink} SE, {Cantiello} M, {Langer} N, {de Koter} A, {Evans} CJ,
  et~al.
\newblock {Rotating massive main-sequence stars. I. Grids of evolutionary
  models and isochrones}.
\newblock \aap. 2011 Jun;530:A115.

\bibitem{choi:16}
{Choi} J, {Dotter} A, {Conroy} C, {Cantiello} M, {Paxton} B, {Johnson} BD.
\newblock {Mesa Isochrones and Stellar Tracks (MIST). I. Solar-scaled Models}.
\newblock \apj. 2016 Jun;823:102.

\bibitem{zahn:97}
{Zahn} JP, {Talon} S, {Matias} J.
\newblock {Angular momentum transport by internal waves in the solar interior.}
\newblock \aap. 1997 Jun;322:320--328.

\bibitem{talon:02}
{Talon} S, {Kumar} P, {Zahn} JP.
\newblock {Angular Momentum Extraction by Gravity Waves in the Sun}.
\newblock \apjl. 2002 Aug;574:L175--L178.

\bibitem{talon:05}
{Talon} S, {Charbonnel} C.
\newblock {Hydrodynamical stellar models including rotation, internal gravity
  waves, and atomic diffusion. I. Formalism and tests on Pop I dwarfs}.
\newblock \aap. 2005 Sep;440:981--994.

\bibitem{talon:08}
{Talon} S.
\newblock {Transport Processes in Stars: Diffusion, Rotation, Magnetic fields
  and Internal Waves}.
\newblock In: {Charbonnel} C, {Zahn} JP, editors. EAS Publications Series.
  vol.~32 of EAS Publications Series; 2008. p. 81--130.

\bibitem{golmurrkum:94}
{Goldreich} P, {Murray} N, {Kumar} P.
\newblock {Excitation of solar p-modes}.
\newblock \apj. 1994 Mar;424:466--479.

\bibitem{ct:05}
{Charbonnel} C, {Talon} S.
\newblock {Influence of Gravity Waves on the Internal Rotation and Li Abundance
  of Solar-Type Stars}.
\newblock Science. 2005 Sep;309:2189--2191.

\bibitem{beck:12}
{Beck} PG, {Montalban} J, {Kallinger} T, {De Ridder} J, {Aerts} C,
  {Garc{\'{\i}}a} RA, et~al.
\newblock {Fast core rotation in red-giant stars as revealed by
  gravity-dominated mixed modes}.
\newblock \nat. 2012 Jan;481:55--57.

\bibitem{deheuvels:12}
{Deheuvels} S, {Garc{\'{\i}}a} RA, {Chaplin} WJ, {Basu} S, {Antia} HM,
  {Appourchaux} T, et~al.
\newblock {Seismic Evidence for a Rapidly Rotating Core in a Lower-giant-branch
  Star Observed with Kepler}.
\newblock \apj. 2012 Sep;756:19.

\bibitem{rogers:13}
{Rogers} TM, {Lin} DNC, {McElwaine} JN, {Lau} HHB.
\newblock {Internal Gravity Waves in Massive Stars: Angular Momentum
  Transport}.
\newblock \apj. 2013 Jul;772:21.

\bibitem{alvan:14}
{Alvan} L, {Brun} AS, {Mathis} S.
\newblock {Theoretical seismology in 3D: nonlinear simulations of internal
  gravity waves in solar-like stars}.
\newblock \aap. 2014 May;565:A42.

\bibitem{spruit:02}
{Spruit} HC.
\newblock {Dynamo action by differential rotation in a stably stratified
  stellar interior}.
\newblock \aap. 2002 Jan;381:923--932.

\bibitem{zahn:07}
{Zahn} JP, {Brun} AS, {Mathis} S.
\newblock {On magnetic instabilities and dynamo action in stellar radiation
  zones}.
\newblock \aap. 2007 Oct;474:145--154.

\bibitem{petrovic:05}
{Petrovic} J, {Langer} N, {Yoon} SC, {Heger} A.
\newblock {Which massive stars are gamma-ray burst progenitors?}
\newblock \aap. 2005 May;435:247--259.

\bibitem{Heger:05}
{Heger} A, {Woosley} SE, {Spruit} HC.
\newblock {Presupernova Evolution of Differentially Rotating Massive Stars
  Including Magnetic Fields}.
\newblock \apj. 2005 Jun;626:350--363.

\bibitem{maedmeyn:05}
{Maeder} A, {Meynet} G.
\newblock {Stellar evolution with rotation and magnetic fields. III. The
  interplay of circulation and dynamo}.
\newblock \aap. 2005 Sep;440:1041--1049.

\bibitem{bouvier:97}
{Bouvier} J, {Forestini} M, {Allain} S.
\newblock {The angular momentum evolution of low-mass stars.}
\newblock \aap. 1997 Oct;326:1023--1043.

\bibitem{matt:12}
{Matt} SP, {MacGregor} KB, {Pinsonneault} MH, {Greene} TP.
\newblock {Magnetic Braking Formulation for Sun-like Stars: Dependence on
  Dipole Field Strength and Rotation Rate}.
\newblock \apjl. 2012 Aug;754:L26.

\bibitem{amard:16}
{Amard} L, {Palacios} A, {Charbonnel} C, {Gallet} F, {Bouvier} J.
\newblock {Rotating models of young solar-type stars. Exploring braking laws
  and angular momentum transport processes}.
\newblock \aap. 2016 Mar;587:A105.

\bibitem{kawaler:88}
{Kawaler} SD.
\newblock {Angular momentum loss in low-mass stars}.
\newblock \apj. 1988 Oct;333:236--247.

\bibitem{chaboyer:95}
{Chaboyer} B, {Demarque} P, {Pinsonneault} MH.
\newblock {Stellar models with microscopic diffusion and rotational mixing. 1:
  Application to the Sun}.
\newblock \apj. 1995 Mar;441:865--875.

\bibitem{boesgaard:86a}
{Boesgaard} AM, {Tripicco} MJ.
\newblock {Lithium in the Hyades Cluster}.
\newblock \apjl. 1986 Mar;302:L49--L53.

\bibitem{thorburn:93}
{Thorburn} JA, {Hobbs} LM, {Deliyannis} CP, {Pinsonneault} MH.
\newblock {Lithium in the Hyades. I - New observations}.
\newblock \apj. 1993 Sep;415:150--173.

\bibitem{talon:03}
{Talon} S, {Charbonnel} C.
\newblock {Angular momentum transport by internal gravity waves. I - Pop I main
  sequence stars}.
\newblock \aap. 2003 Jul;405:1025--1032.

\bibitem{michaud:91}
{Michaud} G, {Charbonneau} P.
\newblock {The lithium abundance in stars}.
\newblock \ssr. 1991 Jul;57:1--58.

\bibitem{talon:98}
{Talon} S, {Charbonnel} C.
\newblock {The Li dip: a probe of angular momentum transport in low mass
  stars}.
\newblock \aap. 1998 Jul;335:959--968.

\bibitem{mmlc:13}
{Maeder} A, {Meynet} G, {Lagarde} N, {Charbonnel} C.
\newblock {The thermohaline, Richardson, Rayleigh-Taylor, Solberg-H{\o}iland,
  and GSF criteria in rotating stars}.
\newblock \aap. 2013 May;553:A1.

\bibitem{viallet:13}
{Viallet} M, {Meakin} C, {Arnett} D, {Moc{\'a}k} M.
\newblock {Turbulent Convection in Stellar Interiors. III. Mean-field Analysis
  and Stratification Effects}.
\newblock \apj. 2013 May;769:1.

\bibitem{am:15}
{Arnett} WD, {Meakin} C, {Viallet} M, {Campbell} SW, {Lattanzio} JC,
  {Moc{\'a}k} M.
\newblock {Beyond Mixing-length Theory: A Step Toward 321D}.
\newblock \apj. 2015 Aug;809:30.

\bibitem{bossini:15}
{Bossini} D, {Miglio} A, {Salaris} M, {Pietrinferni} A, {Montalb{\'a}n} J,
  {Bressan} A, et~al.
\newblock {Uncertainties on near-core mixing in red-clump stars: effects on the
  period spacing and on the luminosity of the AGB bump}.
\newblock \mnras. 2015 Nov;453:2290--2301.

\bibitem{consta:15}
{Constantino} T, {Campbell} SW, {Christensen-Dalsgaard} J, {Lattanzio} JC,
  {Stello} D.
\newblock {The treatment of mixing in core helium burning models - I.
  Implications for asteroseismology}.
\newblock \mnras. 2015 Sep;452:123--145.

\bibitem{elm:16}
{Eggenberger} P, {Lagarde} N, {Miglio} A, {Montalb{\'a}n} J, {Ekstr{\"o}m} S,
  {Georgy} C, et~al.
\newblock {Models of rotating stars constrained by asteroseismic measurements
  of red giants}.
\newblock Astronomische Nachrichten. 2016 Sep;337:832.

\bibitem{elm:16b}
{Eggenberger} P, {Lagarde} N, {Miglio} A, {Montalb{\'a}n} J, {Ekstr{\"o}m} S,
  {Georgy} C, et~al.
\newblock {Constraining the efficiency of angular momentum transport with
  asteroseismology of red giants: the effect of stellar mass}.
\newblock ArXiv e-prints. 2016 Dec;.

\end{thebibliography}

%
%
%
%
%

\end{document}